\documentclass[11pt,a4paper]{article}
\usepackage{a4wide}
\usepackage{graphicx}
\usepackage{color}
\usepackage{latexsym}
\usepackage{amssymb}
\usepackage{amsthm}
\usepackage{amsmath}
\allowdisplaybreaks
\usepackage{url}
\newcommand{\dd}{~\mathrm{d}}
\newcommand{\sgn}{\mathrm{sgn}\,}

\newcommand{\transpose}{^{\top}}

\definecolor{lightgrey}{rgb}{0.8,0.8,0.8}
\theoremstyle{plain}
\newtheorem{proposition}{Proposition}
\theoremstyle{definition}
\newtheorem{definition}{Definition}
\begin{document}
\title{PDE Evolutions for M-Smoothers in One, Two, and Three Dimensions}

\author{Martin Welk\textsuperscript1 and Joachim Weickert\textsuperscript2
\\[2ex]
\textsuperscript1
Institute of Biomedical Image Analysis\\
UMIT -- Private University for Health Sciences, Medical Informatics
and Technology\\
Eduard-Walln{\"o}fer-Zentrum 1, 6060 Hall/Tyrol, Austria\\
Tel.: +43-50-8648-3974\\
\url{martin.welk@umit.at}\\[2ex]
\textsuperscript2
Mathematical Image Analysis Group\\
Faculty of Mathematics and Computer Science, Campus E1.7,\\
Saarland University,
66041 Saarbr\"ucken, Germany\\
Tel.: +49-681-30257340\\
\url{weickert@mia.uni-saarland.de}
}
\date{July 26, 2020}
\maketitle
\begin{abstract}
Local M-smoothers are interesting and important signal and image 
processing techniques with many connections to other methods. 
In our paper we derive a family of partial differential equations (PDEs) 
that result in one, two, and three dimensions as limiting processes 
from M-smoothers 
which are based on local order-$p$ means within a ball the radius of 
which tends to zero. The order $p$ may take any nonzero value $>-1$, 
allowing also negative values. In contrast to results from the 
literature, we show in the space-continuous case that mode filtering 
does not arise for $p \to 0$, but for $p \to -1$. 
Extending our filter class to $p$-values smaller than $-1$ allows to 
include e.g.\ the classical image sharpening flow of Gabor. The PDEs
we derive in 1D, 2D, and 3D show large structural similarities.\\
Since our PDE class is highly anisotropic and may contain backward
parabolic operators, designing adequate numerical methods is difficult.
We present an $L^\infty$-stable explicit finite difference scheme that
satisfies a discrete maximum--minimum principle, offers excellent 
rotation invariance, and employs a splitting into four fractional
steps to allow larger time step sizes. Although it approximates 
para\-bolic PDEs, it consequently benefits from stabilisation concepts 
from the numerics of hyperbolic PDEs.\\
Our 2D experiments show that the PDEs for $p<1$ are of specific interest: 
Their backward parabolic term creates favourable sharpening properties, 
while they appear to maintain the strong shape simplification properties 
of mean curvature motion.

\end{abstract}
\section{Introduction}
\label{sec-intro}

Partial differential equations (PDEs) constitute a natural framework to 
model processes in numerous real-world applications, ranging from physics
over life sciences to economy. Thus, it is not surprising that they
have also contributed substantially to the mathematical foundations
of signal and image analysis. For instance, they appear as
Euler--Lagrange equations when solving continuous optimisation
problems that result from variation models \cite{AK06,CS05a} or
regularisations of ill-posed problems \cite{BPT88}. It has also been
shown that they are the natural setting for scale-spaces \cite{Alvarez-ARMA93}, 
they are successfully used for image enhancement \cite{We97},
inpainting \cite{Sch15}, and image compression \cite{GWWB05}. PDE-based 
models benefit from many decades of research on their
theoretical foundations and efficient numerical algorithms. Since they
are continuous concepts, it is also very easy to incorporate useful 
invariances such as rotation invariance. 

One of the most fascinating aspects of PDE-based image analysis is
its capability to unify a number of existing methods in image
analysis. This has led to deeper structural insights as well as to 
novel algorithms. For instance, PDE formulations and connections to 
PDE-based image analysis are known for Gaussian smoothing \cite{Ii71a}, 
dilation and erosion \cite{Alvarez-ARMA93,AVK93,BM94,BS94}, 
morphological amoebas \cite{WBV11}, wavelet shrinkage \cite{WSMW05},
mean and median filtering \cite{Guichard-sana97}, and mode 
filtering \cite{Griffin-PRSLA00}.
 
Since mean, median and mode filtering are three representatives of 
M-smoothers based on local order-$p$ means, the question arises if
there is a more general PDE formulation that covers the full class of
local order-$p$ mean filtering in signal, image and volumetric data 
processing. This is the topic of our paper. Before we go more deeply 
into our contributions, let us first clarify in more detail 
the concept of M-estimators and M-smoothers based on order-$p$ means which we 
will consider in this work.

\paragraph{M-estimators.}
It has been observed long ago by Legendre \cite{Legendre-Book1805} and
Gau{\ss} \cite{Gauss-Book1809} that the mean
of a finite multiset $\mathcal{X}=\{a_1,a_2,\ldots,a_n\}$
of real numbers can be described as the minimiser of
the sum of squared distances to the given numbers:
\begin{align}
\operatorname{mean}(\mathcal{X}) 
&= \mathop{\operatorname{argmin}}\limits_{\mu\in\mathbb{R}}
\sum\nolimits_{i=1}^n (\mu-a_i)^2\;.
\end{align}
Likewise it has been 
proven by Fechner \cite{Fechner-AKSGW1878}
that the median of $\mathcal{X}$ minimises the
sum of absolute distances:
\begin{align}
\operatorname{median}(\mathcal{X}) 
&= \mathop{\operatorname{argmin}}\limits_{\mu\in\mathbb{R}}
\sum\nolimits_{i=1}^n \lvert \mu-a_i\rvert\;.
\end{align}
This can be generalised to the notion of
\emph{order-$p$ means} given by
\begin{align}
\operatorname{mean}_p(\mathcal{X}) 
&:= \mathop{\operatorname{argmin}}\limits_{\mu\in\mathbb{R}}
\sum\nolimits_{i=1}^n \lvert \mu-a_i\rvert^p
\label{pmean-finite}
\end{align}
for any $p>0$, with $\operatorname{mean}_2\equiv\operatorname{mean}$,
$\operatorname{mean}_1\equiv\operatorname{median}$.
After more restricted formulations by several predecessors, order-$p$
means for general real-valued $p>0$ were discussed by Barral Souto
\cite{BarralSouto-tr38}.
In robust statistics, order-$p$ means belong to the class of
\emph{M-estimators} \cite{Huber-Book81}.

Including the limiting case of the monomials as
\begin{equation}
\lvert z\rvert^0 = \begin{cases}0\;,&z=0\;,\\1\;,&z\ne0\end{cases}
\label{penaliser0}
\end{equation}
Barral Souto \cite{BarralSouto-tr38} also extends
the definition \eqref{pmean-finite} to the case $p=0$ for which the
\emph{mode} of $\mathcal{X}$, i.e., its most frequent value, is obtained.
As is also noted in \cite{BarralSouto-tr38}, the limit $p\to\infty$ yields
what is also called the mid-range value, i.e., the arithmetic mean of the
extremal values of $\mathcal{X}$.

\paragraph{Historical remarks.}
In fact, the concept of order-$p$ means has evolved in steps with increasing
generality over centuries, which we will briefly mention in the following.
The paper \cite{Armatte-JSFS06} was helpful in identifying some of these
steps, and provides some further information.

Consolidating a value from observations by minimisation of the sum
of absolute differences was proposed by Laplace in 1774 \cite{Laplace-Mem1774}; 
however, it seems to have been only with Fechner's 1878 proof 
\cite{Fechner-AKSGW1878} that the
connection to the median of discrete data was clearly established. 

Least-squares optimisation was introduced, and put into relation with
the arithmetic mean, by
Legendre (1805) \cite{Legendre-Book1805} and 
Gauss (1809) \cite{Gauss-Book1809}. 
However, already Gauss discussed in \cite[p.~221pp.]{Gauss-Book1809}
alternatives to least squares: on one hand, the use of even integer exponents
$p>2$, mentioning even the limit case $p\to\infty$; on the other hand, he also
made remarks about Laplace' idea of minimising the sum of absolute differences.
Later on, Ellis (1844) \cite{Ellis-TCPS1844} pointed out that quite general 
penalisers $\psi(\lvert\mu-a_i\rvert)$ could be used, thus actually 
proposing a fairly general class of M-estimators even beyond order-$p$ means.

Fechner (1878) \cite{Fechner-AKSGW1878} introduced the
family of order-$p$ means of discrete data with integer $p\ge0$, including 
the case $p=0$ yielding the mode of a discrete data set.
In 1921, Jackson \cite{Jackson-BAMS21} restated the minimising property of 
the median and introduced order-$p$ means with non-integer $p>1$, focusing on 
the use of the limit $p\to1^+$ as a means to disambiguate the median. Picking 
up Jackson's notion, Jordan (1927) \cite{Jordan-Book1927} stated the
mode as limit case for $p\to0^+$. 
Order-$p$ means with non-integer $p>0$ in their own right made their appearance
in 1938 with Barral Souto's paper \cite{BarralSouto-tr38}.

\paragraph{M-estimators, continuous case.}
It is straightforward to rewrite the definition of order-$p$ means for
continuous distributions (densities) on $\mathbb{R}$ just by replacing sums
with integrals:
Let $\gamma:\mathbb{R}\to\mathbb{R}^+_0$ be a density (integrable in a
suitable sense), then one defines
\begin{align}
\operatorname{mean}_p(\gamma)
&:= \mathop{\operatorname{argmin}}\limits_{\mu\in\mathbb{R}}
\int\nolimits_{-\infty}^{\infty} \gamma(z)\lvert\mu-z\rvert^p\,\mathrm{d}z \;.
\label{pmean-cont}
\end{align}
The general notion of continuous order-$p$ means was investigated in several
papers by Fr{\'e}chet. For $p\ge1$ it is mentioned in 1946 in 
\cite{Frechet-AULSA46} where, however, detailed discussion is restricted to
$p=1$ and $p=2$. A thorough treatment of $p\ge1$ is provided in 1948
in \cite{Frechet-AIHP48}, whereas \cite{Frechet-RIIS48,Frechet-ASENS48},
also from 1948, consider the general case $p>0$ including a thorough discussion
of the cases $p\to0$ and $p\to\infty$. In fact, Fr{\'e}chet analyses that
the mode is obtained in the limit $p\to0$
for discrete distributions (or such with a discrete component)
but not for purely continuous distributions.

The cases of the median and mean had been considered before:
In the case $p=1$, the continuous formulation goes back to 
Laplace (1774) \cite{Laplace-Mem1774} who, unlike in the discrete case,
also identified the minimiser as the median. 
Fr{\'e}chet analysed 
$p=2$ in \cite{Frechet-RIIS43} from 1943.

\paragraph{M-smoothers.}
In image processing, M-estimators are commonly used to build local
filters, see \cite{Tukey-Book71} for the median filter (in signal
processing) and \cite{Torroba-OE94} for order-$p$ means with $p>0$.
In a local filter, one takes at each location the greyvalues
from a neighbourhood (selection step) and computes some common value of
these (aggregation step) that is assigned to the location in the filtered
signal, see e.g.\ \cite{Chu-JASA98,Griffin-PRSLA00}. These filters can be
iterated to generate a series of progressively processed images.

It has been noticed since long that some of these filters behave similar 
to certain image filters based on PDEs.
Mean filters are a spatial discretisation of linear diffusion.
Guichard and Morel \cite{Guichard-sana97} have proven that iterated median 
filtering approximates mean curvature motion \cite{Br78}. 
To this end, they consider a space-continuous version of median filtering,
in which the selection step is based on a disc-shaped neighbourhood.
Sending the radius of the neighbourhood to zero, they show that the 
effect of the median filtering step becomes asymptotically equal to a time
step of an explicit time discretisation of the mean curvature motion PDE.

Griffin \cite{Griffin-PRSLA00} proves similar results for three different
filters, and puts them in the context of order-$p$ means. In addition to the 
median ($p=1$) and the arithmetic mean ($p=2$) he considers for the first time
the mode filter (associating it with $p=0$). 
In contrast to \cite{Guichard-sana97},
the selection step in \cite{Griffin-PRSLA00} is based on a Gaussian window,
i.e., the input value density of the respective means is made up by the values
from the entire image plane but reweighted with a Gaussian function. 
The limit case is constituted by the standard deviation of the
Gaussian window approaching zero. 
In this framework, the mean curvature motion PDE is re-derived as the limit
case of median filtering. For arithmetic mean filtering, the linear diffusion
PDE is obtained. For mode filtering, a PDE is derived that combines
mean curvature motion (diffusion along level sets) with backward diffusion in 
gradient flowline direction, compare Proposition~\ref{prop-pde-gmdisc-mode}
below.

\paragraph{Our contributions.}
The goal of our paper is to complete this picture by deriving the
PDE limit for arbitrary order-$p$ means and introducing a suitable 
numerical algorithm.
Up to a time rescaling, the PDE limits of all three cases of 
\cite{Griffin-PRSLA00} will be contained in our results in the following. 
The reason for the time rescaling is that we use for the selection step 
disc-shaped 
neighbourhoods such as in \cite{Guichard-sana97}. With this choice we aim
at modelling the space-continuous filter in an analogy as close as possible
to the usual setup of discrete local signal and image filters.

Our paper is based on the conference publication \cite{Welk-ssvm19}.
However, these results are presented in more detail and substantially 
extended, covering now also the 1D and 3D setting. We also propose a
novel splitting-based numerical algorithm with improved efficiency 
and better rotation invariance. 

Starting with the case of planar grey-value images,
we derive a family of PDEs associated with M-smoothers based on
order-$p$ means with variable $p$ and vanishing disc radius. In contrast
to results from the literature, we also permit negative $p$-values with
$p>-1$. Compared to \cite{Welk-ssvm19}, the proof of this approximation
result is presented in a more detailed form. Moreover, we discuss
the behaviour near critical points (saddle points and extrema) and critical
curves. We also analyse the effect of staircasing. Using the calculus of
distributions, we can show that the PDEs derived for smooth images remain
valid for step functions.

Reconsidering the relation of order-$p$ means and their corresponding
PDEs to existing image filters, we show that in the space-continuous
setting the mode filter does \emph{not} arise for $p\to0$, as is commonly
assumed \cite{Griffin-PRSLA00} 
(despite the analysis in 
\cite{Frechet-RIIS48,Frechet-ASENS48}), 
but for $p\to-1$. Since the common
assumption countered by our analysis is derived by analogy from discrete
theory, we discuss in this paper also where and why this analogy fails.

In the present work, we extend our results also to 1D signals and
3D grey-value images. 
Table~\ref{tab-pde} at the end of Section~\ref{sec-pde3d} summarises the
PDE approximation results obtained in one, two, and three dimensions.
The PDEs approximated for $p>-1$ are in full analogy
to the 2D case. This is also the case for the mode filter as limiting case
for $p\to-1$ in three dimensions. For 1D signals, the limit $p\to-1$ is
non-uniform, and the mode filter approximates a shock-filter PDE which
was already stated in earlier work \cite{WWG07}.
Continuing our PDE
family to values $p<-1$ allows to cover also the sharpening Gabor flow
\cite{Gabor-LI65,Lindenbaum-PR94}, for which no M-smoothing counterpart
is known. 

In spite of the fact that our PDE family is anisotropic and
may even involve backward parabolic operators, we design an
$L^\infty$-stable numerical scheme that enjoys excellent rotation
invariance and employs operator splitting to improve its efficiency.
Our experiments show that the PDEs for $p<1$ are particularly attractive
since they simultaneously allow image sharpening and shape simplification.

\paragraph{Structure of the paper.} In Section~\ref{sec-pde2d}
we present our theory
that allows us to derive PDE evolutions from M-smoothers, and discuss in 
detail important aspects of the PDE limit such as its behaviour near critical
points and staircasing.
Section~\ref{sec-pde1d} presents the analogous results for 1D signals, 
whereas Section~\ref{sec-pde3d} covers 3D images.
In Section~\ref{sec-disccont} we discuss the relation between discrete and 
continuous M-smoothing and explain why the discrete result about approximation
of the mode filter for $p\to0$ cannot be transferred to the continuous
situation. Our numerical algorithm is discussed in Section~\ref{sec-num}, 
and Section~\ref{sec-exp} is devoted to an experimental evaluation.
The paper is concluded with a summary in Section~\ref{sec-conc}.
Two appendices provide additional material: The detailed proofs of the 
results in Sections~\ref{sec-pde2d}--\ref{sec-pde3d} are collected in 
Appendix~\ref{app-proofs}, whereas Appendix~\ref{app-toy} presents an 
illustrative example to support the discussion in Section~\ref{sec-disccont}.

\section{M-Smoothers, Mode and Partial Differential Equations for 2D Images}
\label{sec-pde2d}

In this section, we derive PDEs for M-smoothers and the mode filter in the
case of 2D greyvalue images, and discuss some of their properties.

\subsection{Generalised Order-$p$ Means}

In the following, M-smoothers are
based on order-$p$ means with $p>-1$, $p\ne0$. As this range for $p$
goes beyond the usual $p>0$, let us first extend the definition of
order-$p$ means of continuous-scale distributions accordingly.

\begin{definition}
\label{def-pmean}
Let $z$ be a real random variable with the bounded, piecewise continuous
density $\gamma:\mathbb{R}\to\mathbb{R}$.
For $p\in(-1,+\infty)\setminus\{0\}$, define the \emph{order-$p$ mean} of
$\gamma$ as
\begin{equation}
\operatorname{mean}_p(\gamma) =
\mathop{\operatorname{argmin}}
\limits_{\mu\in\mathbb{R}}\int\limits_{\mathbb{R}}
\gamma(z) \, \operatorname{sgn}(p) \lvert \mu-z\rvert^p \,\mathrm{d}z \;.
\label{eq-defpmean}
\end{equation}
\end{definition}

As $\lvert z\rvert^p$ is monotonically increasing on
$\mathbb{R}^+_0$ (the set of all nonnegative real numbers) for $p>0$, but 
monotonically decreasing on $\mathbb{R}^+$ (the set of all positive real
numbers) for
$p<0$, the $\operatorname{sgn}(p)$ factor in~\eqref{eq-defpmean} ensures
that in both cases an increasing penalty function is used.

For $p>0$ the requirement of continuity of $\gamma$ in Def.~\ref{def-pmean}
can be relaxed;
by modelling a discrete density as a weighted sum of delta peaks, the
discrete order-$p$ means as in \cite{BarralSouto-tr38} can be
included in this definition.

The continuity is, however, essential for $p<0$:
In this case, the penalty function has a pole at
$z=0$ such that an improper integral is obtained; for $p>-1$ this integral
exists provided that $\gamma$ is continuous, i.e., no delta peaks are
allowed. In particular, we cannot define an order-$p$ mean with $-1<p<0$
for discrete distributions as considered in \cite{BarralSouto-tr38}.

\subsection{Infinitesimal Limits of M-Smoothers}
\label{ssec-pde2d}

We turn now to derive partial differential equations approximated by
M-smoothers applied to 2D images.
The proofs of the following propositions are given in 
Appendix~\ref{app-proofs}.
When speaking of smooth functions we always mean $\mathrm{C}^\infty$ functions
although a weaker hypothesis could be sufficient for some results.
The first proposition contains our first main result.

\begin{proposition}[2D PDE limit for $p>-1$]\label{prop-pde-gmdisc-generic}
Let a smooth image $u:\mathbb{R}^2\to\mathbb{R}$ be given, and let
$\boldsymbol{x}_0=(x_0,y_0)$ be a regular point,
$\lvert\boldsymbol{\nabla}u(\boldsymbol{x}_0)\rvert>0$.
One step of order-$p$ mean filtering of $u$ with a disc-shaped window
$\mathrm{D}_\varrho(\boldsymbol{x}_0)$ and $p>-1$, $p\ne0$
approximates for $\varrho\to0$
a time step of size $\tau=\varrho^2/(2p+4)$ of an
explicit time discretisation of the PDE
\begin{equation}
u_t = u_{\xi\xi} + (p-1)\,u_{\eta\eta}
\label{pde-p}
\end{equation}
where $\eta$ and $\xi$ are geometric coordinates referring at each image
location to the direction of the positive gradient, and the level-line
direction, respectively:
\begin{align}
\kern2em&\kern-2em
\operatorname{mean}_p\{u(x,y)~|~(x,y)\in\mathrm{D}_{\varrho}(x_0,y_0)\} 
- u(x_0,y_0)
\notag \\*
&= \frac{\varrho^2}{2(p+2)} 
\bigl( u_{\xi\xi}(x_0,y_0) + (p-1) u_{\eta\eta}(x_0,y_0) \bigr)
+ \mathcal{O}(\varrho^{(\min\{p,0\}+5)/2})\;.
\label{pde-gmdisc-generic}
\end{align}
At a local minimum (maximum) of $u$, i.e., $\boldsymbol{x}_0$ with
$\lvert\boldsymbol{\nabla}u(\boldsymbol{x}_0)\rvert=0$ where the Hessian
$\mathbf{D}^2u(\boldsymbol{x}_0)$ is positive (negative) semidefinite,
the same filtering step fulfils for $\varrho\to0$ the inequality
$\operatorname{mean}_p\{u(x,y)~|~(x,y)\in\mathrm{D}_{\varrho}(x_0,y_0)\} 
- u(x_0,y_0) \ge0$ ($\le 0$), thus approximates an evolution $u_t\ge0$
($u_t\le0$).
\end{proposition}
The approximation order in \eqref{pde-gmdisc-generic} is
$\mathcal{O}(\varrho^{1/2})$ for positive $p$
but reduces to $\mathcal{O}(\varrho^{(p+1)/2})$ for negative $p$.

For $p=2$ and $p=1$ the proposition yields the same
PDEs as \cite{Griffin-PRSLA00}
except for a time rescaling which is due to the choice of a Gaussian window
in \cite{Griffin-PRSLA00}.
\medskip

Under analogous assumptions as in Prop.~\ref{prop-pde-gmdisc-generic},
one can also derive the PDE limit for the
mode filter, where the mode is not obtained by a minimisation in the
sense of \eqref{pmean-cont} but directly as the maximum of the density
of values in $\{u(x,y)~|~(x,y)\in\mathrm{D}_{\varrho}(x_0,y_0)\}$.

\begin{proposition}[2D PDE limit for mode filtering]
\label{prop-pde-gmdisc-mode}
Let $u$ and $\boldsymbol{x}_0$ be as in
Proposition~\ref{prop-pde-gmdisc-generic}.
One step of mode filtering of $u$ with a disc-shaped window
$\mathrm{D}_\varrho(\boldsymbol{x}_0)$
approximates for $\varrho\to0$
a time step of size $\tau=\varrho^2/2$ of an
explicit time discretisation of the PDE $u_t=u_{\xi\xi}-2u_{\eta\eta}$
with $\eta$, $\xi$ as in Proposition~\ref{prop-pde-gmdisc-generic}.
At a local minimum (maximum), mode filtering approximates
$u_t\ge0$ ($u_t\le0$).
\end{proposition}

The PDE for mode filtering coincides with the one given
in \cite{Griffin-PRSLA00}, again up to time rescaling.
We see, however,
that \eqref{pde-gmdisc-generic} for $p\to0$ does not yield the
PDE from Proposition~\ref{prop-pde-gmdisc-mode}
but $u_t=u_{\xi\xi}-u_{\eta\eta}$. Instead, the mode filtering PDE
is obtained for $p\to-1$.
Inserting $p=-2$ into \eqref{pde-gmdisc-generic} yields
$u_t=u_{\xi\xi}-3u_{\eta\eta}$ which was stated as an image sharpening
PDE that has been proposed by Gabor already in 1965 
\cite{Gabor-LI65,Lindenbaum-PR94}.

Remarkably, the PDEs for mean ($p=2$), median ($p=1$) and mode ($p=-1$) also
match the often-stated empirical rule noted first by 
Pearson \cite[p.~376]{Pearson-PRSLA1895} according to which the median in a
large class of skew densities is located at two-thirds the way between mode
and mean (which, however, is not a general law).

\subsection{Discussion of PDE Evolutions Near Critical Points}

Propositions~\ref{prop-pde-gmdisc-generic} and \ref{prop-pde-gmdisc-mode}
state PDEs approximated by the respective M-smoothers at regular points,
and inequalities that hold at local minima and maxima. Let us briefly
discuss how these results determine uniquely the evolutions of the entire
image $u$ (including critical points) approximated by the M-smoothers.

\subsubsection{Regions of Critical Points}
\label{sssec-plateau}

Let us consider first the case of a connected critical region, i.e., a closed
set in $\mathbb{R}^2$ consisting entirely of critical points, with
nonempty interior. In such a region, the inequalities for minima and maxima 
together imply $u_t=0$, which is also consistent with the obvious limit
of any M-smoother in all interior points of the region.

\subsubsection{Isolated Critical Points}
\label{sssec-icp}

Let us now consider the case of an isolated critical point, i.e., a
point $\boldsymbol{x}_0$ with 
$\boldsymbol{\nabla}u(\boldsymbol{x}_0)=\boldsymbol{0}$
but $\boldsymbol{\nabla}u(\boldsymbol{x})\ne\boldsymbol{0}$ for all
other points $\boldsymbol{x}$ within an open neighbourhood of 
$\boldsymbol{x}_0$.

A direct calculation of the limit for vanishing window size of
an M-smoother at $\boldsymbol{x}_0$ would suggest an approximation
that differs substantially from that in regular points. 
We will argue in the following that this naive limit is irrelevant for the
time-continuous image evolution approximated by iterated M-smoothing.

On one hand, limit calculations at regular points 
(see the proofs in Appendix~\ref{app-proofs})
require a neighbourhood that contains no critical points at all.
Thus, on approaching a critical point of $u$, the admissible neighbourhood
radius $\varrho$ around regular points tends to zero.
Therefore, the PDE limit 
within any open region of the plane that does not contain critical points
is not uniform if the boundary of that region contains a critical point. The 
result of the naive application of the same limit procedure at a critical point
can thus not be expected to fit smoothly into the evolution of the regular
points around.

On the other hand, for an initial-boundary value problem describing an image
evolution, it is in general sufficient for the PDE to be prescribed everywhere
except at isolated points. Assuming viscosity solutions as a solution concept,
the solution of the initial-boundary value problem will
fill in the evolution at the exceptional points.

Revisiting the evolution from Proposition~\ref{prop-pde-gmdisc-generic},
we notice first that the PDE \eqref{pde-p} in regular points of $u$
can be rewritten with
the Laplacian $\Delta u = u_{xx}+u_{yy}=u_{\xi\xi}+u_{\eta\eta}$ as
\begin{align}
u_t &= (2-p) u_{\xi\xi} 
+ (p-1) \Delta u 
\;,
\end{align}
a linear combination of homogeneous diffusion and
curvature motion.
Given the smoothness of $u$, the diffusion term $u_{xx}+u_{yy}$ can obviously
be continued smoothly to isolated critical points.

The curvature motion term is more difficult. At an \emph{isotropic critical
point}, i.e., $\boldsymbol{x}_0=(x_0,y_0)$ with
$u_x(\boldsymbol{x}_0)=u_y(\boldsymbol{x}_0)=0$,
$u_{xx}(\boldsymbol{x}_0)=u_{yy}(\boldsymbol{x}_0)$,
$u_{xy}(\boldsymbol{x}_0)=0$,
also this term has a unique limit for $(x,y)\to(x_0,y_0)$, namely
$u_{xx}(\boldsymbol{x}_0)$.
In contrast, when approaching an \emph{anisotropic} critical
point $(x_0,y_0)$ (where the Hessian $\mathrm{D}^2u$ is not a multiple 
of the unit matrix) from different directions,
one obviously obtains different limits such that no unique value can be filled
in at this critical point.

To understand the effect of the evolution near a critical point, assume that
$\boldsymbol{x}_0=\boldsymbol{0}$ is a local minimum of $u$.
For simplicity, we neglect higher order terms of the Taylor expansion and
assume at a given time
$u(x,y)=\beta x^2+\delta y^2$ with $\beta\ge\delta>0$
in a neighbourhood $N$
of $\boldsymbol{x}_0$. For $\boldsymbol{x}=(x,y)\transpose\in N$ one has 
then
\begin{gather}
\boldsymbol{\nabla}u(\boldsymbol{x})=2(\beta x,\delta y)\transpose\;,\quad
\mathrm{D}^2u(\boldsymbol{x})=\begin{pmatrix}2\beta&0\\0&2\delta\end{pmatrix}
\;,\\
\eta = \frac{(\beta x,\delta y)\transpose}{\sqrt{\beta^2x^2+\delta^2y^2}}\;,
\quad
\xi = \frac{(-\delta x,\beta y)\transpose}{\sqrt{\beta^2x^2+\delta^2y^2}}\;,
\\
u_{\eta\eta} = \frac{2\beta^3x^2+2\delta^3y^2}{\beta^2x^2+\delta^2y^2}\;,\\
u_{\xi\xi} = \frac{2\beta^2\delta x^2+2\beta\delta^2y^2}
{\beta^2x^2+\delta^2y^2}\;.
\end{gather}
Both $u_{\xi\xi}$ and $u_{\eta\eta}$, and thus also 
$u_t=u_{\xi\xi}+(p-1)u_{\eta\eta}$, are constant along radial lines through 
$\boldsymbol{0}$. 

For an isotropic minimum ($\beta=\delta>0$), the evolution 
speed is uniform in the neighbourhood $N$, ensuring that the isotropy of
the minimum is preserved during evolution. In particular, for $p\ge0$, 
this evolution speed is positive such that the inequality 
$u_t(\boldsymbol{x}_0)\ge0$ is automatically preserved.

For an anisotropic minimum ($\beta>\delta>0$), the evolution speeds $u_t$
along different radial lines differ, with $u_t(z,0)<u_t(0,z)$, such that
the anisotropy is reduced by the evolution. For $p\ge1-\delta/\beta$ 
all evolution speeds are nonnegative, so $u_t(\boldsymbol{x}_0)\ge0$ is still
automatically satisfied, and the anisotropic minimum is converted into
an isotropic minimum by the evolution. The position of the minimum can move
due to the evolution.

For $1-\beta/\delta<p<1-\delta/\beta$, one has $u_t(z,0)<0<u_t(0,z)$. 
In this case, the inequality $u_t(\boldsymbol{x}_0)\ge0$ acts to constrain 
the evolution near the $x$ axis, implying the immediate formation of a 
critical line (see Section~\ref{sssec-cc}) or plateau 
(see Section~\ref{sssec-plateau}) around $\boldsymbol{x}_0$.

If, finally, $p<1-\beta/\delta$, the speed $u_t$ is negative throughout $N$,
which implies that a critical plateau is formed immediately.

Analogous considerations apply to local maxima. Finally, a saddle point
(with indefinite Hessian) remains a saddle point, and as such is well
constrained by the surrounding regular points from above and below.
Therefore the evolution at isolated critical points for $p\ge1$ is fully
determined by filling in the evolution from the surrounding regular points,
whereas for $-1<p<1$ it is fully determined by filling in combined with the
inequality constraints $u_t\ge0$ at minima and $u_t\le0$ at maxima.

\subsubsection{Critical Curves}
\label{sssec-cc}

The considerations from Section~\ref{sssec-icp} can be extended
to regular curves consisting of critical points. If such a curve is formed
by local minima, any point on this curve is a maximally anisotropic local
minimum, $\beta>\delta=0$, yielding evolution speeds (up to higher order
terms) $u_t\approx2(p-1)\beta$ for nearby regular points. 

If $p>1$, one has $u_t>0$, thus the differential inequality at critical
points is automatically satisfied. The critical curve is preserved as a
critical curve or may be broken up into isolated critical points.

If $p<1$, the regular points in $N$ evolve with $u_t>0$, which leads to
an immediate expansion of the critical curve into a plateau.

\subsection{Staircasing and Analysis of the PDE for Step Functions}

For $p<1$, the PDE \eqref{pde-p} involves a backward parabolic term in 
gradient flowline direction. In evolutions of this kind staircasing effects
are common, i.e., the evolving function turns into a step
function which is only piecewise smooth with jumps between the smooth
segments. Indeed, staircasing can also observed in numerical experiments with 
\eqref{pde-p}. Unfortunately, the staircasing undermines the smoothness
assumption underlying the approximation result of 
Proposition~\ref{prop-pde-gmdisc-generic}. 
Therefore we dedicate this section to discuss how our approximation results
extend to the situation of step functions. First, we will use the
calculus of distributions to generalise the image filtering PDE \eqref{pde-p}.
Afterwards we will discuss order-$p$ mean M-smoothers for step functions.
Although we do not possess, at the time being, a full asymptotic analysis
of this case, we will consider a simplified case and demonstrate by a 
combination of analytic and numeric evidence that the behaviour of M-smoothers
is still comparable to that of the PDE.

\subsubsection{Distributional Analysis}
\label{sssec-distrib}

In the following, we will consider piecewise smooth step functions, i.e.,
functions over $\mathbb{R}^n$ which are smooth except on a set of smooth 
hypersurfaces that decompose the space $\mathbb{R}^n$ into connected segments
$\varOmega_1,\ldots,\varOmega_k$.

A natural way to analyse the effect of the PDE evolution \eqref{pde-p}
on a step function $u$ is to apply the PDE to a series of smoothed 
functions that converge to $u$, and consider the limit of the so obtained 
evolutions. 
For example, $u$ could be convolved with Gaussians $G_\sigma$ of decreasing
standard deviation $\sigma$, yielding the desired result for $\sigma\to0$.
As $G_\sigma$ for $\sigma\to0$ weakly converges to a Dirac delta distribution,
the calculus of distributions \cite{Vladimirov-Book79en,Vladimirov-Book67ge} 
allows to calculate
the evolution of interest in a more compact form without explicitly
carrying out the limiting procedure. Technically, a distribution, or 
generalised function, $f\in\mathcal{D}'(\mathbb{R}^n)$ is a functional
that acts on smooth basic functions $\varphi$ by the scalar product of
functions 
$(f,\varphi):=
\int\nolimits_{\mathbb{R}^n}f(\boldsymbol{x})\varphi(\boldsymbol{x})
\,\mathrm{d}\boldsymbol{x}$.

Assuming that the jump set of $f$ consists of just one smooth
hypersurface $S$ dividing $\mathbb{R}^n$ into domains $\varOmega_1$ and 
$\varOmega_2$, we notice that $f$ is differentiable in distributional sense,
and we recall the following essential formula from 
\cite[II, \S6]{Vladimirov-Book67ge}:
\begin{align}
\frac{\partial f}{\partial x_i} &= 
\left\{\frac{\partial f}{\partial x_i}\right\}
+[f]_S\cos(\boldsymbol{n}_S,x_i)\delta_S
\label{d-distr}
\end{align}
for $i=1,\ldots,n$.
Here, $\{\partial f/\partial x_i\}$ denotes the \emph{regular part} of
the derivative, i.e., essentially an ordinary function.
The vector
$\boldsymbol{n}_S$ is the unit outer normal vector of $S$ at a given point,
and $[f]_S$ the jump height of $u$ at this point in direction 
$\boldsymbol{n}_S$.
The \emph{single-layer distribution} $\delta_S$ is a generalisation of the
one-dimensional Dirac delta distribution, behaving like the delta distribution
on crossing the hypersurface $S$ in normal direction.

Furthermore, second derivatives of $f$ can be written as
\cite[II, \S6]{Vladimirov-Book67ge}
\begin{align}
\frac{\partial^2 f}{\partial x_i\partial x_j} &=
\left\{\frac{\partial^2 f}{\partial x_i\partial x_j} \right\}
+ \frac{\partial}{\partial x_j}
\bigl([f]_S\cos(\boldsymbol{n}_S,x_i)\delta_S\bigr)
+ \left[\left\{\frac{\partial f}{\partial x_i}\right\}\right]_S
\cos(\boldsymbol{n}_S,x_j)\delta_S
\label{dd-distr}
\end{align}
for $i,j=1,\ldots,n$
where in the second summand a \emph{double-layer distribution} occurs as the
derivative of a single-layer distribution; 
$\partial\delta_S/\partial\boldsymbol{n}_S$ is a generalisation of
the derivative of the one-dimensional delta distribution, behaving like
$\delta'$ on crossing $S$ in normal direction.

To analyse our example, we consider the evolution \eqref{pde-p} with a
step function $u_0$ as initial condition. The evolution $u$ will be
described by a function $u$ over $\mathbb{R}^2\times[0,T]$ which is
smooth except on a jump set consisting of regular surfaces. Outside
the jump set, our previous analysis applies. The case of interest is
therefore a point $(x,y,t)$ on a jump surface $S$.
Simplifying further, we assume that the normal vector $\boldsymbol{n}_S$
at $(x,y,t)$ is in the $x$-$t$ plane and in positive $x$ direction, i.e.,
\begin{align}
\boldsymbol{n}_S &= \frac{1}{\sqrt{1+v^2}}\bigl(1, 0, -v)\transpose
\end{align}
where $v$ is the speed at which the jump moves in $x$ direction as the
time $t$ progresses.
Then \eqref{pde-p} becomes $u_t=u_{yy}+(p-1)u_{xx}$.
From \eqref{d-distr} with $\cos(\boldsymbol{n}_S,t)=-v/\sqrt{1+v^2}$
we calculate
\begin{align}
u_t (x,y,t) &= \{u_t(x,y,t)\}
[u]_S(x,y,t)\frac{-v}{\sqrt{1+v^2}} \delta_S \;.
\label{ut-distr}
\end{align}
Similarly, we obtain from \eqref{dd-distr} with
$\cos(\boldsymbol{n}_S,y)=0$ and\\
$\partial_y\cos(\boldsymbol{n}_S,y)=-\sin(\boldsymbol{n}_S,y)
\cos(\partial_y\boldsymbol{n}_S,y)=-\kappa_S/\sqrt{1+v^2}$
\begin{align}
u_{yy}(x,y,t) &= \{u_{yy}(x,y,t)\} 
-\frac{\partial}{\partial y}
\bigl([u]_S(x,y,t)\cos(\boldsymbol{n}_S,y)\delta_S\bigr)
\notag\\
&=\{u_{yy}(x,y,t)\} 
+ [u]_S(x,y,t)\frac{\kappa_S(x,y,t)}{\sqrt{1+v^2}}\delta_S
\label{uyy-distr}
\end{align}
where $\kappa_S(x,y,t)$ denotes the curvature of $S$ in the $x$-$y$ plane
at $(x,y,t)$. Finally,
\begin{align}
u_{xx}(x,y,t) &= \{u_{xx}(x,y,t)\}
+\frac1{\sqrt{1+v^2}}[u]_S\frac{\partial}{\partial x}\delta_S
+\frac1{\sqrt{1+v^2}}[\{u_x\}]_S\delta_S
\;.
\label{uxx-distr}
\end{align}
Inserting \eqref{ut-distr}, \eqref{uyy-distr} and \eqref{uxx-distr}
into \eqref{pde-p}, 
we have by equating the $\delta_S$ contributions
\begin{align}
v &= -\kappa_S - (p-1) \frac{[\{u_x\}]_S}{[u]_S}\,,
\label{step-speed}
\end{align}
which describes the speed at which 
the interface between the two smooth segments of $u$ moves in $x$ direction.
The contribution $-\kappa_S$ is in full agreement with the behaviour
of the mean curvature motion part of \eqref{pde-p} for smooth functions,
whereas $-(p-1)[\{u_x\}]_S/[u]_S$ modifies this speed by accelerating or
slowing down the inward motion of the interface depending on $p$ and whether
the gradient of $u$ is greater or smaller on the outside or inside of the
evolving interface. If $p<1$, the effect is to push the evolution towards
increasing contrast at the interface, thus encouraging staircasing.
For $p>1$, the evolution is biased towards reducing contrast at the interface,
thus counteracting staircasing.

We notice finally that on the right hand side of \eqref{uxx-distr} also
a double-layer term $\partial_x\delta_S$ appears. When equating with 
the first-order time derivative $u_t$ in \eqref{pde-p}, this term has
no implication for the evolution of $u$ as time integration across $S$
integrates it to zero. However, it indicates that no well-defined regular
function value can be assigned to $u$ on the interface $S$ itself.

\subsubsection{M-Smoothing a Step Function}
\label{sssec-step-pmean}

For a full theoretical analysis of an M-smoothing step with order-$p$ means
for step functions, the limit $\varrho\to0$ of $\operatorname{mean}_p(u)-u$
at a fixed location of a given step function is of little help
because it just reproduces the result for smooth functions everywhere outside
the jump set, and isolated values on the jump set itself are meaningless.
To validate \eqref{step-speed} in the asymptotic case, it would be necessary
instead to determine the displacement of the interface itself by one 
M-smoothing step for positive $\varrho$, and consider the asymptotic 
behaviour of this displacement. This appears substantially more complicated
than the proof of Proposition~\ref{prop-pde-gmdisc-generic}, and no analysis
of this kind is available at the moment.

In the following, we study instead a simple case of a step function for
fixed $\varrho$ by combining analytical with numerical arguments.
Let the step function $u:\mathbb{R}^2\to\mathbb{R}$ be given as
\begin{align}
u(x,y) &= \alpha (x + \delta y^2) + h \mathbb{1}(x+\delta y^2+\theta>0)
\label{stepfn-u}
\end{align}
where $\mathbb{1}(x+\delta y^2+\theta>0)$ is the Heaviside function of
$x+\delta y^2+\theta$. The jump set of $u$ is the parabola 
$x+\delta y^2=-\theta$, and we have $u_x=\alpha$, $u_{xx}=u_{xy}=0$,
$u_{yy}=\tfrac12\alpha\delta$ everywhere outside the jump set.

Let $\mu$ be the order-$p$ mean of $u$ within the disc $\mathrm{D}_\varrho$
centred at $(0,0)$, for some $p\in(-1,0)$. 
We have
\begin{align}
\mu &= \mathop{\operatorname{argmin}}\limits_{\mu}
\operatorname{sgn}(p)\iint\nolimits_{\mathrm{D}_\varrho}
\lvert\mu-u(x,y)\rvert^p\,\mathrm{d}x\,\mathrm{d}y
\end{align}
which can be simplified to
\begin{align}
\mu 
&=
\mathop{\operatorname{argmin}}\limits_{\mu}
\operatorname{sgn}(p)
\left(
\int\nolimits_{-\varrho}^{-\theta}
y^*(x)\,\lvert\mu-\alpha x\rvert^p\,\mathrm{d}x
+
\int\nolimits_{-\theta}^{\varrho}
y^*(x)\,\lvert\mu-\alpha x - h\rvert^p\,\mathrm{d}x
\right)
\label{mu-step-theta}
\end{align}
where (for sufficiently small $\delta$)
\begin{align}
y^*(x) &= \sqrt{\varrho^2 - \frac1{4\delta^2}\left(
1-\sqrt{1-4\delta x+4\delta^2\varrho^2}\right)^2}
\end{align}
assigns to each $x\in[-\varrho,\varrho]$ the $y$ coordinate of the two
points $\bigl(x,\pm y^*(x)\bigr)$ where the level line of $u$ going
through $(x,0)$ hits the boundary of $\mathrm{D}_\varrho$.

Equation~\eqref{step-speed} together with the step size $\tau=\varrho^2/(2p+4)$
from Proposition~\ref{prop-pde-gmdisc-generic} suggests that $\mu$ should
be in the range of the lower part of $u$ (i.e., $\mu\approx0$) for 
$\theta < -\delta/(p+2)$ and in the upper part of $u$ (i.e., $\mu\approx h$)
for $\theta > -\delta/(p+2)$.

To check this numerically, we fixed $\varrho=1$ and 
evaluated \eqref{mu-step-theta} for a set of randomly chosen values of
$p\in[-0.99,-0.1]$, $\alpha\in[0.03,0.15]$, $\delta\in[-0.2,0.2]$ and for
different jump heights $h=0.01$, $h=0.1$, $h=1$. 

\begin{figure}[t]
\unitlength0.005\textwidth
\begin{picture}(200,70)
\put( 0, 0){\raisebox{70\unitlength}{\rotatebox{270}{%
\includegraphics[height=100\unitlength]{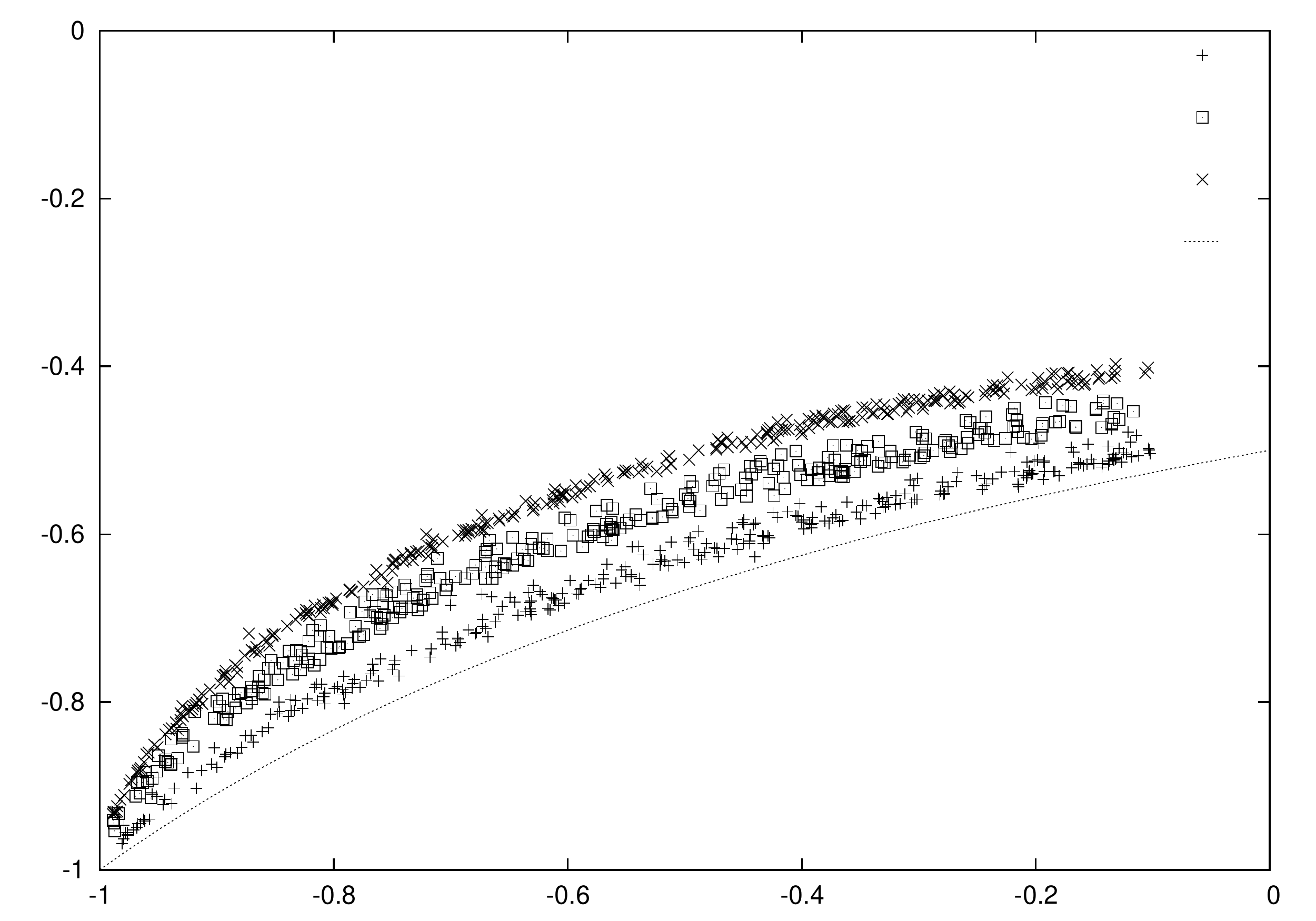}}}}
\put(79,65){\tiny$h=0.01$}
\put(79,60.3){\tiny$h=0.1$}
\put(79,55.6){\tiny$h=1$}
\put(75,50.9){\tiny$-1/(p+2)$}
\put(50,0){\tiny$p$}
\put(0,35){\tiny$\theta/\delta$}
\put(105,0){\parbox[b]{95\unitlength}{\bfseries%
\caption{\normalfont Jump values of $\theta/\delta$ for the example function 
\eqref{stepfn-u}
in the range $p\in[-0.99,-0.1]$ for three jump heights $h$ and randomly
chosen $\alpha$, $\delta$. Curve shows the theoretical value 
$\theta/\delta=-1/(p+2)$ for comparison.}
\label{fig-stepfn-theta}}}
\end{picture}
\end{figure}
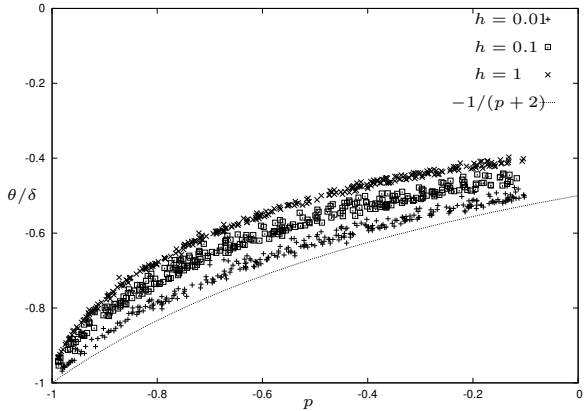

As is evident from the results visualised in Figure~\ref{fig-stepfn-theta},
the values of $\theta/\delta$ at which $\mu$ jumps from the lower to the higher
segment of $u$ are close to the predicted ones for small $h$ and deviate
somewhat more for larger $h$, but the overall behaviour of the interface
displacement speed is consistent with the analysis of the PDE.

\section{M-Smoother PDEs for 1D Signals}
\label{sec-pde1d}

A similar analysis as in the 2D case can be carried out for 1D signals.
The proofs of the next two results are found in Appendix~\ref{app-proofs}.

\begin{proposition}[1D PDE limit for $p>-1$]\label{prop-pde1d-p}
Let a smooth signal $u:\mathbb{R}\to\mathbb{R}$ be given, and let
$x_0$ be a regular point,
$u_x\ne0$.
One step of order-$p$ mean filtering of $u$ with a box window
$\mathrm{I}_\varrho(x_0):=[x_0-\varrho,x_0+\varrho]$ and $p>-1$, $p\ne0$
approximates for $\varrho\to0$
a time step of size $\tau=\varrho^2/(2p+2)$ of an
explicit time discretisation of the PDE
\begin{equation}
 u_t = (p-1)\,u_{xx}
 \label{pde-1d-p}
\end{equation}
in the sense that
\begin{align}
\operatorname{mean}_p\{u(x)~|~x\in\mathrm{I}_{\varrho}(x_0)\} 
- u(x_0)
&= \frac{\varrho^2}{2(p+1)} 
(p-1) u_{xx}(x_0)
+ \mathrm{o}(\varrho^2)\;.
\label{pde-1d-p-step}
\end{align}
At a local minimum (maximum) of $u$, i.e., $x_0$ with
$u_x=0$ where $u_{xx}$ is nonnegative (nonpositive),
the same filtering step fulfils for $\varrho\to0$ the inequality
$\operatorname{mean}_p\{u(x)~|~x\in\mathrm{I}_{\varrho}(x_0)\} 
- u(x_0) \ge0$ ($\le 0$), thus approximates an evolution $u_t\ge0$
($u_t\le0$).
\end{proposition}

\begin{proposition}[1D PDE limit for mode filtering]\label{prop-pde1d-mode}
Let $u$ be as in
Proposition~\ref{prop-pde1d-p}, and $x_0$ any point in its domain.
One step of mode filtering of $u$ with a box window
$\mathrm{I}_\varrho(x_0)$
approximates for $\varrho\to0$
a time step of size $\tau=\varrho$ of an
explicit time discretisation of the shock filter PDE \cite{OR90} given by
\begin{equation}
 u_t=-\sgn(u_{xx})\,\lvert u_x\rvert \;.
\label{pde-1d-mode}
\end{equation}
\end{proposition}

This relation between local mode filtering of 1D signals and shock
filtering extends a result for discrete signals from \cite{WWG07}.

Unlike in the 2D case, the PDE approximated by the mode filter does not fit
in as the limit $p\to-1$ of the PDE for order-$p$ means, and the approximation
in the mode filter case is with time step size $\mathcal{O}(\varrho)$ instead
of $\mathcal{O}(\varrho^2)$. To understand this, notice that in 
\eqref{pde-1d-p-step} the coefficient in front of $u_{xx}$ goes to infinity
for $p\to-1^+$, which means that the approximation holds for ever smaller
$\varrho$ as $p$ approaches $-1$, such that there is no uniform limit of
the approximation \eqref{pde-1d-p-step} for any fixed positive radius 
$\varrho$.

\section{M-Smoother PDEs for 3D Images}
\label{sec-pde3d}

In this section, we extend our previous results also to the case of
three-dimensional, i.e., volume images. The results are similar to those in 
two dimensions. However, now the smoothing in level set direction takes place
in a surface with two geometric coordinates $\xi$ and $\chi$ referring to
mutually orthogonal tangential directions of the level set. We give again
two propositions referring to $p\in(-1,\infty)\setminus\{0\}$ and the
mode case; both proofs are found in Appendix~\ref{app-proofs}.

\begin{proposition}[3D PDE limit for $p > -1$]\label{prop-pde3d-p}
Let a smooth volume image $u:\mathbb{R}^3\to\mathbb{R}$ be given, and let
$\boldsymbol{x}_0=(x_0,y_0,z_0)$ be a regular point,
$\lvert\boldsymbol{\nabla}u(\boldsymbol{x}_0)\rvert>0$.
One step of order-$p$ mean filtering of $u$ with a ball-shaped window
$\mathrm{B}_\varrho(\boldsymbol{x}_0)$ and $p>-1$, $p\ne0$
approximates for $\varrho\to0$
a time step of size $\tau=\varrho^2/(2p+6)$ of an
explicit time discretisation of the PDE
\begin{equation}
u_t = u_{\xi\xi} + u_{\chi\chi} + (p-1)\,u_{\eta\eta}
\label{pde-3d-p}
\end{equation}
where $\eta$, $\xi$ and $\chi$ are geometric coordinates referring at each 
image location to the direction of the positive gradient, and two 
mutually orthogonal directions tangent to the level surface, respectively:
\begin{align}
\kern2em&\kern-2em
\operatorname{mean}_p
\{u(x,y,z)~|~(x,y,z)\in\mathrm{B}_{\varrho}(x_0,y_0,z_0)\} \
- u(x_0,y_0,z_0)
\notag \\*
&= \frac{\varrho^2}{2(p+3)} 
\bigl( u_{\xi\xi}(x_0,y_0,z_0) + u_{\chi\chi} (x_0,y_0,z_0)
+ (p-1) u_{\eta\eta}(x_0,y_0,z_0) \bigr)
\notag \\* & \quad {}
+ \mathcal{O}(\varrho^{(\min\{p,0\}+5)/2})\;.
\label{pde-3d-p-step}
\end{align}
At a local minimum (maximum) of $u$, i.e., $\boldsymbol{x}_0$ with
$\lvert\boldsymbol{\nabla}u(\boldsymbol{x}_0)\rvert=0$ where the Hessian
$\mathbf{D}^2u(\boldsymbol{x}_0)$ is positive (negative) semidefinite,
the same filtering step fulfils for $\varrho\to0$ the inequality
$\operatorname{mean}_p
\{u(x,y,z)~|~(x,y,z)\in\mathrm{B}_{\varrho}(x_0,y_0,z_0)\} 
- u(x_0,y_0,z_0) \ge0$ ($\le 0$), thus approximates an evolution $u_t\ge0$
($u_t\le0$).
\end{proposition}
Like in the 2D case (but in contrast to the 1D case) the mode filter in 3D
again fits in as $p\to-1$ into the general case.
\begin{proposition}[3D PDE limit for mode filtering]\label{prop-pde3d-mode}
Let $u$ and $\boldsymbol{x}_0$ be as in
Proposition~\ref{prop-pde3d-p}.
One step of mode filtering of $u$ with a ball-shaped window
$\mathrm{B}_\varrho(\boldsymbol{x}_0)$
approximates for $\varrho\to0$
a time step of size $\tau=\varrho^2/4$ of an
explicit time discretisation of the PDE 
$u_t=u_{\xi\xi}+u_{\chi\chi}-2u_{\eta\eta}$
with $\eta$, $\xi$, $\chi$ as in Proposition~\ref{prop-pde3d-p}.
At a local minimum (maximum), mode filtering approximates
$u_t\ge0$ ($u_t\le0$).
\end{proposition}

We summarise the PDE approximation results from 
Propositions~\ref{prop-pde-gmdisc-generic}--\ref{prop-pde3d-mode} in
Table~\ref{tab-pde}. It shows a very systematic behaviour w.r.t.\ the
influence of the dimension, such that it is straightforward to come
up with a conjecture for arbitrary dimensions larger than 3: Depending 
on the order $p$, we expect a PDE that has forward ($p>1$) or backward 
($p<1$) parabolic behaviour in the gradient direction $\boldsymbol{\eta}$, 
combined with forward parabolic smoothing orthogonal to it:
\begin{eqnarray}
 u_t &=& \Delta u - u_{\eta\eta} + (p-1) u_{\eta\eta}%
     \;=\;
     \Delta u + (p-2) u_{\eta\eta} \;.  
\end{eqnarray}

\begin{table*}[b]
{\bfseries\caption{\normalfont
PDE approximation results for order-$p$ mean filters and mode 
filtering in one to three dimensions.}
\label{tab-pde}}
\bigskip

\begin{tabular}{@{~}c@{\quad}c@{\quad}c@{\quad}c@{\quad}c@{~}}
Dimension&
PDE for $p>-1$, $p\ne0$&
\parbox[t]{18mm}{\centering Time step\\ size $\tau$}&
PDE for mode filter&
\parbox[t]{18mm}{\centering Time step\\ size $\tau$\vphantom{$a^f_f$}}\\
\hline
\rule{0pt}{1.5em}%
1D&$u_t=(p-1)u_{xx}$&
\scalebox{1.15}{$\frac{\varrho^2}{2p+2}$}&
$u_t=-\mathrm{sgn}(u_{xx})\lvert u_x\rvert$&
$\varrho$\\
\rule{0pt}{1.5em}%
2D&$u_t=u_{\xi\xi}+(p-1)u_{\eta\eta}$&
\scalebox{1.15}{$\frac{\varrho^2}{2p+4}$}&
$u_t=u_{\xi\xi}-2u_{\eta\eta}$&
\scalebox{1.15}{$\frac{\varrho^2}{2}$}\\
\rule{0pt}{1.5em}%
3D&$u_t=u_{\xi\xi}+u_{\chi\chi}+(p-1)u_{\eta\eta}$&
\scalebox{1.15}{$\frac{\varrho^2}{2p+6}$}&
$u_t=u_{\xi\xi}+u_{\chi\chi}-2u_{\eta\eta}$&
\scalebox{1.15}{$\frac{\varrho^2}{4}$}
\end{tabular}
\end{table*}

\section{Discrete Versus Continuous M-Smoothing}
\label{sec-disccont}

The previous results rise the question about the relation between the
discrete and continuous situation. For discrete distributions the
mode is approximated according to \cite{BarralSouto-tr38} by
order-$p$ means for $p\to0$. Negative orders $p$ in the sense of
Section~\ref{sec-pde2d} cannot be applied at all for discrete distributions.
In contrast, in the case of
densities over a continuous range the mode is obtained for $p\to-1$.
The limit $p\to0$ for continuous distributions instead results in a mean
(that could be called order-$0$ mean to close the gap of definition) 
that does in relevant cases not coincide with the mode. 
As this situation is difficult to grasp intuitively, we present in 
Appendix~\ref{app-toy} a worked-out example of a simple density function
(a cut-off quadratic function) for which the order-$p$ mean and mode can
be calculated in closed form, so one can clearly see the
discrepancy between order-$0$ mean and mode as well as the convergence to
the mode for $p\to-1$.

Looking at the continuous case first, it is clear that the penaliser function
$\sgn(p)\lvert z\rvert^p$ converges for $p\to+0$ (from the positive side)
to the function \eqref{penaliser0}, with the convergence being non-uniform
around $z=0$. From the negative side, one has non-uniform convergence
to the function
\begin{equation}
\lim\limits_{p\to-0}\sgn(p)\lvert z\rvert^p = 0\quad \text{defined for}~z\ne0
\;.
\label{penaliser0m}
\end{equation}
As constants are negligible in penalisers for ``means'' of continuous
distributions anyway, \eqref{penaliser0} and
\eqref{penaliser0m} have not only the same effect but they act simply as
constants, as the removable discontinuity at $0$ is without influence
under integration. Thus they do not give rise to an ``order-$0$ mean''
whatsoever.

To model the mode of a continuous distribution, a penaliser is needed that
under integration gives positive weight to a single location; thus the
penaliser must be a distribution with a (negative, for the mode to arise
as minimiser) delta peak at $0$. Indeed this is the limit of
$\sgn(p)\lvert z\rvert^p$ for $p\to-1^+$.

To transfer a continuous penaliser $\varPsi$
to the discrete case, the correct way
would be to use not sampling,
\begin{equation}
\psi_k := \varPsi (kh)\;,
\label{penaliser-sampling}
\end{equation}
(where $h>0$ is the step width between sampling locations)
but a finite-volume discretisation
which is essentially the composition of a box-kernel convolution (other
low-pass kernels would be possible) with sampling,
\begin{align}
\psi_k &:= \frac1h \int\limits_{(k-\frac12)h}^{(k+\frac12)h}\varPsi(x)\dd x
= (\varPsi * \mathrm{B}_h) (kh)\;,
\label{penaliser-finvol}
\\
\mathrm{B}_h(z)&:=\frac1h\chi_{[-h/2,h/2]}(z)\;.
\end{align}
The width $h$ of the box kernel $\mathrm{B}_h$
takes the role of the bin width of a histogram into which data
are aggregated.
For originally discrete distributions (finite multisets of data points)
one might omit making this step explicit as $h$ may be chosen
arbitrarily small such that \eqref{penaliser-finvol} approximates
\eqref{penaliser-sampling} with arbitrary accuracy (as long as
$\varPsi$ is integrable in each sampling interval). As soon as $h$ is smaller
than the minimal distance between two locations of the discrete distribution,
further reduction of $h$ does not increase the number of histogram bins
with positive weights.

In contrast, when discretising an originally continuous distribution,
the histogram bin width becomes relevant, and it is essential
to use \eqref{penaliser-finvol}.

Let us regard now the penaliser $\varPsi_p(z)=\lvert z\rvert^p$ for $p>0$.
In order for a finite-volume discretisation of $\varPsi_p$ to converge for 
$p\to+0$ to the naive sampling of \eqref{penaliser0}, i.e.,
\begin{equation}
\psi_k=\begin{cases}0\;,&k=0\;,\\1&\text{else}\end{cases}
\label{discretepenaliser0-naive}
\end{equation}
the bin width $h$ needs to be sent to zero along with $p$. For any fixed
bin width $h$, the finite-volume discretisation of $\varPsi_p$ 
converges to the constant unity function, $\psi_k=1$ for all $k$, instead.

On the other hand, the finite-volume discretisation of
$\varPsi_p(z)=-\lvert z\rvert^p$ for $-1<p<0$ converges for $p\to-1^+$
exactly to $\psi_0=-1$ and $\psi_k=0$ for $k\ne0$, i.e., 
\eqref{discretepenaliser0-naive} up to an irrelevant constant offset. 
In this sense, the case
$p=0$ as defined in \cite{BarralSouto-tr38} for discrete distributions
does indeed correspond to the discretisation of the limit $p\to-1$
of the continuous setting.

As a final remark, we point out that in an image filtering context
the discretisation of distributions as discussed in this section is
in fact applied to the intensity domain, thus takes the role of
quantisation. The mere spatial discretisation of an image leaves the
intensity domain continuous,
and one should try to approximate the concepts for continuous
distributions numerically as accurate as possible.
However, the spatial discretisation creates a discrete sample from the
continuous distribution, and (at least for $p<1$) filtering this discrete
sample as a finite set will not be a proper approximation of the
continuous filter. For example, the mode of the discrete distribution
will be meaningless as in generic cases finite samples from continuous
distributions consist of distinct values, each with trivial frequency $1$.
It is therefore necessary to design a numerical process that in the one
or other way estimates the continuous distribution from the set of
discrete sample values.

\section{Numerical Scheme for the 2D PDE Limit}
\label{sec-num}

Next we discuss an numerical algorithm for approximating
our two-dimensional PDE limit \eqref{pde-p} in an adequate way.
The 2D setting is practically most important, and it contains all 
essential difficulties that also arise in higher dimensions.
Our 2D PDE gives rise to two major numerical problems: 
\begin{itemize}
\item It involves the anisotropic expressions $u_{\xi\xi}$ and 
      $u_{\eta\eta}$.
      To reproduce their qualitative properties adequately, one has to take 
      care that the discretisation approximates rotationally invariant
      behaviour well and that it satisfies a discrete maximum--minimum
      principle which prevents over- and undershoots. 
\item For $p<1$, the sign in front of the operator $u_{\eta\eta}$
      becomes negative, which results in a backward parabolic operator.
      Such operators are known to be ill-posed. They require additional
      stabilisation in the model and the numerics. 
\end{itemize}
These challenges show that great care must be invested in the design of 
appropriate numerical algorithms. Thus, let us have a deeper look into 
our efforts along these lines. 

\paragraph{Reformulation.}
Using $\,u_{\eta\eta} = \Delta u - u_{\xi\xi}\,$ and 
$\,u_{\xi\xi} = \mathrm{curv}(u)\lvert\boldsymbol{\nabla} u\rvert\,$ with the 
isophote curvature
$\mathrm{curv}(u)$ we rewrite \eqref{pde-p} in a numerically more convenient 
form:
\begin{equation}
 u_t \;=\; 
 (2\!-\!p)\,\mathrm{curv}(u)\lvert\boldsymbol{\nabla} u\rvert \,+\, 
 (p\!-\!1)\,\Delta u\,.    
 \label{eq:mforw}
\end{equation}
If $p \ge 1$, we apply this equation in all locations, including extrema.

\noindent
For $p<1$, the second term describes backward diffusion, which we
stabilise by freezing its action in extrema where 
$\lvert\boldsymbol{\nabla} u\rvert$ vanishes: 
\begin{equation} 
 u_t \;=\; (2\!-\!p)\, \mathrm{curv}(u)\lvert\boldsymbol{\nabla} u\rvert \,+\, 
           (p\!-\!1) \, 
                \mathrm{sgn}(\lvert\boldsymbol{\nabla} u\rvert) \, \Delta u\,.
 \label{eq:mback}
\end{equation}
In practice, our image domain is finite and of rectangular size. 
This motivates us to equip the equations \eqref{eq:mforw} and 
\eqref{eq:mback} with reflecting (i.e., homogeneous Neumann) boundary 
conditions.
Both evolutions \eqref{eq:mforw} and \eqref{eq:mback} are replaced by 
finite difference schemes         
on a regular grid of size $h$ in $x$- and $y$-direction and time step
size $\tau$. By $u_{i,j}$ we denote an approximation of $u$ in pixel
$(i,j)$.

\paragraph{Space discretisation of forward diffusion.}
If $p \ge 1$, we discretise $\Delta u$ in \eqref{eq:mforw} with a 
nine-point stencil. It is a weighted average of an approximation 
aligned along the $x$- and $y$-axis with one aligned along the diagonal 
directions:
\renewcommand{\arraystretch}{1.2}%
\begin{eqnarray}
  & & \frac{1\!-\!\nu}{h^2} \;\,
  \begin{array}{|c|c|c|}
    \hline
    0 &  1 &  0 \\ \hline
    \;1\; & \;-4 \; & \;1\; \\ \hline
    0 &  1 &  0 \\
    \hline
  \end{array}
  \;+\;
  \frac{\nu}{(\sqrt{2}\,h)^2} \;\,
    \begin{array}{|c|c|c|}
    \hline
    1 &  0 &  1 \\ \hline
    \;0\; & \;-4 \; & \;0\; \\ \hline
    1 &  0 &  1 \\
    \hline
  \end{array} \nonumber \\[2mm]
  & & =\; 
  \frac{1}{2h^2} \;\,
  \begin{array}{|c|c|c|} 
    \hline
    \nu &  2\!-\!2\nu &  \nu \\ \hline
    \;2\!-\!2\nu\; & \;4\nu\!-\!8\; & \;2\!-\!2\nu\; \\ \hline
    \nu &   2\!-\!2\nu &  \nu \\ 
    \hline
  \end{array}  \;,
  \label{eq:nine}
\end{eqnarray}
\renewcommand{\arraystretch}{1}%
where the weight $\nu \in[0,1]$ is used to optimise the rotation
invariance of the stencil. Since the stencil has an axial size of
$3 h$ and a diagonal one of $3\sqrt{2} h$, we choose
$\nu:=\sqrt{2} - 1$. This leads to the weight ratio
$(1-\nu):\nu=\sqrt{2}:1$, which compensates for the different sizes.
Our experiments will show that in this way, rotation invariance is 
approximated very well. 

\paragraph{Space discretisation of backward diffusion.}
For $p<1$, the term 
$(p\!-\!1)\,\mathrm{sgn}(\lvert\boldsymbol{\nabla} u\rvert) \, \Delta u\,$ in
\eqref{eq:mback} creates stabilised backward diffusion. Here 
we base our finite difference approximation on a minmod discretisation of
Osher and Rudin \cite{OR91}, but improve its rotation invariance again
by a weighted averaging with its diagonally aligned counterpart with
weight $\nu=\sqrt{2} - 1$.
We denote the forward differences in $x$-, $y$-, and the diagonal 
directions $\,\boldsymbol{d}=(1,1)\,$ and $\,\boldsymbol{e}=(1,-1)\,$ by 
\begin{align}
&&
 \! u^x_{i,j} &:= \frac{u_{i+1,j}\!-\!u_{i,j}}{h}, \! &
 \! u^y_{i,j} &:= \frac{u_{i,j+1}\!-\!u_{i,j}}{h}, \! &&\\*
&&
 \! u^d_{i,j} &:= \frac{u_{i+1,j+1}\!-\!u_{i,j}}{\sqrt{2}\,h}, \!&
 \! u^e_{i,j} &:= \frac{u_{i+1,j-1}\!-\!u_{i,j}}{\sqrt{2}\,h}. \! &&
\end{align}
By $\mathrm{M}\,(a,b,c)$ we abbreviate the minmod function of three arguments
which chooses the argument of minimal modulus if the arguments have the same 
sign, and yields $0$ otherwise:
\begin{equation}
 \mathrm{M}\,(a,b,c) \;:=\;
  \begin{cases}
    \mathop{\operatorname{argmin}}\limits_{z\in\{a,b,c\}}\lvert z\rvert & 
      \text{if $ab\ge0$ and $ac\ge0$}\,,\\
    0&\text{else}\,.
  \end{cases}
\end{equation}
With these notations we approximate 
$\,\mathrm{sgn}(\lvert\boldsymbol{\nabla} u\rvert) \, \Delta u\,$ in pixel 
$(i,j)$ by 
\begin{eqnarray}
 &{{\textstyle\frac{1-\nu}{h}}} & \big( \;\: 
    \mathrm{M}\,(u^x_{i+1,j}, u^x_{i,j}, u^x_{i-1,j}) %
      - \: \mathrm{M}\,(u^x_{i,j}, u^x_{i-1,j}, u^x_{i-2,j})
    \nonumber \\
  & & + \:
    \mathrm{M}\,(u^y_{i,j+1}, u^y_{i,j}, u^y_{i,j-1}) %
      - \: \mathrm{M}\,(u^y_{i,j}, u^y_{i,j-1}, u^y_{i,j-2})
  \; \big) \nonumber\\
 + &{{\textstyle\frac{\nu}{\sqrt{2}\,h}}} & \big( \;\: 
    \mathrm{M}\,(u^d_{i+1,j+1}, u^d_{i,j}, u^d_{i-1,j-1}) %
      - \: \mathrm{M}\,(u^d_{i,j}, u^d_{i-1,j-1}, u^d_{i-2,j-2})
    \nonumber \\
  & & + \:
    \mathrm{M}\,(u^e_{i+1,j-1}, u^e_{i,j}, u^e_{i-1,j+1}) %
      - \: \mathrm{M}\,(u^e_{i,j}, u^e_{i-1,j+1}, u^e_{i-2,j+2})
  \; \big)\,. 
 \label{eq:mm}
\end{eqnarray}

\medskip
\paragraph{Space discretisation of mean curvature motion.}
Let us now discuss our approximation of the mean curvature term
$\,(2\!-\!p)\,\mathrm{curv}(u)\lvert\boldsymbol{\nabla} u\rvert\,$.
The isophote curvature
\begin{equation}
 \mathrm{curv}(u) \;=\; \frac{u_x^2u_{yy} -2 u_x u_y u_{xy} + u_y^2u_{xx} }
                             {(u_x^2+u_y^2)^{3/2}} 
 \label{eq:curv}
\end{equation}
can be discretised in a straightforward way with central differences. To 
avoid a potential singularity in the denominator, we regularise by adding
$\epsilon=10^{-10}$ to $\,u_x^2+u_y^2\,$. 
Moreover, note that the isophote curvature $\mathrm{curv}(u)$ describes the 
inverse radius of the osculating circle to the level line. Since a discrete 
image does not have structures that are smaller than a single pixel, the 
smallest practically relevant radius is $\frac{h}{2}$. Thus, we impose a
curvature limiter that restricts the computed result to the range 
$[-\frac{2}{h}, \frac{2}{h}]$.
\medskip

Depending on the sign of $\,(2\!-\!p) \, \mathrm{curv}(u)$,\, we may 
interpret $\,(2\!-\!p)\,\mathrm{curv}(u)\lvert\boldsymbol{\nabla} u\rvert$ 
either as a dilation term (for positive
sign) or an erosion term (for negative sign) with a disc-shaped structuring
element of radius $\,\lvert(2\!-\!p) \, \mathrm{curv}(u)\rvert$; see e.g.\
\cite{Alvarez-ARMA93}.
For a stable discretisation of $\lvert\boldsymbol{\nabla}u\rvert$, we use the
Rouy-Tourin upwind scheme \cite{RT92}. In contrast to our conference paper
\cite{Welk-ssvm19}, we again improve its rotation invariance by a weighted 
averaging of axial and diagonal discretisations with weight $\nu=\sqrt{2} - 1$.
In the dilation case, this comes down to
\begin{eqnarray}
 \lvert\boldsymbol{\nabla}u\rvert_{i,j} &\;\approx\;& (1-\nu)  
 \left(\Bigl(\max\bigl(-u^x_{i-1,j},\, u^x_{i,j}, \,0\bigr)\Bigr)^2%
  + %
  \Bigl(\max\bigl(-u^y_{i,j-1},\, u^y_{i,j}, \,0\bigr)\Bigr)^2
  \right)^{1/2}\nonumber \\
 & & + \; \nu
 \left(\Bigl(\max\bigl(-u^d_{i-1,j-1},\, u^d_{i,j}, \,0\bigr)\Bigr)^2%
  + %
  \Bigl(\max\bigl(-u^e_{i-1,j+1},\, u^e_{i,j}, \,0\bigr)\Bigr)^2
  \right)^{1/2}
\end{eqnarray}
and in the erosion case to
\begin{eqnarray}
 \lvert\boldsymbol{\nabla}u\rvert_{i,j} &\;\approx\;& (1-\nu)
 \left(\Bigl(\max\bigl(-u^x_{i,j},\, u^x_{i-1,j}, \,0\bigr)\Bigr)^2%
  + %
 \Bigl(\max\bigl(-u^y_{i,j},\, u^y_{i,j-1}, \,0\bigr)\Bigr)^2
 \right)^{1/2}\nonumber \\
 & & + \; \nu
 \left(\Bigl(\max\bigl(-u^d_{i,j},\, u^d_{i-1,j-1}, \,0\bigr)\Bigr)^2%
  + %
 \Bigl(\max\bigl(-u^e_{i,j},\, u^e_{i-1,j+1}, \,0\bigr)\Bigr)^2
 \right)^{1/2}.
\end{eqnarray}

\paragraph{Operator splitting.}
The space discretisations we have discussed convert our PDEs \eqref{eq:mforw} 
and \eqref{eq:mback} to systems of ordinary differential equations (ODEs). 
Their general structure is given by
\begin{eqnarray} 
 \frac{d\boldsymbol{u}}{dt} &\,=\,& 
      (1-\nu) \, \boldsymbol{D}_+(\boldsymbol{u}) 
      \,+\, \nu \,\boldsymbol{D}_\times(\boldsymbol{u})%
      \,+\, (1-\nu) \, \boldsymbol{M}_+(\boldsymbol{u}) 
      \,+\, \nu \,\boldsymbol{M}_\times(\boldsymbol{u})\,,
 \label{eq:ode}
\end{eqnarray} 
where the vector $\boldsymbol{u}$ assembles the function values of $u$ at all
grid points in our discretised domain. The expressions 
$\boldsymbol{D}_+(\boldsymbol{u})$ and $\boldsymbol{D}_\times(\boldsymbol{u})$ 
stand for axial and diagonal discretisations of the
diffusion terms. For $p>1$ they refer to the forward term $(p\!-\!1)\Delta u$,
and for $p<1$ to the stabilised backward term 
$(p\!-\!1)\sgn(\lvert\boldsymbol{\nabla}u\rvert)\Delta u$.
Likewise, $\boldsymbol{M}_+(\boldsymbol{u})$ and 
$\boldsymbol{M}_\times(\boldsymbol{u})$ denote
our axial and diagonal discretisations of the mean curvature motion
term $\,(2\!-\!p)\,\mathrm{curv}(u)\lvert\boldsymbol{\nabla} u\rvert$.
All discrete operators take into account the homogeneous Neumann boundary
conditions by mirroring one or two layers of boundary pixels. For $t=0$,
the ODE system uses the discretised original image $\boldsymbol{f}$ as initial
condition:
\begin{equation}
 \boldsymbol{u}(0) = \boldsymbol{f}\,.
\end{equation}

\medskip
For the time discretisation of \eqref{eq:ode}, we proceed in four explicit 
fractional steps.
Denoting the time step size by $\tau$, and $\boldsymbol{u}$ at time level 
$k\tau$ by $\boldsymbol{u}^k$, our scheme is given by
\begin{eqnarray}
 \label{eq:frac1}
 \boldsymbol{u}^{k+1/4} &\;=\;& \boldsymbol{u}^k \:+\: 
              \tau\,(1\!-\!\nu)\,\boldsymbol{D}_+(\boldsymbol{u}^k)\,,\\
 \label{eq:frac2}
 \boldsymbol{u}^{k+1/2} &\;=\;& \boldsymbol{u}^{k+1/4} \:+\: \tau \, \nu \,
              \boldsymbol{D}_\times(\boldsymbol{u}^{k+1/4})\,,\\
 \label{eq:frac3}
 \boldsymbol{u}^{k+3/4} &\;=\;& \boldsymbol{u}^{k+1/2} \:+\: 
              \tau\,(1\!-\!\nu)\,\boldsymbol{M}_+(\boldsymbol{u}^{k+1/2})\,,\\
 \label{eq:frac4}
 \boldsymbol{u}^{k+1}   &\;=\;& \boldsymbol{u}^{k+3/4} \:+\: \tau \, \nu \,
              \boldsymbol{M}_\times(\boldsymbol{u}^{k+3/4})\,.
\end{eqnarray}
We will see that compared to an unsplit explicit scheme as was used in our 
conference paper
\cite{Welk-ssvm19}, the split variant allows substantially larger
time step sizes. For more information on operator splitting we refer the
reader to the classical literature \cite{Mar90,MG80,Ya71}.
 
\paragraph{Consistency.}
Since our resulting explicit scheme uses various one-sided -- and thus 
first order -- finite difference approximations within its upwind and 
minmod strategies, if follows that its general consistency order outside 
extrema is ${\mathcal{O}}(h\!+\!\tau)$. For the pure forward diffusion 
case $p=2$, however, the second order stencil
\eqref{eq:nine} gives ${\mathcal{O}}(h^2\!+\!\tau)$.

\paragraph{Stability.}
Stability of a numerical algorithm typically refers to the discrete 
preservation of an essential property of the continuous process. By 
design, all M-smoothers satisfy a maximum--minimum principle, which 
states that maxima must not become larger during filtering, and minima 
not smaller. Since our PDEs of interest have been derived as limits of 
M-smoothers, it is natural that they should obey a maximum--minimum 
principle as well. This is also supported by the fact that all our 
evolutions under consideration satisfy $u_t \le 0$ in maxima and 
$u_t \ge 0$ in minima. This motivates us to study the stability of
our algorithm in terms of a discrete maximum--minimum principle,
which obviously also implies $L^\infty$-stability. 
To this end we show that all fractional 
steps \eqref{eq:frac1}--\eqref{eq:frac4} in our explicit scheme are 
designed to satisfy a discrete maximum--minimum principle for 
suitably chosen time step sizes. Since the details are somewhat 
cumbersome and do not give more general insights, we sketch only 
the basic ideas.

For $p>1$ and $\nu < 1$, the first fractional step \eqref{eq:frac1} 
approximates the equation $u_t=(1-\nu)(p\!-\!1)\Delta u$ with an
explicit axial scheme. This leads to a stencil with weight sum $1$.
All noncentral weights are nonnegative, and the central stencil weight 
is given by $1 - 4(1-\nu)(p\!-\!1) \frac{\tau}{h^2}$. It becomes 
nonnegative for
\begin{equation}
 \tau \;\le\; \frac{h^2}{4(1-\nu)\,|p\!-\!1|} \;=:\; \tau_1\,.
 \label{eq:stab1}
\end{equation}
In this case the scheme computes a convex combination of data from
the previous time step, which implies a maximum--minimum principle.\\ 
Osher and Rudin \cite{OR91} report the same stability limit for their
minmod scheme for stabilised backward diffusion as one gets for the forward
process, and they emphasise that their scheme does neither increase local
maxima no does it decrease local minima. Thus, the step size restriction
\eqref{eq:stab1} also holds for the stabilised backward PDE 
$u_t=(1-\nu)(p\!-\!1)\mathrm{sgn}(\lvert\boldsymbol{\nabla} u\rvert)\Delta u$ 
that is approximated by Step \eqref{eq:frac1} for $p<1$. Note that
\eqref{eq:stab1} formally becomes singular for $p=1$ or $\nu=1$, when 
the evolution equation
in the first fractional step degenerates to $u_t = 0$. In this case
the fractional step does nothing at all, such that its stability
limit could be seen as $\tau_1 = \infty$. 
The same considerations also apply for the step size limits of the 
other fractional steps that we discuss below. 

The second fractional step \eqref{eq:frac2} approximates 
the forward diffusion PDE $u_t=\nu(p\!-\!1)\Delta u$ for
$p>1$, or the stabilised backward process 
$u_t=\nu(p\!-\!1)\mathrm{sgn}(\lvert\boldsymbol{\nabla} u\rvert)\Delta u$ for
$p < 1$, but in both cases with a diagonal stencil.
Hence, we can use the same reasoning as in the first step, if we 
exchange $1-\nu$ by $\nu$ and $h$ by $h\sqrt{2}$. This leads to 
the stability limit
\begin{equation}
 \tau \;\le\; \frac{h^2}{2\nu\,|p\!-\!1|} \;=:\; \tau_2\,.
 \label{eq:stab2}
\end{equation}

For an axial stencil, the classical Rouy-Tourin scheme for the 
dilation/erosion evolutions $\,u_t = \pm \lvert\boldsymbol{\nabla} 
u\rvert\,$ is well known to satisfy a maximum-minimum principle if
its time step size obeys  
$\tau \le \frac{h}{2}\sqrt{2}$\,; see e.g.\ \cite{BW06}. Thus, the 
third fractional step \eqref{eq:frac3}, which approximates 
$\,u_t = (1-\nu) (2-p) \mathrm{curv}(u) \lvert\boldsymbol{\nabla}u\rvert\,$ 
with a curvature limiter interval $[-\frac{2}{h},\frac{2}{h}]$, must
satisfy the step size restriction 
\begin{equation}
 \tau \;\le\; \frac{h^2}{2\sqrt{2}\,(1-\nu)\,\lvert2\!-\!p\rvert} 
 \;=:\; \tau_3\,.
 \label{eq:stab3}
\end{equation}

Similar arguments can be used for the fourth fractional step \eqref{eq:frac4}.
Since it approximates the equation
$\,u_t = \nu (2-p) \mathrm{curv}(u) \lvert\boldsymbol{\nabla}u\rvert\,$ on a
diagonal stencil, we obtain the time step size restriction
\begin{equation}
 \tau \;\le\; \frac{h^2}{2 \,\nu\,\lvert2\!-\!p\rvert} \;=:\; \tau_4\,.
 \label{eq:stab4}
\end{equation}

\noindent
These considerations immediately lead to the following stability result: 
\begin{proposition}[Numerical Stability]
 Let the splitting scheme (\ref{eq:frac1})--(\ref{eq:frac4}) be equipped
 with mirrored boundary layers and initialisation 
 $\boldsymbol{u}^0=\boldsymbol{f}$.
 Moreover, let its time step size $\tau$ satisfy
 \begin{equation}
  \tau \;\le\; \mathrm{min} \{\tau_1, \,\tau_2, \,\tau_3, \,\tau_4\}
  \label{eq:stab}
 \end{equation}
 with $\tau_1$,...,$\tau_4$ from (\ref{eq:stab1})--(\ref{eq:stab4}).\\
 Then the scheme is $L^\infty$-stable, 
 \begin{equation}
  \lVert\boldsymbol{u}^{k}\rVert_\infty \le 
  \lVert\boldsymbol{u}^{k-1}\rVert_\infty \qquad 
  \forall k \ge 1\,,
 \end{equation}
 and respects the discrete maximum--minimum principle
 \begin{equation}
  \min_{i,j} f_{i,j} \le u_{n,m}^k \le \max_{i,j} f_{i,j} 
  \qquad \forall n,m, \; \forall k \ge 1\,.
 \end{equation}
\end{proposition}

In practice the step size limit \eqref{eq:stab} is not very restrictive: 
With $h:=1$ and $\nu=\sqrt{2}-1$, it comes down to $\,\tau \le 0.4267\,$ 
for the diffusion evolution ($p=2$), to $\,\tau \le 0.6035\,$ for mean 
curvature motion ($p=1$), to $\,\tau \le 0.2011\,$ for the mode 
equation ($p=-1$), and to $\,\tau \le 0.1422\,$ for the Gabor flow ($p=-2$).
These limits are larger than the ones in our conference
paper \cite{Welk-ssvm19}, and they allow efficient numerical approximations
of PDE evolutions for M-smoothers.

\section{Experiments}
\label{sec-exp}

In our experiments, we evaluate the PDE \eqref{pde-p} with five different 
settings for $p$: a temporally rescaled midrange evolution ($p \to \infty$) 
using $u_t=u_{\eta\eta}$ with $\tau=0.25$, 
the mean evolution leading to homogeneous diffusion ($p=2$, $\tau=0.25$), 
the median evolution yielding mean curvature motion ($p=1$, $\tau=0.25$),
the mode evolution ($p=-1$, $\tau=0.1$), 
and the Gabor flow ($p=-2$, $\tau=0.1$). 
Unless stated otherwise, we use the diagonal weight 
$\nu=\sqrt{2}-1$. The first two experiments recompute results from our 
conference paper \cite{Welk-ssvm19} by using our novel algorithm that 
has been improved w.r.t.\ rotation invariance and efficiency. 

\begin{figure*}[p]

\begin{center}
\begin{tabular}{ccc}
 original \hspace{1mm} ($256 \times 256$) & 
 midrange \hspace{1mm} ($t=8$) & 
 mean \hspace{1mm} ($t=5$)\\[0.5mm]
 \includegraphics[width=0.24\linewidth] {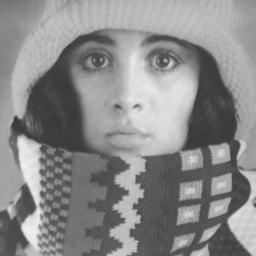} &
 \includegraphics[width=0.24\linewidth] {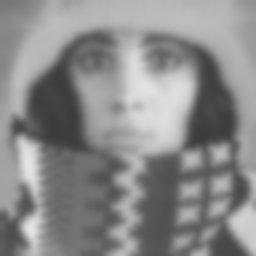} &
 \includegraphics[width=0.24\linewidth] {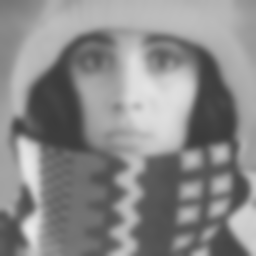} \\[3mm] 
 \includegraphics[width=0.24\linewidth] {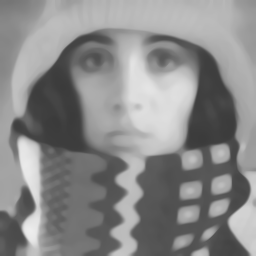} &
 \includegraphics[width=0.24\linewidth] {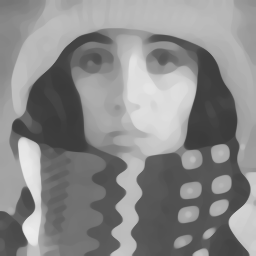} &   
 \includegraphics[width=0.24\linewidth] {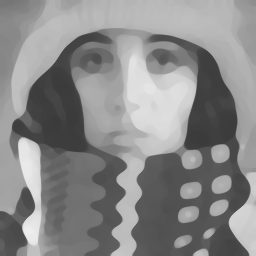} \\
 median \hspace{1mm} ($t=5$) & 
 mode   \hspace{1mm} ($t=3$) & 
 Gabor  \hspace{1mm} ($t=2.5$)
\end{tabular}
\end{center}

\vspace{-2mm}
{\bfseries\caption{\label{fig:trui}\normalfont
 Smoothing effect of the different evolution equations on the test image
 \emph{trui}. 
 Recomputed from \cite{Welk-ssvm19} with our improved algorithm.
}}

\end{figure*}

\medskip
Fig.~\ref{fig:trui} illustrates the effect of these equations on the 
real-world test image \emph{trui}. The CPU times for computing each of 
these results on a contemporary laptop are in the order of half a second.
We observe that the midrange filter
produces fairly jagged results, although it has a clear smoothing effect.
Homogeneous diffusion does not suffer from jagged artifacts, but blurs
also important structures such as edges. The median evolution is designed
to smooth only along isolines which results in a smaller deterioration of 
edge-like structures. The mode and the Gabor evolutions are very similar. 
They produce the sharpest results and may even enhance edges due to their 
backward parabolic term $(p-1)\,u_{\eta\eta}$. 

\begin{figure*}[p]

\begin{center}
\begin{tabular}{ccc}
 original \hspace{1mm} ($256 \times 256$) &
 midrange \hspace{1mm} ($t=100$) & 
 mean     \hspace{1mm} ($t=100$)\\[0.5mm]
 \includegraphics[width=0.24\linewidth] {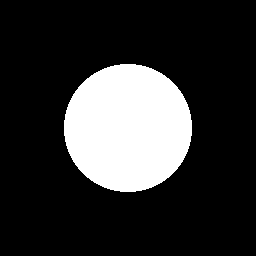} &
 \includegraphics[width=0.24\linewidth] {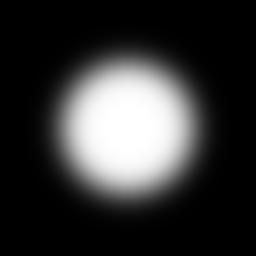} &
 \includegraphics[width=0.24\linewidth] {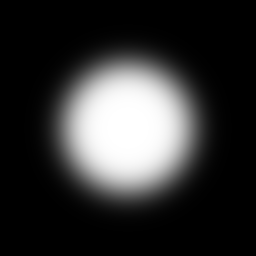} \\[3mm]
 \includegraphics[width=0.24\linewidth] {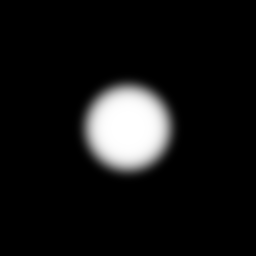} &
 \includegraphics[width=0.24\linewidth] {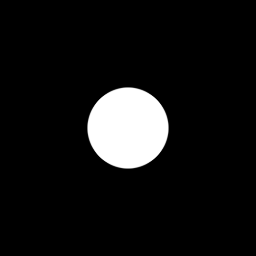} & 
 \includegraphics[width=0.24\linewidth] {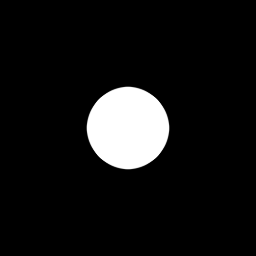} \\
 median \hspace{1mm} ($t=1200$) & 
 mode   \hspace{1mm} ($t=400$) & 
 Gabor  \hspace{1mm} ($t=300$) 
\end{tabular}
\end{center}

{\bfseries\caption{\label{fig:disk}\normalfont
 Effect of the different evolution equations on a disk. 
 Recomputed from \cite{Welk-ssvm19} with our improved algorithm.
}}

\end{figure*}

\medskip
Fig.~\ref{fig:disk} allows to judge if our numerical algorithm is
capable of reproducing the rotationally invariant behaviour of its
underlying PDE \eqref{pde-p}. We observe excellent rotation invariance.
Moreover, we see that the mode and Gabor evolutions have comparable 
shrinkage properties as mean curvature motion. However, they differ 
from mean curvature motion by their backward term $(p-1)\,u_{\eta\eta}$,
which can compensate dissipative artifacts that are caused by the
discretisations of the forward parabolic term $u_{\xi\xi}$.

\begin{figure*}[t]

\begin{center}
\begin{tabular}{ccccc}
 original & $\nu=0$ & $\nu=\sqrt{2}-1$ \\[0.5mm]
 \includegraphics[width=0.24\linewidth] {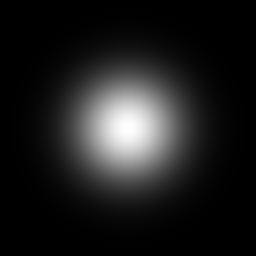} &
 \includegraphics[width=0.24\linewidth] {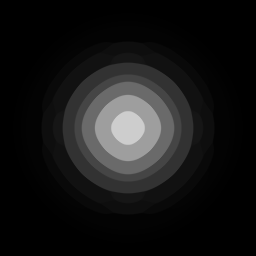} &
 \includegraphics[width=0.24\linewidth] {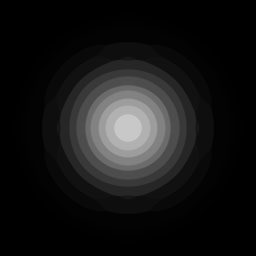} \\[3mm] 
 \includegraphics[width=0.24\linewidth] {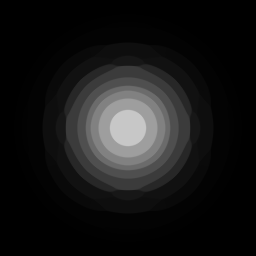} &
 \includegraphics[width=0.24\linewidth] {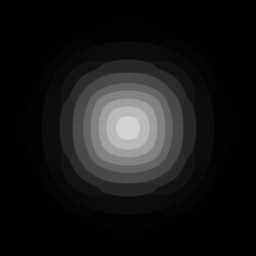} &\\
 $\nu=0.5$ & $\nu=1$ & \\[0.5mm]
\end{tabular}
\end{center}

{\bfseries\caption{\label{fig:gauss}\normalfont
 Staircasing effect of the mode evolution on a Gaussian test image, and 
 influence of the diagonal weight $\nu$ on the rotation invariance. 
 Image size: $256 \times 256$. Evolution time: $t=100$.
}}

\end{figure*}

\medskip
Fig.~\ref{fig:gauss} illustrates the staircasing behaviour of the mode
evolution. As already mentioned, staircasing is a common phenomenon for
PDEs that enhance images by means of some backward parabolic
concepts. It has been observed for the Perona--Malik filter \cite{PM90}, 
for forward-and-backward (FAB) diffusion \cite{GSZ02a}, 
and for shock filters \cite{KB75,OR90}. Staircasing becomes 
pronounced if a smoothly varying image structure is to be enhanced.
Therefore, we have chosen a Gaussian-like test image, which also
allows to judge the rotation invariance of our algorithm for different
values of the diagonal weight $\nu$. We observe that $\nu$ also has 
some impact on the number and size of the evolving stairs: 
Since backward parabolic processes are very sensitive w.r.t.\ the data 
and corresponding algorithms, such a behaviour is not unnatural. 
The discretisation with $\nu=0$ produces the coarsest stairs, while
the ones for $\nu=\sqrt{2}-1$ are particularly small. Regarding 
rotation invariance, Fig.~\ref{fig:gauss} shows that a pure axial 
($\nu=0$) or a pure diagonal approximation ($\nu=1$) perform relatively
bad, which is to be expected.
We see that the proposed value of $\nu=\sqrt{2}-1\approx 0.4142$ 
yields the most favourable result. It also outperforms the result for 
$\nu=0.5$. The latter parameter was used in the discretisation of the 
diffusion term in our conference paper \cite{Welk-ssvm19}.

\begin{figure*}[p!]

\begin{center}
\scalebox{0.95}{\colorbox{lightgrey}{
\vspace*{4mm}
\begin{tabular}{cccc}
 original & $t=70$ & $t=300$ & $t=800$ \\[1mm]
 \includegraphics[width=0.17\linewidth] {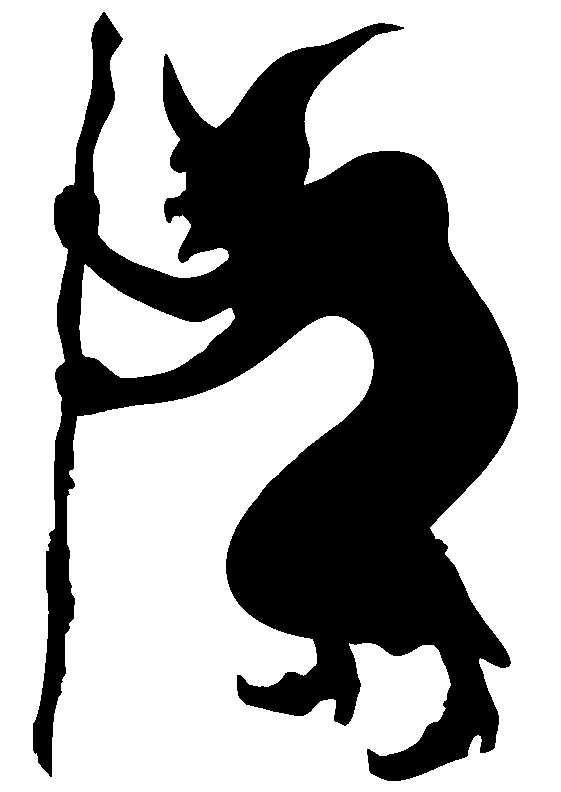} &
 \includegraphics[width=0.17\linewidth] {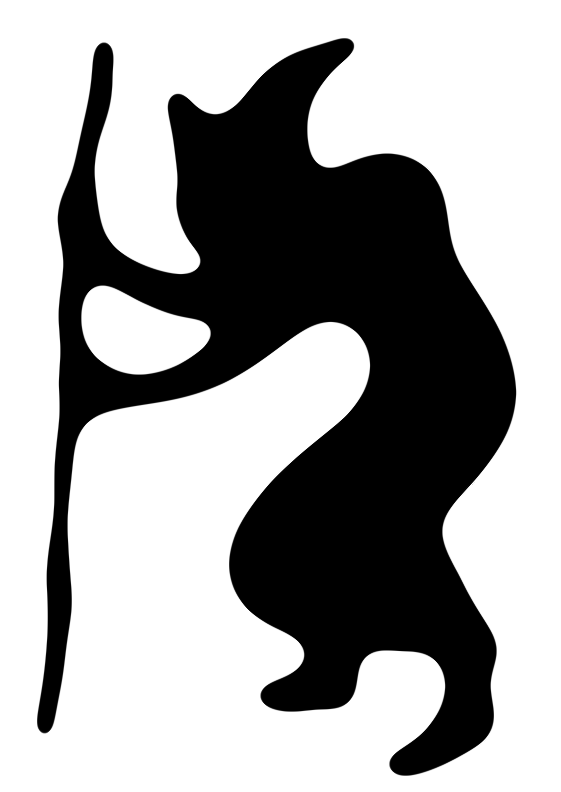} &
 \includegraphics[width=0.17\linewidth] {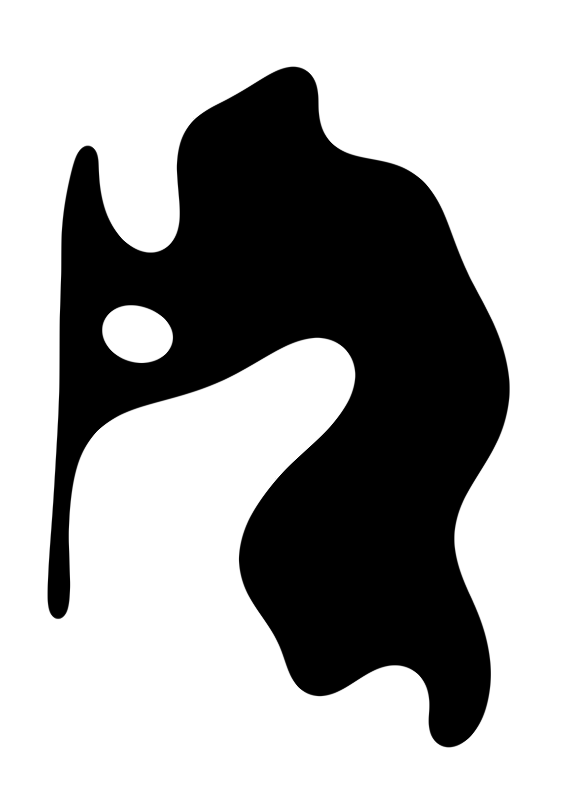} &
 \includegraphics[width=0.17\linewidth] {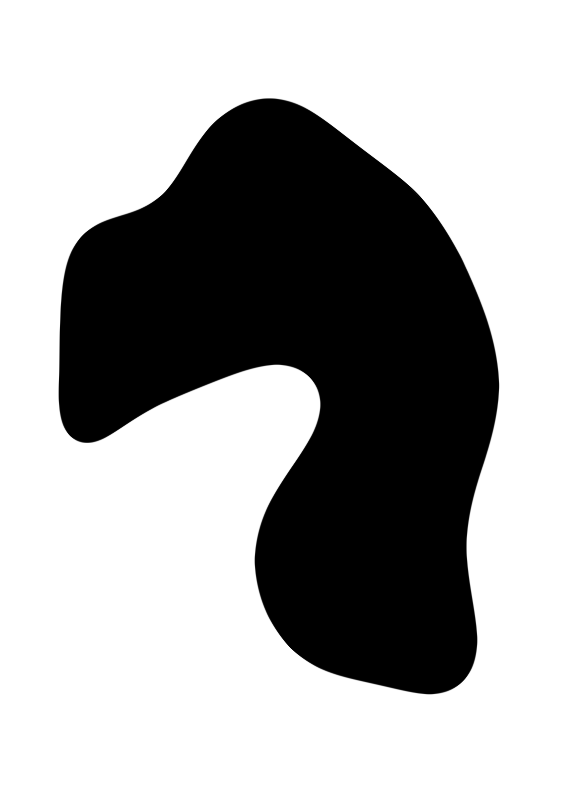} \\[3mm] 
 \includegraphics[width=0.17\linewidth] {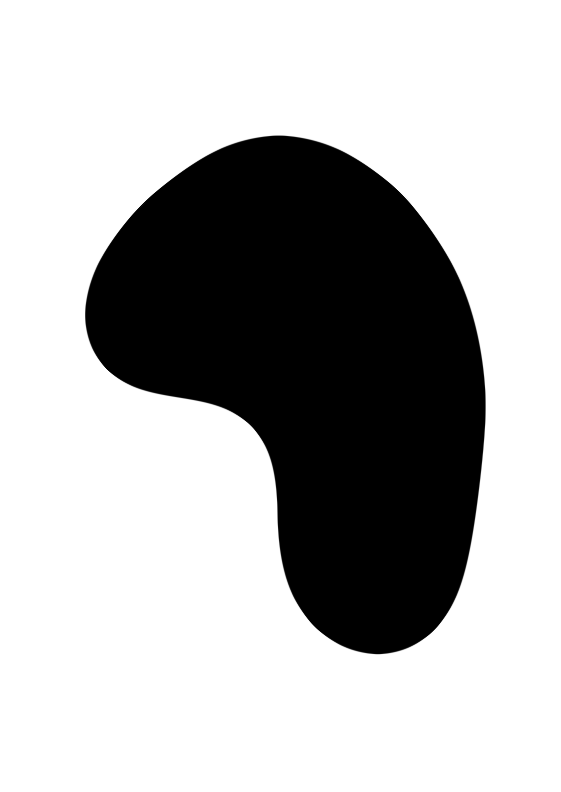} & 
 \includegraphics[width=0.17\linewidth] {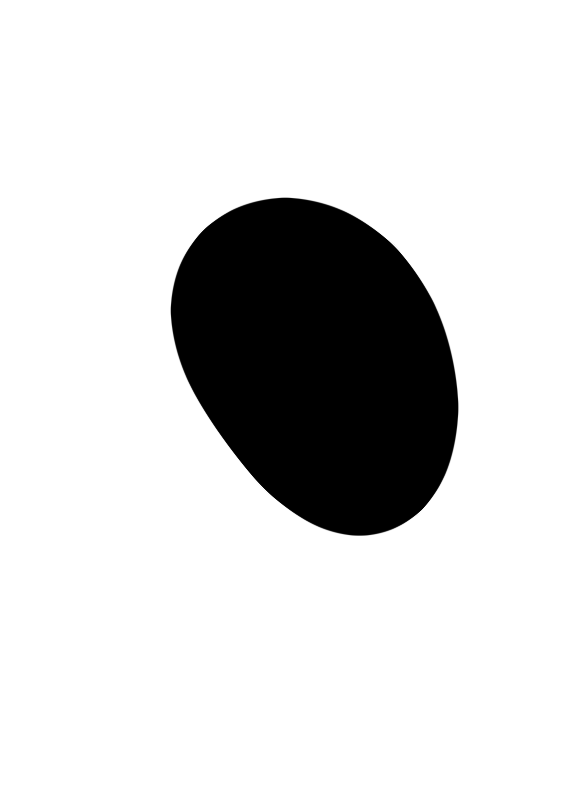} &
 \includegraphics[width=0.17\linewidth] {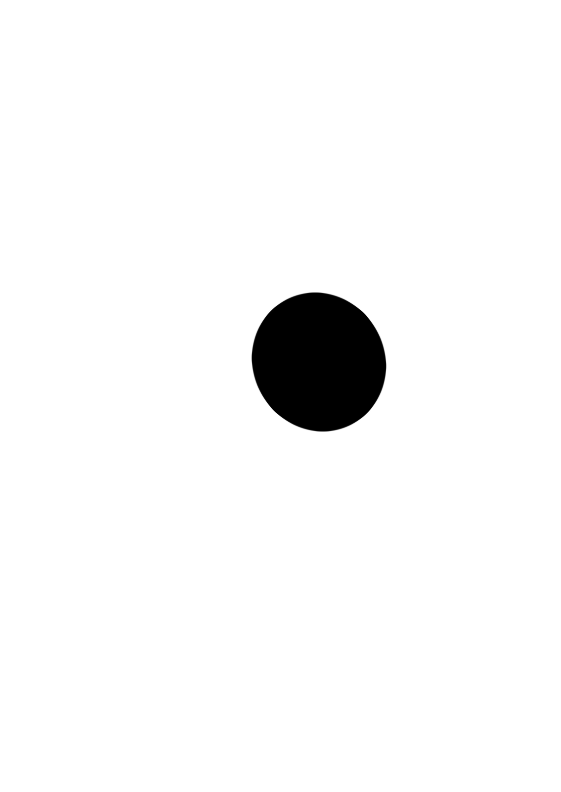} &
 \includegraphics[width=0.17\linewidth] {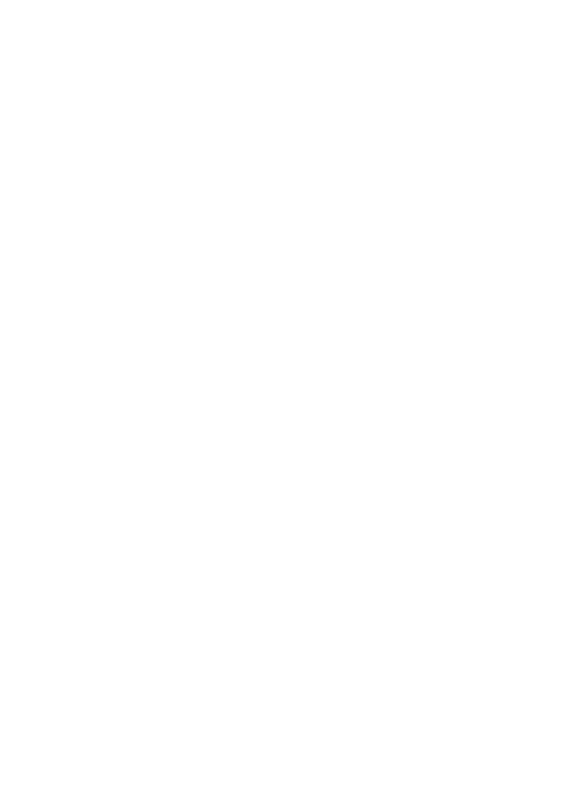} \\
 $t=2000$ & $t=5000$ & $t=8000$ & $t=10000$
\end{tabular}
}}
\end{center}

{\bfseries\caption{\label{fig:witch}\normalfont
 Shape simplification properties of the mode evolution. Image size:
 $561 \times 792$. Source of binarised original image:
 {https://www.kissclipart.com/best-priced-decals-halloween-decor-witch-and-brew-rmm1nq/}.
}}

\end{figure*}

\medskip
In Fig.~\ref{fig:witch}, we study the shape simplification properties
of the mode evolution by applying it to the binary image of a witch.
We observe that under the mode evolution, connected components remain 
connected. It shrinks the shape in such a way that highly curved 
structures evolve faster than less curved ones, resulting in an 
evolution where nonconvex shapes become convex and vanish in finite 
time by shrinking to a so-called circular point. Thus, the mode 
evolution appears to enjoy experimentally the same binary shape 
simplification qualities as the theory states for mean curvature 
motion. This may surprise at first glance when looking only at 
the PDEs: Mean curvature motion is a morphologically invariant 
geometric PDE in the sense of Alvarez et al.~\cite{Alvarez-ARMA93},
while the mode evolution is not.
The M-smoother interpretation can shed some light on this:
While the mode evolution is designed to reproduce the qualities of 
mode filtering, mean curvature motion is related to median filtering. 
For binary data, we face a specific scenario where median and mode 
coincide. Moreover, both median and mode filters preserve the binary 
nature. Finite difference approximations for mean curvature motion,
however, suffer from dissipative artifacts which result in unwanted 
blurring that destroys the binary nature. Because of its backward 
parabolic term, the mode evolution does not suffer from these 
dissipative artifacts. Fig.~\ref{fig:witch} shows that it can preserve 
the binary nature of the data very well. This property constitutes a 
distinctive advantage over mean curvature motion and makes the binary 
mode evolution attractive for shape analysis problems.

\begin{figure*}[p!]

\begin{center}
\scalebox{0.95}{\colorbox{lightgrey}{
\vspace*{4mm}
\begin{tabular}{cccc}
 original & $8$ iterations & $32$ iterations & $87$ iterations\\[1mm]
 \includegraphics[width=0.17\linewidth] {witch} &
 \includegraphics[width=0.17\linewidth] {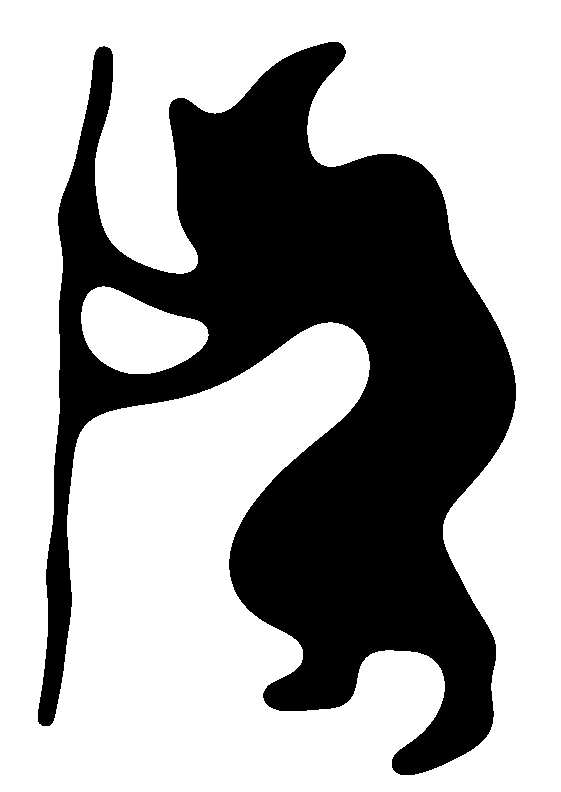} &
 \includegraphics[width=0.17\linewidth] {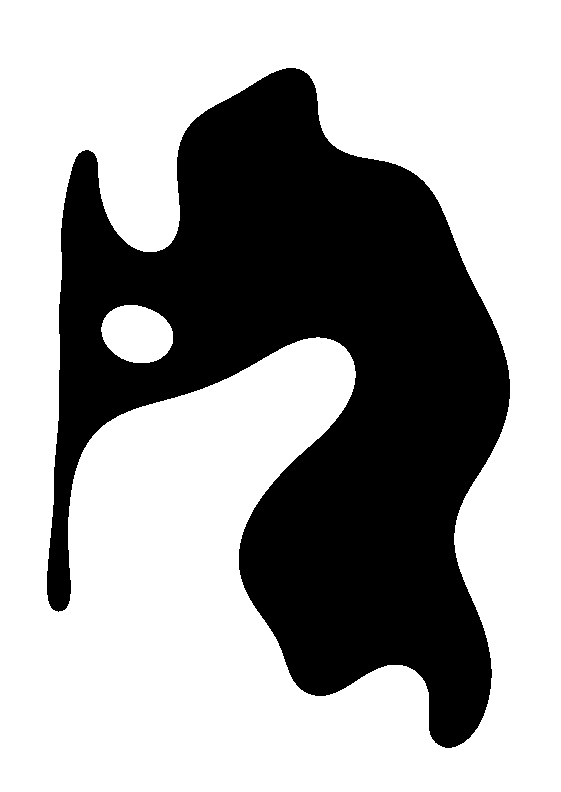} &
 \includegraphics[width=0.17\linewidth] {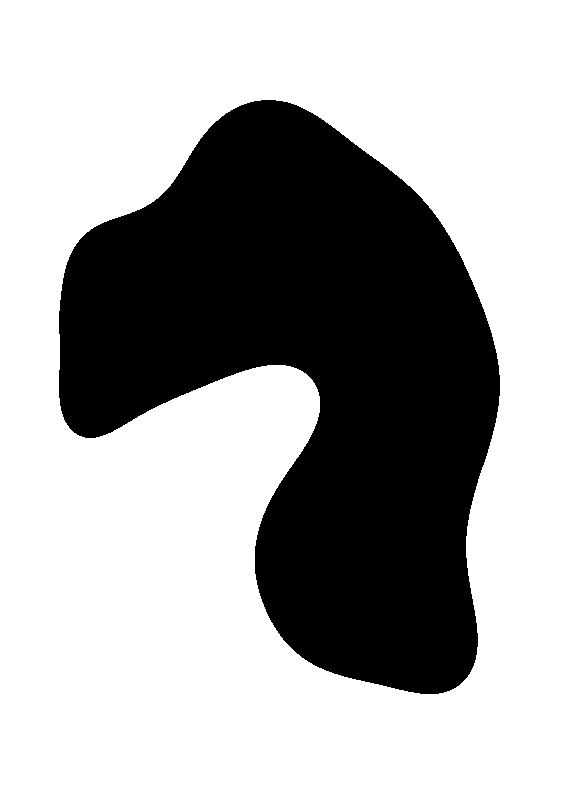} \\[3mm]
 \includegraphics[width=0.17\linewidth] {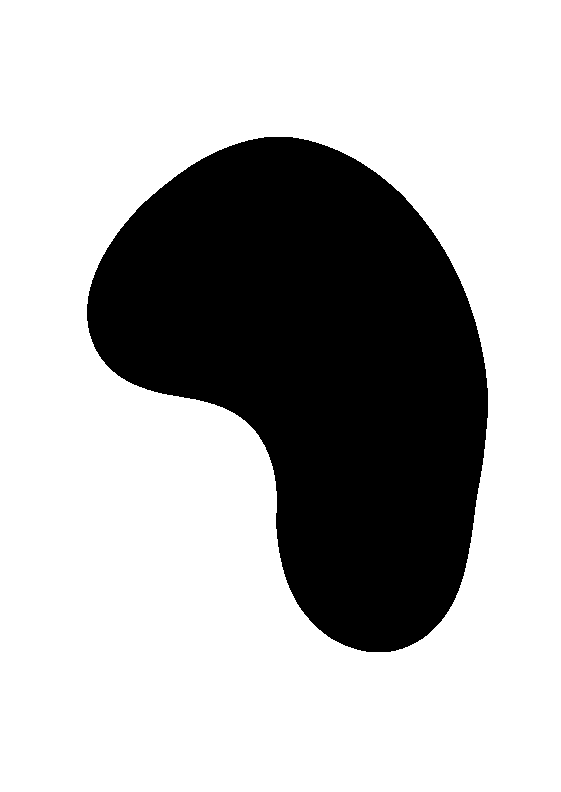} &
 \includegraphics[width=0.17\linewidth] {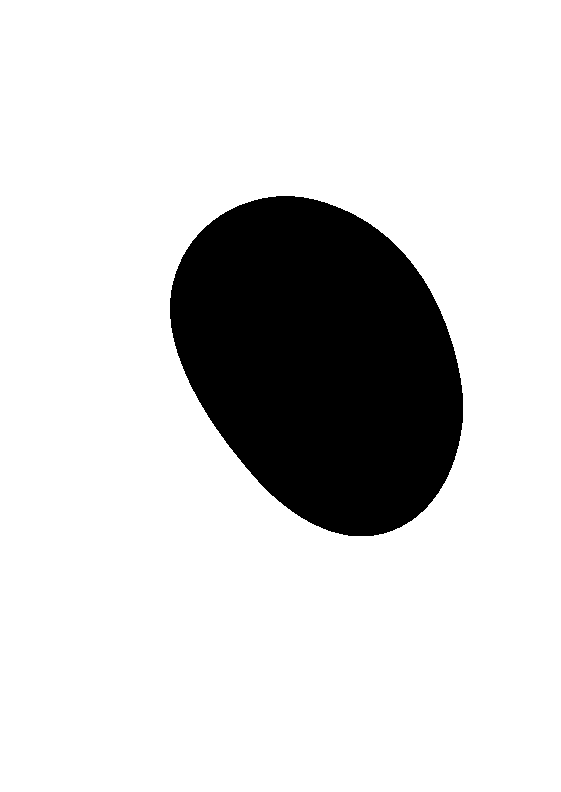} &
 \includegraphics[width=0.17\linewidth] {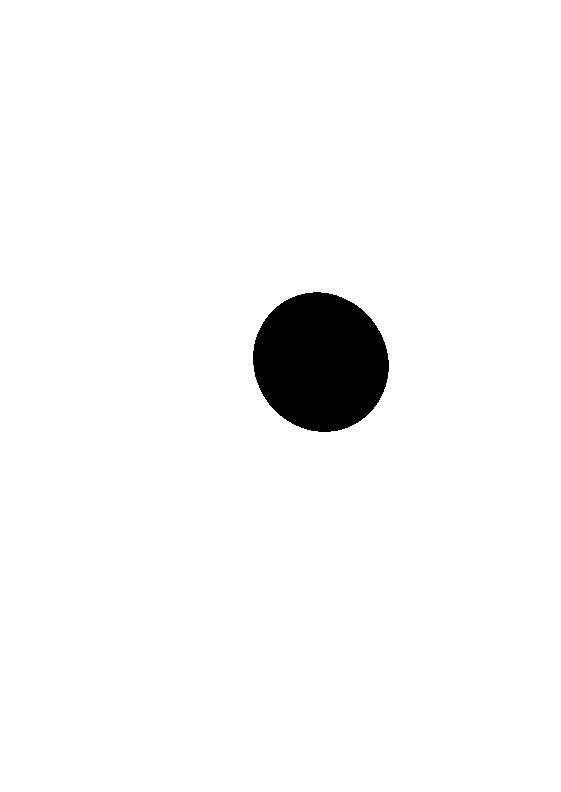} &
 \includegraphics[width=0.17\linewidth] {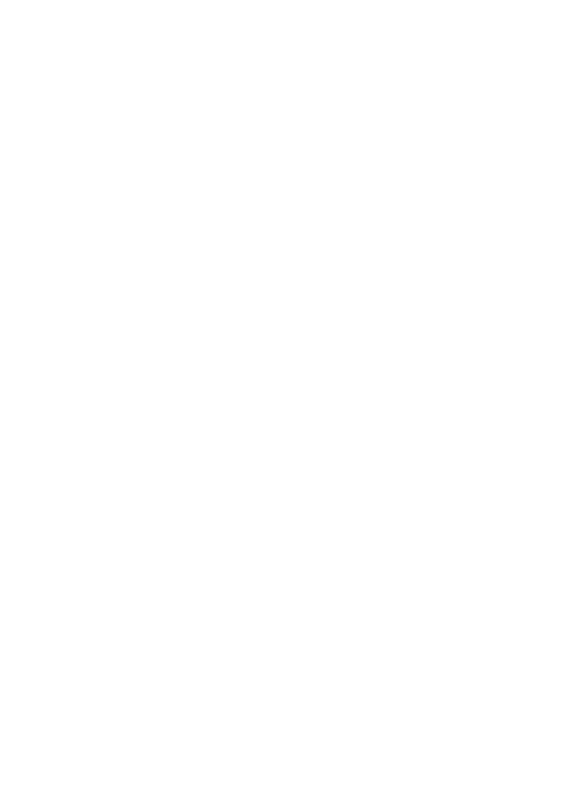} \\
 $220$ iterations & $540$ iterations & $870$ iterations & $1100$ iterations
\end{tabular}
}}
\end{center}

{\bfseries\caption{\label{fig:witch2}\normalfont
 Shape simplification properties of iterated histogram-based mode 
 filtering with a disk of radius $13$. Image size: $561 \times 792$. 
}}

\end{figure*}

\medskip
The results in Fig.~\ref{fig:witch} can be juxtaposed to the ones
in Fig.~\ref{fig:witch2}. The latter one shows the effect of a 
histogram-based implementation of iterative mode filtering: In every 
iteration it 
replaces each pixel by its mode within a disk-shaped neighbourhood 
of radius $13$ pixels. Although our PDE limit has been obtained only 
for vanishing radii and although its numerical scheme approximates the
PDE only with first order consistency, we observe a large qualitative
agreement of Figs.~\ref{fig:witch} and~\ref{fig:witch2}. This
confirms the validity of the PDE limit.

\section{Summary and Conclusions}
\label{sec-conc}

We have established a comprehensive analysis that identifies
the PDE limit for the full class of iterated M-smoothers with order-$p$
means. Our discussion was not restricted to the two-dimensional case which
constitutes the most natural setting in image analysis: We have also 
derived analog results in the one- and three-dimensional case. This 
allows to gain deeper structural insights into the general behaviour
of this filter class.

In the 2D setting, our analysis does not only reproduce known results 
for mean and median filtering, but also corrects a common misconception 
in the literature:
We have shown the surprising fact that in the continuous limit, mode
filtering does not correspond to $p=0$, but results from the limit
$p \to -1$. Moreover, our filter class
$\,u_t=u_{\xi\xi}+(p\!-\!1)\,u_{\eta\eta}\,$ can also be extended to
models that have no interpretation within the setting of M-smoothers,
e.g.~Gabor's classical method for $p=-2$.

At the time being, our results are restricted to grey-value images. An
extension to multivariate data such as colour images or diffusion tensor
fields would be interesting but is not straightforward, and has to be left
to future research. Available results on multivariate median filters 
\cite{Welk-JMIV16,Welk-tr19} indicate that substantial work will be required
for such a generalisation.

Since adequate histogram-based implementations of some M-smoothers such 
as mode filtering can become highly nontrivial when using small local 
histograms \cite{GL03,KS10}, we have proposed a novel numerical algorithm 
in 2D that can handle the PDE evolution for arbitrary values of $p$.
Although these evolutions can be highly anisotropic and may even
exhibit backward parabolic behaviour, we managed to come up with an
$L^\infty$-stable finite difference scheme that is efficient,
satisfies a maximum--minimum principle and shows excellent rotation
invariance. This has been partly achieved by employing and adapting
powerful stabilisation concepts from the numerics of hyperbolic PDEs,
such as upwinding, minmod functions, and curvature limiters.

It should be emphasised that our numerical algorithm is applicable to 
any stable evolution of type $\,u_t = a \, u_{\xi\xi} + b \, u_{\eta\eta}\,$,
where $a$ and $b$ may have arbitrary sign. Thus, it is of very general
nature and covers also numerous applications beyond M-smoothing, including
image interpolation \cite{CMS98}, adaptive filter design \cite{ALM92,CZ96},
many level set methods \cite{OP03}, as well as other second-order PDEs 
in gauge coordinates such as $p$-Laplacian evolutions \cite{CFG19,Ku09}. 

Our experiments indicate that the PDEs for $p<1$, such as the
mode evolution, are particularly appealing: They combine strong
shape simplification properties with pronounced sharpening qualities.
They clearly deserve more research.

Connecting the class of M-smoothers to the family of PDE-based methods
contributes one more mosaic stone to the mathematical foundations of
image analysis. Since M-smoothers themselves are related to many other
approaches \cite{MWB06,BW02a,WAHM99}, including W-smoothers, bilateral
filters, mean-shift and robust estimation, our results can help
to gain a broader and more coherent view on the entire field.

\medskip
\begin{small}
\subsection*{Acknowledgements}
This project has received funding from the European Research Council (ERC)
under the European Union's Horizon 2020 research and innovation programme
(grant agreement No.\ 741215, ERC Advanced Grant INCOVID). We thank
Luis Alvarez (University of Las Palmas de Gran Canaria) for interesting
and inspiring discussions on this topic.
\par
\end{small}
\medskip

\appendix

\section{Proofs of PDE Approximation Results}
\label{app-proofs}

\subsection{Proof of Proposition~\ref{prop-pde-gmdisc-generic}}
\label{proof-prop-pde-gmdisc-generic}

\subsubsection{Preliminaries: Some Important Integrals}

We start by collecting some definite integrals that will be useful in the
following.
We define for $\varrho\in(0,1)$ and $q\in\mathbb{R}$
\begin{align}
I_q &:= \int\nolimits_{\sqrt{\vphantom{X}\varrho}}^1
\sqrt{1-\xi^2}\,\xi^q\dd\xi\;,\\
S_q &:= \int\nolimits_{\arcsin\sqrt{\vphantom{X}\varrho}}^{\pi/2}\sin^q\varphi
\dd\varphi \;.
\end{align}
With the additional abbreviation
\begin{align}
R_{q} &:= \varrho^{q/2}\sqrt{1-\varrho}
\end{align}
we can derive via substituting $\xi=\sin\varphi$ and integration by parts
(integrating $\sin^q\varphi\cos\varphi$ and differentiating $\cos\varphi$)
\begin{align}
I_q &= \int\nolimits_{\arcsin\sqrt{\vphantom{X}\varrho}}^{\pi/2}
\sin^q\varphi\cos^2\varphi\dd\varphi
= \frac1{q+1}\bigl[\sin^{q+1}\varphi\cos\varphi
\bigr]_{\arcsin\sqrt{\vphantom{X}\varrho}}^{\pi/2}
+\frac1{q+1}S_{q+2}
\notag\\*&
= \frac{-1}{q+1}R_{q+1}+\frac1{q+1}S_{q+2}
\label{Iq-generic}
\end{align}
for $q\ne-1$.
Moreover, we have by $1=\sin^2\varphi+\cos^2\varphi$
\begin{align}
S_q &= 
\int\nolimits_{\arcsin\sqrt{\vphantom{X}\varrho}}^{\pi/2}
\sin^q\varphi\cos^2\varphi\dd\varphi + S_{q+2}
=\frac{-1}{q+1}R_{q+1}+\frac{q+2}{q+1}S_{q+2}
\label{Sq-generic}
\end{align}
for $q\ne-1$ which allows to transform $S_q$ into $S_{q+2}$ and vice versa.

From \eqref{Iq-generic} we can obtain thereby
\begin{align}
I_{p-4} &= \frac{-1}{p-3}R_{p-3} + \frac{-1}{(p-1)(p-3)}R_{p-1}
+ \frac{-p}{(p+1)(p-1)(p-3)}R_{p+1} 
\notag\\*&\quad{}
+ \frac{(p+2)p}{(p+1)(p-1)(p-3)}S_{p+2}\;,
\label{Ipm4}
\\
I_{p-2} &= \frac{-1}{p-1}R_{p-1} + \frac{-1}{(p+1)(p-1)}R_{p+1}
+\frac{p+2}{(p+1)(p-1)}S_{p+2}\;,
\label{Ipm2}
\\
I_p &= \frac{-1}{p+1}R_{p+1} + \frac{1}{p+1}S_{p+2}\;,
\label{Ip}
\\
I_{p+2} &= \frac{-1}{p+4}R_{p+3} + \frac{1}{p+4}S_{p+2}\;,
\label{Ipp2}
\end{align}
for real $p$ with exception of some odd integers. Note that also for the
exceptional values (where some of the denominators become zero) the
integrals exist.

\subsubsection{Regular Points: Ansatz via Taylor Expansion}

Let the image $u$ and mean order $p$ be given as in the proposition.
Assume w.l.o.g.\ that
the regular location $\boldsymbol{x}_0$ is $(0,0)$ with $u(0,0)=0$, and that
the gradient of $u$ at $(0,0)$ is in the positive $x$ direction, i.e.,
$u_x>0$, $u_y=0$.
Let a neighbourhood radius $\varrho>0$ be given, and denote the
closed (Euclidean) $\varrho$-neighbourhood of $(0,0)$ by
$\mathrm{D}_{\varrho}$.

Using Taylor expansion of $u$ up to third order, we can write
for $(x,y)\in\mathrm{D}_\varrho$
\begin{align}
u(x,y) &= \alpha\bigl(x + \beta x^2 + \gamma xy + \delta y^2
+ \varepsilon_0x^3+\varepsilon_1x^2y
+\varepsilon_2xy^2+\varepsilon_3y^3\bigr)
+ \mathcal{O}\bigl((x+y)^4\bigr)
\label{u-taylor}
\end{align}
where $\alpha=u_x$, $2\beta = u_{xx}/u_x$, $\gamma = u_{xy}/u_x$,
$2\delta=u_{yy}/u_x$.

We assume that $\varrho$ is chosen small enough such that $u_x$ is positive
everywhere in $\mathrm{D}_\varrho$,
each level set of $u$ within the disc
$\mathrm{D}_\varrho$ is either a smooth line connecting two points at the
circular boundary of the disc, or one of two single points on the boundary
of $\mathrm{D}_\varrho$ where $u$ takes its maximum and minimum on
$\mathrm{D}_\varrho$, respectively.

The order-$p$ mean of $u$ within $\mathrm{D}_\varrho$ is the minimiser
of
\begin{equation}
E_0(\mu) := \sgn(p)
\iint\nolimits_{\mathrm{D}_\varrho}\lvert u(x,y)-\mu\rvert^p\dd y\dd x \;.
\end{equation}
By some rough estimates one can conclude that for $\varrho\to0$,
$\mu\sim\varrho^2$.
We substitute therefore
\begin{gather}
x = \varrho\xi\;, \quad
y = \varrho\eta\;, \quad
\mu = \varrho^2\alpha\kappa\;, 
\quad
u(x,y) = \varrho\alpha\omega(\xi,\eta)
\label{D1substitution}
\end{gather}
and obtain
\begin{align}
E_0(\mu) &= \sgn(p) \varrho^{p+2}\alpha^p E(\kappa) \;,
\\
E(\kappa) &= \iint\nolimits_{\mathrm{D}_1}\lvert\omega-\kappa\varrho\rvert^p
\dd\eta\dd\xi \;,
\label{Ekappa2D}
\\
\omega(\xi,\eta)&=\xi+\beta\xi^2\varrho+\gamma\xi\eta\varrho+\delta\eta^2\varrho
+\varepsilon_0\xi^3\varrho^2+\varepsilon_1\xi^2\eta\varrho^2
+\varepsilon_2\xi\eta^2\varrho^2+\varepsilon_3\eta^3\varrho^2
\notag\\*&\quad{}
+\mathcal{O}\bigl(\varrho^3(\xi+\eta)\bigr)
\;.
\label{omega-taylor}
\end{align}
In the following we focus therefore on finding the extremum of $E$
(minimum for $p>0$, maximum for $p<0$).

\subsubsection{Separation of the Integral}
\label{proof-prop-gmdisc-generic-3}

The integral $E$ from \eqref{Ekappa2D} can be reorganised into a nested
integration where the inner integral integrates
along a level line of $\omega$ going through $(\xi,0)$, and the outer integral
then integrates along the $\xi$ axis. We have
\begin{align}
E(\kappa) &= \int\nolimits_{-1}^{1} 
\left(\,
\int\nolimits_{\eta^*_-(\xi)}^{\eta^*_+(\xi)}
\frac{1}
     {\frac{\partial\omega}{\partial\xi}\bigl(\tilde{\xi}(\eta),\eta\bigr)}
\dd\eta
\right) 
\lvert\omega(\xi,0)-\kappa\varrho\rvert^p
\frac{\partial\omega}{\partial\xi}(\xi,0)
\dd\xi
+\mathcal{O}(\varrho^3)
\label{EkappaLL}
\end{align}
where $\tilde{\xi}$ is a function of $\eta$ that describes the level line of
$\omega$ that goes through $(\xi,0)$, and reaches the boundary of
$\mathrm{D}_1$ at $\eta^*_+>0$ and $\eta^*_-<0$. (Note that the fact that
$\omega_\xi$ is positive throughout $\mathrm{D}_1$ implies that the level
line through $(\xi,0)$ can be described in this way.)

The error term $\mathcal{O}(\varrho^3)$ results from the neglection of
those level lines near the maximum and minimum of $\omega$ within
$\mathrm{D}_1$ that do not reach the $\xi$ axis within $\mathrm{D}_1$.

In \eqref{EkappaLL},
the inner integral
\begin{align}
V(\xi) &:= \int\nolimits_{\eta^*_-(\xi)}^{\eta^*_+(\xi)}
\frac{1}
     {\frac{\partial\omega}{\partial\xi}\bigl(\tilde{\xi}(\eta),\eta\bigr)}
\dd\eta
\label{Vxi}
\end{align}
measures the density of the value $\omega(\xi,0)$ in the
overall distribution of $\omega$ values within $\mathrm{D}_1$ by
integrating along the level line $\tilde{\xi}(\eta)$ with $\eta$ as
integration parameter the inverse density of level lines in $\xi$ direction.
It is important here that the inverse density of level lines is measured
in a direction perpendicular to that of integration.
The density of level lines in $\xi$ direction is exactly the derivative
$\partial\omega/\partial\xi$ taken at the point $(\tilde{\xi},\eta)$, i.e.,
the denominator of the integrand.

Integrating the quantity $V$ multiplied with the penaliser
$\lvert\omega-\kappa\varrho\rvert^p$ would directly yield $E(\kappa)$
if the integration were carried out w.r.t.\ $\omega$.
We prefer, however, to keep the integration over $\xi$ in order to avoid
plugging in the inverse function of $\omega(\xi)\equiv\omega(\xi,0)$ 
everywhere in the
expressions. This is compensated by the factor
$(\partial\omega/\partial\xi)(\xi,0)$ placed at the end of the integrand
of \eqref{EkappaLL} that represents just the substitution of $\omega$
with $\xi$ (along the $\xi$ axis $\eta=0$) as integration variable.

For ease of evaluation, we combine in the following the substitution factor
with the weight $V(\xi)$ in one single expression:
\begin{align}
W(\xi) &:= \frac{\partial\omega}{\partial\xi}(\xi,0) \, V(\xi)
= \int\nolimits_{\eta^*_-(\xi)}^{\eta^*_+(\xi)}
\frac{\frac{\partial\omega}{\partial\xi}(\xi,0)}
     {\frac{\partial\omega}{\partial\xi}\bigl(\tilde{\xi}(\eta),\eta\bigr)}
\dd\eta
\label{EkappaLLi}
\end{align}

\subsubsection{Evaluation of the Inner (Weight) Integral}
\label{proof-prop-gmdisc-generic-4}

To evaluate \eqref{EkappaLLi},
we determine first the level line function $\tilde{\xi}(\eta)$ for given
$\xi=\tilde{\xi}(0)$ by using the Taylor expansion \eqref{omega-taylor}:
\begin{align}
\omega(\xi,0) &= \omega(\tilde{\xi}(\eta),\eta)\\*
\xi+\beta\xi^2\varrho+\varepsilon_0\xi^3\varrho^2 &=
\tilde{\xi}
+\bigl(\beta\tilde{\xi}^2+\gamma\tilde{\xi}\eta+\delta\eta^2\bigr)\varrho 
\notag\\*&\quad{}
+\bigl(\varepsilon_0\tilde{\xi}^3+\varepsilon_1\tilde{\xi}^2\eta
+\varepsilon_2\tilde{\xi}\eta^2+\varepsilon_3\eta^3\bigr)\varrho^2
+\mathcal{O}(\varrho^3) \\
\tilde{\xi}(\eta) &= \xi - (\gamma\xi+\delta\eta)\eta\varrho 
\notag\\*&\quad{}
+\bigl((2\beta\xi+\gamma\eta)(\gamma\xi+\delta\eta)
-\varepsilon_1\xi^2-\varepsilon_2\xi\eta-\varepsilon_3\eta^2\big)\eta\varrho^2
\notag\\*&\quad{}
+\mathcal{O}(\varrho^3) \;.
\label{xiLL}
\end{align}
The $\eta$ coordinates $\eta^*_{\pm}$ of the end points of the level line
are obtained from the condition $\tilde{\xi}^2+{\eta^*}^2=1$ as
\begin{align}
\eta^*_{\pm} &= \pm\sqrt{1-\xi^2}
+\bigl(\gamma\xi^2\pm\delta\xi\sqrt{1-\xi^2}\bigr)\varrho
+\bigl(\chi(\xi)\pm\psi(\xi)\sqrt{1-\xi^2}\bigr)\varrho^2
+\mathcal{O}(\varrho^3) 
\label{etapm}
\end{align}
where
\begin{align}
\chi(\xi)&=\chi_0+\chi_1\xi+\chi_2\xi^2+\chi_3\xi^3+\chi_4\xi^4\;,\\
\psi(\xi)&=\psi_0+\psi_1\xi+\psi_2\xi^2+\psi_3\xi^3
\label{polypsi}
\end{align}
are polynomials in $\xi$ the exact coefficients
of which are not further needed.

Based on the Taylor expansion \eqref{omega-taylor} we obtain
\begin{align}
\frac{\partial\omega}{\partial\xi}(\xi,\eta)
&=1+\bigl(2\beta\xi+\gamma\eta\bigr)\varrho
+\bigl(3\varepsilon_0\xi^2+2\varepsilon_1\xi\eta+\varepsilon_2\eta^2)\varrho^2
+\mathcal{O}(\varrho^3) \;, 
\label{domega-dxi-taylor}
\\
\frac{\partial\omega}{\partial\xi}(\xi,0)
&=1+2\beta\xi\varrho
+3\varepsilon_0\xi^2\varrho^2
+\mathcal{O}(\varrho^3) \;,
\label{domega-dxi-taylor-xiaxis}
\end{align}
and with \eqref{xiLL}
\begin{align}
\frac{\partial\omega}{\partial\xi}(\tilde{\xi},\eta)
&=1+\bigl(2\beta\xi+\gamma\eta\bigr)\varrho
+\bigl(-2\beta\gamma\xi\eta-2\beta\delta\eta^2
+3\varepsilon_0\xi^2+2\varepsilon_1\xi\eta+\varepsilon_2\eta^2\bigr)\varrho^2
+\mathcal{O}(\varrho^3) \;.
\label{domega-dxi-taylor-LL}
\end{align}
Combining \eqref{domega-dxi-taylor-xiaxis} and \eqref{domega-dxi-taylor-LL}
we have
\begin{align}
\frac{\frac{\partial\omega}{\partial\xi}(\xi,0)}
     {\frac{\partial\omega}{\partial\xi}(\tilde{\xi},\eta)}
&=
1-\gamma\eta\varrho
+\bigl(4\beta\gamma\xi\eta+2\beta\delta\eta^2+\gamma^2\eta^2
-2\varepsilon_1\xi\eta-\varepsilon_2\eta^2\bigr)\varrho^2
+\mathcal{O}(\varrho^3)
\label{domega-dxi-axisLLquotient}
\end{align}
and therefore
\begin{align}
W(\xi)
&=
\int\nolimits_{\eta^*_-}^{\eta^*_+} \dd\eta
+\bigl(-\gamma\varrho+4\beta\gamma\xi\varrho^2-2\varepsilon_1\xi\varrho^2\bigr)
\int\nolimits_{\eta^*_-}^{\eta^*_+} \eta \dd\eta
\notag\\&\quad{}
+\bigl(2\beta\delta+\gamma^2-\varepsilon_2\bigr)\varrho^2
\int\nolimits_{\eta^*_-}^{\eta^*_+} \eta^2 \dd\eta
+\mathcal{O}(\varrho^3)
\notag\\
&=
\bigl(\eta^*_+-\eta^*_-\bigr)
+\frac12
\bigl(-\gamma+4\beta\gamma\xi\varrho-2\varepsilon_1\xi\varrho\bigr)\varrho
\bigl({\eta^*_+}^2-{\eta^*_-}^2\bigr)
\notag\\*&\quad{}
+\frac13
\bigl(2\beta\delta+\gamma^2-\varepsilon_2\bigr)\varrho^2
\bigl({\eta^*_+}^3-{\eta^*_-}^3\bigr)
+\mathcal{O}(\varrho^3)\;.
\label{Wxi-2}
\end{align}
From \eqref{etapm} one sees that
\begin{align}
\eta^*_+-\eta^*_-
&=
2\sqrt{1-\xi^2}+2\delta\xi\sqrt{1-\xi^2}\varrho
+2\psi(\xi)\sqrt{1-\xi^2}\varrho^2
+\mathcal{O}(\varrho^3)\;,
\\
{\eta^*_+}^2-{\eta^*_-}^2
&=
4\gamma\xi^2\sqrt{1-\xi^2}\varrho
+\mathcal{O}(\varrho^2)\;,
\\
{\eta^*_+}^3-{\eta^*_-}^3
&=
2(1-\xi^2)^{3/2} 
+\mathcal{O}(\varrho)\;,
\end{align}
which allows to continue \eqref{Wxi-2} into
\begin{align}
W(\xi) &=
2\sqrt{1-\xi^2}+2\delta\xi\sqrt{1-\xi^2}\varrho
+2\psi(\xi)\sqrt{1-\xi^2}\varrho^2
-
2\gamma^2\xi^2\sqrt{1-\xi^2}\varrho^2
\notag\\*&\quad{}
+
\frac23\bigl(2\beta\delta+\gamma^2-\varepsilon_2\bigr)(1-\xi^2)^{3/2}
\varrho^2
+\mathcal{O}(\varrho^3)
\notag \\
&=
\left(
2
+\left(
\frac43\beta\delta+\frac23\gamma^2-\frac23\varepsilon_2
+2\psi_0
\right)
\varrho^2
\right)
\sqrt{1-\xi^2}
+
\left(
2\delta
+2\psi_1\varrho
\right)
\varrho
\xi\sqrt{1-\xi^2}
\notag\\*&\quad{}
+
\left(
-2\gamma^2
-\frac43\beta\delta-\frac23\gamma^2+\frac23\varepsilon_2
+2\psi_2
\right)
\varrho^2
\xi^2\sqrt{1-\xi^2}
\notag\\*&\quad{}
+
2\psi_3\varrho^2
\xi^3\sqrt{1-\xi^2}
+\mathcal{O}(\varrho^3)
\notag \\
&= \Bigl(
(w_{0,0}+w_{0,2}\varrho^2)
+w_1\varrho\xi
+w_2\varrho^2\xi^2
+w_3\varrho^2\xi^3\Bigr) \sqrt{1-\xi^2}
+\mathcal{O}(\varrho^3)
\label{Wxi-final}
\end{align}
with
\begin{equation}
\left.
\begin{aligned}
w_{0,0} &= 2\;, \\
w_{0,2} &= \frac43\beta\delta+\frac23\gamma^2-\frac23\varepsilon_2 +2\psi_0\;,
\\
w_1 &= 2\delta +2\psi_1\varrho \;, \\
w_2 &= -2\gamma^2
-\frac43\beta\delta-\frac23\gamma^2+\frac23\varepsilon_2
+2\psi_2 \;,
\\
w_3 &= 2\psi_3 \;.
\end{aligned}
\quad
\right\}
\label{w0thru3}
\end{equation}

\subsubsection{Domain Splitting of the Outer Integral}
\label{proof-prop-gmdisc-generic-5}

The outer integral of \eqref{EkappaLL}, i.e., the integration of
$W(\xi)$ with the penaliser function
$\lvert\omega-\kappa\varrho\rvert^p$,
is now split into four parts.

First, we split the integration interval at $\xi=\nu\varrho$ where
$\omega(\nu\varrho)=\kappa\varrho$ to reduce
$\lvert\omega-\kappa\varrho\rvert$ to either $\omega-\kappa\varrho$
or $-\omega+\kappa\varrho$ in each subinterval.
By \eqref{omega-taylor} one has $\nu=\kappa+\mathcal{O}(\varrho^2)$.

Second, the density term $W(\xi)$ contains $\sqrt{1-\xi^2}$ which is
not differentiable at $\pm1$, precluding
Taylor expansion of this term near the outer interval boundaries.
On the other hand, the $p$-th power penaliser is for $p\le1$ not
differentiable at $0$ and can therefore not be treated by Taylor
expansion at the boundary $\nu$ between the two integration intervals.
For this reason, we split each of the two intervals
again at $\lvert\xi\rvert=\sqrt{\vphantom{X}\varrho}$.
This allows to simplify the integrals in later steps by applying
Taylor expansion to either $W(\xi)$ or the penaliser function,
safely avoiding the critical regions of each.

As a result, we have
\begin{align}
E(\kappa) 
&= F_-(\kappa)+G_-(\kappa)+G_+(\kappa)+F_+(\kappa)
+ \mathcal{O}(\varrho^3)
\;,
\label{Ekappa-split}
\\
F_-(\kappa) &= \int\nolimits_{-1}^{-\sqrt{\vphantom{X}\varrho}}
W(\xi)\,\bigl(-\omega(\xi)+\kappa\varrho\bigr)^p \dd\xi 
= \int\nolimits_{\sqrt{\vphantom{X}\varrho}}^1
W(-\xi)\,\bigl(-\omega(-\xi)+\kappa\varrho\bigr)^p \dd\xi
\;,
\\
G_-(\kappa) &= \int\nolimits_{-\sqrt{\vphantom{X}\varrho}}^{\nu\varrho}
W(\xi)\,\bigl(-\omega(\xi)+\kappa\varrho\bigr)^p \dd\xi 
\notag \\ &
= \int\nolimits_0^{\sqrt{\vphantom{X}\varrho}+\nu\varrho}
W(-(\xi-\nu\varrho))
\bigl(-\omega(-(\xi-\nu\varrho))+\kappa\varrho\bigr)^p
\dd\xi
\;,
\\
G_+(\kappa) &= \int\nolimits_{\nu\varrho}^{\sqrt{\vphantom{X}\varrho}}
W(\xi)\,\bigl(\omega(\xi)-\kappa\varrho\bigr)^p \dd\xi
\notag \\ &
= \int\nolimits_0^{\sqrt{\vphantom{X}\varrho}-\nu\varrho}
W(\xi+\nu\varrho)
\,
\bigl(\omega(\xi+\nu\varrho)-\kappa\varrho\bigr)^p
\dd\xi
\;,
\\
F_+(\kappa) &= \int\nolimits_{\sqrt{\vphantom{X}\varrho}}^1
W(\xi)\,\bigl(\omega(\xi)-\kappa\varrho\bigr)^p \dd\xi
\;.
\end{align}

\subsubsection{Evaluation of the Outer Integral I}
\label{proof-prop-gmdisc-generic-6}

We start by evaluating the integrals $F_{\mp}$. In the following the
upper signs refer to $F_-$, the lower ones to $F_+$. In expanding the
power $(1+\ldots)^p$ by a Taylor series, it is important to note that
$\varrho/\xi$ is of order $\mathcal{O}\bigl(\sqrt{\vphantom{X}\varrho}\bigr)$
due to the lower integral bound.
\begin{align}
F_{\mp} 
&= \int\nolimits_{\sqrt{\vphantom{X}\varrho}}^1
W(\mp\xi)\,\bigl(\mp\omega(\mp\xi)\pm\kappa\varrho\bigr)^p \dd\xi
\notag \\
&= \int\nolimits_{\sqrt{\vphantom{X}\varrho}}^1
W(\mp\xi)
\,
\bigl(\xi\mp\beta\xi^2\varrho\pm\kappa\varrho
+\varepsilon_0\xi^3\varrho^2+\mathcal{O}(\varrho^3\xi)\bigr)^p \dd\xi
\notag \\
&= \int\nolimits_{\sqrt{\vphantom{X}\varrho}}^1
W(\mp\xi)
\xi^p\,\left(1\pm\kappa\frac{\varrho}{\xi}\mp\beta\xi\varrho
+\varepsilon_0\xi^2\varrho^2+\mathcal{O}(\varrho^3)\right)^p \!\!\dd\xi
\notag \\
&= \int\nolimits_{\sqrt{\vphantom{X}\varrho}}^1
W(\mp\xi)\,\xi^p
\,
\left(1 \vphantom{\binom{p}{0}}
\pm p\kappa\frac{\varrho}{\xi}\mp p\beta\xi\varrho+p\varepsilon_0\xi^2\varrho^2
+\binom{p}{2}\kappa^2\frac{\varrho^2}{\xi^2}-2\binom{p}{2}\beta\kappa\varrho^2
+\binom{p}{2}\beta^2\xi^2\varrho^2
\right. \notag\\&\qquad\qquad\qquad\qquad\left.{}
\mp\binom{p}{3}\kappa^3\frac{\varrho^3}{\xi^3}
+\binom{p}{4}\kappa^4\frac{\varrho^4}{\xi^4}
+\mathcal{O}(\varrho^{5/2})
\right)
\dd\xi
\notag \\
&= \int\nolimits_{\sqrt{\vphantom{X}\varrho}}^1
\Bigl(
(w_{0,0}+w_{0,2}\varrho^2)
\mp w_1\varrho\xi
+w_2\varrho^2\xi^2
\mp w_3\varrho^2\xi^3 
+ \mathcal{O}(\varrho^3) 
\Bigr) \xi^p\sqrt{1-\xi^2} %
\notag\\&\qquad{}\times
\left(1
\pm p\kappa\frac{\varrho}{\xi} %
\mp p\beta\xi\varrho %
+\binom{p}{2}\kappa^2\frac{\varrho^2}{\xi^2} %
\mp\binom{p}{3}\kappa^3\frac{\varrho^3}{\xi^3} %
+p\varepsilon_0\xi^2\varrho^2 %
-2\binom{p}{2}\beta\kappa\varrho^2 %
+\binom{p}{2}\beta^2\xi^2\varrho^2 %
\right.\notag\\&\qquad\quad\left.{}
+\binom{p}{4}\kappa^4\frac{\varrho^4}{\xi^4} %
+ \mathcal{O}(\varrho^{5/2})
\right)
\dd\xi\;.
\label{FmFp-intermed1}
\end{align}
This gives
\begin{align}
F_-+F_+ &=
2\int\nolimits_{\sqrt{\vphantom{X}\varrho}}^1
w_{0,0}
\xi^p\sqrt{1-\xi^2} 
\left(1+\binom{p}{2}\kappa^2\frac{\varrho^2}{\xi^2} %
+p\varepsilon_0\xi^2\varrho^2 %
-2\binom{p}{2}\beta\kappa\varrho^2 %
+\binom{p}{2}\beta^2\xi^2\varrho^2 %
\right.\notag\\&\qquad\quad\left.{}
+\binom{p}{4}\kappa^4\frac{\varrho^4}{\xi^4} %
+ \mathcal{O}(\varrho^{5/2})
\right)
\dd\xi
\notag \\ &\quad{} 
+2\int\nolimits_{\sqrt{\vphantom{X}\varrho}}^1
\Bigl(
w_{0,2}
+w_2\xi^2
\Bigr) \varrho^2 \xi^p\sqrt{1-\xi^2}
\left(1
+ \mathcal{O}(\varrho^{1/2})
\right)
\dd\xi
\notag\\&\quad{} 
+2\int\nolimits_{\sqrt{\vphantom{X}\varrho}}^1
\Bigl(
w_1\varrho\xi
+ \mathcal{O}(\varrho^2) 
\Bigr) \xi^p\sqrt{1-\xi^2}
\left(
-p\kappa\frac{\varrho}{\xi} %
+p\beta\xi\varrho %
+ \mathcal{O}(\varrho^{3/2})
\right)
\dd\xi
\notag\\
&= 
2w_{0,0}\binom{p}{4}\kappa^4\varrho^4
I_{p-4}
+2
w_{0,0}\binom{p}{2}\kappa^2\varrho^2
I_{p-2}
\notag\\&\quad{}
+2\left(
w_{0,0} - 2w_{0,0}\binom{p}{2}\beta\kappa\varrho^2
+ w_{0,2}\varrho^2 
- w_1p\kappa\varrho^2
\right) I_p
\notag\\*&\quad{}
+2\left(
w_{0,0}p\varepsilon_0\varrho^2+w_{0,0}\binom{p}{2}\beta^2\varrho^2
+ w_2\varrho^2
+w_1p\beta\varrho^2
\right) I_{p+2}
+ \mathcal{O}(\varrho^{5/2})
\label{FmFp-intermed}
\end{align}
and by \eqref{Ipm4}--\eqref{Ipp2} we obtain
\begin{align}
F_-+F_+ 
&=
2w_{0,0}\binom{p}{4}\kappa^4\varrho^4
\left(\frac{-R_{p-3}}{p-3} + \frac{-R_{p-1}}{(p-1)(p-3)}
+ \frac{-p\,R_{p+1}}{(p+1)(p-1)(p-3)} 
\right.\notag\\&\qquad\qquad\qquad\qquad\left.{}
+ \frac{(p+2)p\,S_{p+2}}{(p+1)(p-1)(p-3)}\right)
\notag\\*&\quad{}
+2\left(
w_{0,0}\binom{p}{2}\kappa^2\varrho^2
\right) 
\left(\frac{-R_{p-1}}{p-1} 
+ \frac{-R_{p+1}}{(p+1)(p-1)}
+\frac{(p+2)\,S_{p+2}}{(p+1)(p-1)}\right)
\notag\\&\quad{}
+2\left(
w_{0,0} - 2w_{0,0}\binom{p}{2}\beta\kappa\varrho^2
+ w_{0,2}\varrho^2 
- w_1p\kappa\varrho^2
\right) 
\left(\frac{-R_{p+1}}{p+1} + \frac{S_{p+2}}{p+1}\right)
\notag\\&\quad{}
+2\left(
w_{0,0}p\varepsilon_0\varrho^2+w_{0,0}\binom{p}{2}\beta^2\varrho^2
+ w_2\varrho^2
+w_1p\beta\varrho^2
\right) 
\left(\frac{-R_{p+4}}{p+4} + \frac{S_{p+2}}{p+4}\right)
\notag\\&\quad{}
+\mathcal{O}(\varrho^{5/2})
\label{Fpm-intermed}
\\
&=
w_{0,0}\frac{2}{p+1}
S_{p+2}
+
\left(
w_{0,0}\frac{(p+2)p}{(p+1)}\kappa^2
-w_{0,0}\frac{2p(p-1)}{p+1}\beta\kappa
+w_{0,2}\frac{1}{p+1}
-w_1\frac{p}{p+1}\kappa
\right.\notag\\&\qquad\qquad\qquad\left.{}
+w_{0,0}\frac{2p}{p+4}\varepsilon_0
+w_{0,0}\frac{p(p-1)}{(p+4)}\beta^2
+w_2\frac{2}{p+4}
+w_1\frac{2p}{p+4}\beta
\right)
\varrho^2 S_{p+2}
\notag\\*&\quad{}
-
w_{0,0}\frac{2}{p+1}
\varrho^{(p+1)/2}\sqrt{1-\varrho}
-
w_{0,0}p\kappa^2
\varrho^{(p+3)/2}\sqrt{1-\varrho}
\notag\\&\quad{}
-
\left(
w_{0,0}\frac{p(p-1)(p-2)}{12}\kappa^4
+w_{0,0}\frac{p}{(p+1)}\kappa^2
-w_{0,0}\frac{2p(p-1)}{p+1}\beta\kappa
\right)
\varrho^{(p+5)/2}\sqrt{1-\varrho}
\notag\\&\quad{}
+\mathcal{O}(\varrho^{5/2})\;.
\label{Fpm-final}
\end{align}
In the intermediate step \eqref{Fpm-intermed} the factors $p-1$, $p-3$ occur
in the denominators of some terms, which would necessitate the exclusion of
$p=1$ and $p=3$. However, we see in \eqref{FmFp-intermed}
that the coefficient $\binom{p}{4}$ in front of $I_{p-4}$ vanishes
for $p=1$ and $p=3$, and similarly $\binom{p}{2}$ in front of $I_{p-2}$
vanishes for $p=1$, thus sparing the expansion of the respective integrals
via \eqref{Ipm4} and \eqref{Ipm2}. With this consideration, \eqref{Fpm-final}
can be obtained also in these cases.

\subsubsection{Evaluation of the Outer Integral II}
\label{proof-prop-gmdisc-generic-7}

We turn now to evaluating $G_\mp$. After expanding $\omega$ in the
penaliser function and cancelling terms due to
$\nu=\kappa+\mathcal{O}(\varrho^2)$ we substitute
$\xi=\sqrt{\vphantom{X}\varrho}\,\zeta$.
Using furthermore the Taylor expansion of $\omega$ in $\xi$ direction
around $\nu\varrho$,
\begin{align}
\omega(\nu\varrho+\xi)&=\kappa\varrho+(1+2\,\beta\nu\varrho^2)\xi
+\beta\varrho\xi^2
+\mathcal{O}(\varrho^3\xi)\;,
\label{omega-taylor-shifted}
\end{align}
we obtain
\begin{align}
G_\mp 
&= 
\int\nolimits_0^{\sqrt{\vphantom{X}\varrho}\pm\nu\varrho}
W(\mp\xi+\nu\varrho)
\bigl(\mp\omega(\mp\xi+\nu\varrho)\pm\kappa\varrho\bigr)^p
\dd\xi
\notag \\
&= 
\int\nolimits_0^{\sqrt{\vphantom{X}\varrho}\pm\nu\varrho}
W(\mp\xi+\nu\varrho)
\bigl(\xi\mp\nu\varrho\mp\beta\xi^2\varrho+2\beta\xi\nu\varrho^2
+\mathcal{O}(\varrho^3\xi)
\pm\kappa\varrho\bigr)^p
\dd\xi
\notag \\
&= 
\int\nolimits_0^{\sqrt{\vphantom{X}\varrho}\pm\nu\varrho}
W(\mp\xi+\nu\varrho)
\bigl(\xi\mp\beta\xi^2\varrho+2\beta\xi\nu\varrho^2
+\mathcal{O}(\varrho^3\xi)
\bigr)^p
\dd\xi
\notag \\
&=
\sqrt{\vphantom{X}\varrho}
\int\nolimits_0^{1\pm\nu\sqrt{\vphantom{X}\varrho}}
W\bigl(\mp\zeta\sqrt{\vphantom{X}\varrho}+\nu\varrho\bigr)
\left(\zeta\sqrt{\vphantom{X}\varrho}\mp\beta\zeta^2\varrho^2
+2\beta\zeta\nu\varrho^{5/2}+\mathcal{O}(\varrho^{7/2}\zeta)
\right)^p
\dd\zeta
\notag \\
&=
\varrho^{(p+1)/2}
\int\nolimits_0^{1\pm\nu\sqrt{\vphantom{X}\varrho}}
W\bigl(\mp\zeta\sqrt{\vphantom{X}\varrho}+\nu\varrho\bigr)
\left(1\mp\beta\zeta\varrho^{3/2}
+\mathcal{O}(\varrho^2)
\right)^p
\zeta^p
\dd\zeta
\notag \\
&\stackrel{\kern-1em\eqref{Wxi-final}\kern-1em}{=}
\varrho^{(p+1)/2}
\int\nolimits_0^{1\pm\nu\sqrt{\vphantom{X}\varrho}}
\left(\bigl(
w_{0,0}
\mp w_1\varrho^{3/2}\zeta
\bigr) 
\vphantom{\sqrt{1-(\zeta\sqrt{\vphantom{X}\varrho}\mp\nu\varrho)^2}}
\sqrt{1-(\zeta\sqrt{\vphantom{X}\varrho}\mp\nu\varrho)^2}
+\mathcal{O}(\varrho^2) \right)
\notag\\&\qquad\qquad\qquad{}\times
\left(1\mp\beta\zeta\varrho^{3/2}
+\mathcal{O}(\varrho^2)
\right)^p
\zeta^p
\dd\zeta
\notag \\
&\stackrel{\kern-1em\eqref{w0thru3}\kern-1em}{=}
\varrho^{(p+1)/2}
\int\nolimits_0^{1\pm\nu\sqrt{\vphantom{X}\varrho}}
\biggl(
2
\bigl(
1 \mp \delta\varrho^{3/2}\zeta
\bigr) 
\left(
1-\frac12\zeta^2\varrho\pm\zeta\nu\varrho^{3/2}
\right)
\left(1\mp p\beta\zeta\varrho^{3/2}\right)
+\mathcal{O}(\varrho^2)
\biggr)
\zeta^p
\dd\zeta
\notag \\
&=
\varrho^{(p+1)/2}
\int\nolimits_0^{1\pm\nu\sqrt{\vphantom{X}\varrho}}
2
\biggl(
1 
\mp \delta\varrho^{3/2}\zeta
-\frac12\zeta^2\varrho\pm\zeta\nu\varrho^{3/2}
\mp p\beta\zeta\varrho^{3/2}
\biggr)
\zeta^p
\dd\zeta
+\mathcal{O}(\varrho^{(p+5)/2})
\notag \\
&=
2\varrho^{(p+1)/2}
\left(
\int\nolimits_0^{1\pm\nu\sqrt{\vphantom{X}\varrho}} \zeta^p \dd\zeta
\mp
(\delta+p\beta) \varrho^{3/2}
\int\nolimits_0^{1\pm\nu\sqrt{\vphantom{X}\varrho}} \zeta^{p+1} \dd\zeta
-\frac12
\varrho
\int\nolimits_0^{1\pm\nu\sqrt{\vphantom{X}\varrho}} \zeta^{p+2} \dd\zeta
\right)
\notag\\&\quad{}
+\mathcal{O}(\varrho^{(p+5)/2})
\notag \\
&=
2\varrho^{(p+1)/2}
\left(
\frac1{p+1}(1\pm\nu\sqrt{\vphantom{X}\varrho})^{p+1}
\mp
\frac1{p+2}
(\delta+p\beta) \varrho^{3/2}
(1\pm\nu\sqrt{\vphantom{X}\varrho})^{p+2}
\right.\notag\\&\qquad\qquad\qquad\left.{}
-
\frac1{2(p+3)}
\varrho
(1\pm\nu\sqrt{\vphantom{X}\varrho})^{p+3}
\right)
+\mathcal{O}(\varrho^{(p+5)/2})
\notag \\
&=
2\varrho^{(p+1)/2}
\left(
\frac1{p+1}
\left(
1\pm(p+1)\nu\sqrt{\vphantom{X}\varrho}
+\binom{p+1}{2}\nu^2\varrho
\pm\binom{p+1}{3}\nu^3\varrho^{3/2}
\right)
\right.\notag\\&\qquad\qquad\qquad\left.{}\vphantom{\binom{p+1}{2}}
\mp
\frac1{p+2}
(\delta+p\beta) \varrho^{3/2}
-
\frac1{2(p+3)}
\varrho
\pm(p+3)\nu\varrho^{3/2}
\right)
+\mathcal{O}(\varrho^{(p+5)/2})
\\
G_-&+G_+
=
2\varrho^{(p+1)/2}
\left(
\frac2{p+1}
+p\nu^2\varrho
-
\frac1{p+3}
\varrho
\right)
+\mathcal{O}(\varrho^{(p+5)/2})\;.
\label{Gpm-final}
\end{align}

\subsubsection{Extremum of the Combined Integral}

Combining \eqref{Ekappa-split}, \eqref{Fpm-final} and \eqref{Gpm-final},
applying \eqref{w0thru3} and $\nu=\kappa+\mathcal{O}(\varrho^2)$
we obtain
\begin{align}
E(\kappa)
&=
\frac{4}{p+1}
S_{p+2}
+
\left(
2\frac{(p+2)p}{p+1}\kappa^2
-2\frac{2p(p-1)}{p+1}\beta\kappa
+\left(\frac43\beta\delta+\frac23\gamma^2-\frac23\varepsilon_2+2\psi_0\right)
\frac{1}{p+1}
\right.\notag\\&\qquad\qquad\left.{}
-2\delta\frac{p}{p+1}\kappa
+2\frac{2p}{p+4}\varepsilon_0
+2\frac{p(p-1)}{p+4}\beta^2
\right.\notag\\&\qquad\qquad\left.{}
+\left(-2\gamma^2-\frac43\beta\delta-\frac23\gamma^2+\frac23\varepsilon_2
+2\psi_2\right)
\frac{2}{p+4}
+2\delta\frac{2p}{p+4}\beta
\right)
\varrho^2 S_{p+2}
\notag\\&\quad{}
-
\frac{4}{p+1}
\varrho^{(p+1)/2}\sqrt{1-\varrho}
-
2p\kappa^2
\varrho^{(p+3)/2}\sqrt{1-\varrho}
\notag\\*&\quad{}
+2\varrho^{(p+1)/2}
\left(
\frac2{p+1}
+p\kappa^2\varrho
-
\frac1{p+3}
\varrho
\right)
+\mathcal{O}(\varrho^{5/2})
+\mathcal{O}(\varrho^{(p+5)/2})
\notag \\
&=
\mathrm{const}(\kappa)
\notag\\&\quad{}
+ \left(
-\frac{-4p(p-1)}{p+1}\beta\varrho^2S_{p+2}
-\frac{2p}{p+1}\delta\varrho^2S_{p+2}
\right) \kappa
\notag\\&\quad{}
+ \left(
\frac{2(p+2)p}{p+1}\varrho^2S_{p+2}
-2p\varrho^{(p+3)/2}\sqrt{1-\varrho}
+2p\varrho^{(p+3)/2}
\right) \kappa^2
\notag\\&\quad{}
+ \mathcal{O}\bigl(\varrho^{\min\{(p+5)/2,5/2\}}\bigr)
\notag \\
&=
\mathrm{const}(\kappa)
+ \mathcal{O}\bigl(\varrho^{\min\{(p+5)/2,5/2\}}\bigr)
\notag\\&\quad{}
+ \left(
-\frac{-4p(p-1)}{p+1}\beta\varrho^2S_{p+2}
-\frac{2p}{p+1}\delta\varrho^2S_{p+2}
\right) \kappa
+ \left(
\frac{2(p+2)p}{p+1}\varrho^2S_{p+2}
\right) \kappa^2 
\;.
\label{quadfnkappa}
\end{align}
It is worth noting that the $\varrho^{(p+3)/2}$ contributions cancelling
in the last step belong to the integration boundaries of $F_{\mp}$ and
$G_{\mp}$ at $\pm\sqrt{\vphantom{X}\varrho}$.

For $\varrho\to0$, the last expression \eqref{quadfnkappa}
is a quadratic function of $\kappa$ with its apex at
\begin{align}
\kappa 
&= - \frac{-\frac{-4p(p-1)}{p+1}\beta\varrho^2S_{p+2}
-\frac{2p}{p+1}\delta\varrho^2S_{p+2}} {2\frac{2(p+2)p}{p+1}\varrho^2S_{p+2}}
+ \mathcal{O}(\varrho^{\min\{(p+1)/2,1/2\}})
\notag \\
&= \frac{p-1}{p+2}\beta + \frac{1}{p+2}\delta 
+ \mathcal{O}(\varrho^{\min\{(p+1)/2,1/2\}})
\;.
\label{kappaapex}
\end{align}
Due to the sign of the $\kappa^2$ coefficient in \eqref{quadfnkappa} the
apex is a minimum for $p>0$ and a maximum for $p<0$. The $\sgn(p)$ factor
in the original energy function $E_0$ compensates for this such that
$E_0$ is always minimised.

\subsubsection{Conclusion for Regular Points}

From \eqref{kappaapex}
the claim of the proposition for regular points follows by substituting back
$\kappa\alpha\varrho^2=\mu$, $\alpha\beta=u_{xx}/2$,
$\alpha\delta=u_{yy}/2$, and noticing that by our ansatz $u_x>0$, $u_y=0$
the coordinates $x$, $y$ conincide with the geometric coordinates
$\eta$, $\xi$ as used in the proposition.

\subsubsection{Critical Points}

The inequalities for local minima (maxima) are obvious consequences of the
fact that for any $\varrho>0$ the mean-$p$ filter value is in the convex
hull of values $u(\boldsymbol{x})$,
$\boldsymbol{x}\in\mathrm{D}_\varrho(\boldsymbol{x}_0)$.
\hfill$\Box$

\subsection{Proof of Proposition~\ref{prop-pde-gmdisc-mode}}
\label{proof-prop-pde-gmdisc-mode}

With the same substitutions as in the previous proof, the mode of $\omega$
is given by the maximiser of $V(\xi)$.
By a slight modification of the calculations of the previous proof one finds
\begin{equation}
V(\xi) = 2\, (1+\delta\xi\varrho-2\beta\xi\varrho)\sqrt{1-\xi^2}
  +\mathcal{O}(\xi^2\varrho^2) \,.
\end{equation}
Equating $V'(\xi)$ to zero yields
$\omega(\xi)=(\delta-2\beta)\varrho+\mathcal{O}(\varrho^2)$ for the mode.
For local minima (maxima), the same reasoning as in the previous
proof applies.
\hfill$\Box$

\subsection{Proof of Proposition~\ref{prop-pde1d-p}}

We proceed largely analogous to the proof of 
Proposition~\ref{prop-pde-gmdisc-generic} in 
Appendix~\ref{proof-prop-pde-gmdisc-generic}.
However, the integral decomposition gets simpler since no infinite ascents
of the weighting at the integral boundaries $\pm1$ need to be controlled.

\subsubsection{Regular Points: Ansatz via Taylor Expansion}

Let the signal $u$ and mean order $p$ be given as in the proposition.
Assume w.l.o.g.\ that the regular location $x_0$ is $0$
with $u(0)=0$, and that the derivative of $u$ at $0$ is positive, $u_x(0)>0$.
Let a neighbourhood radius $\varrho$ be given.

Using Taylor expansion of $u$ up to third order, we obtain for
$-\varrho\le x\le\varrho$ the following expression:
\begin{align}
u(x) &= \alpha(x + \beta x^2 + \varepsilon x^3)
+ \mathcal{O}(x^4)
\label{u-taylor_1D}
\end{align}
where $\alpha=u_x$, $2\beta = u_{xx}/u_x$.

We assume that $\varrho$ is chosen small enough so $u_x$ is positive 
throughout $[-\varrho,\varrho]$, i.e., $u$ is strictly monotonic within this
interval.
The order-$p$ mean of $u$ within $[-\varrho,\varrho]$ is the minimiser of
\begin{equation}
E_0(\mu) := \sgn(p)
\int\nolimits_\varrho^\varrho\lvert u(x)-\mu\rvert^p\dd x \;.
\end{equation}
By rough estimates one can again conclude that for $\varrho\to0$,
$\mu\sim\varrho^2$.
We substitute therefore
\begin{gather}
x = \varrho\xi\;, \quad
\mu = \varrho^2\alpha\kappa\;, 
\quad
u(x) = \varrho\alpha\omega(\xi)
\label{D1substitution_1D}
\end{gather}
and obtain
\begin{align}
E_0(\mu) &= \sgn(p) \varrho^{p+1}\alpha^p E(\kappa) \;,
\\
E(\kappa) &= \int\nolimits_\varrho^\varrho\lvert\omega-\kappa\varrho\rvert^p
\dd\xi \;,
\label{Ekappa_1D}
\\
\omega(\xi)&=\xi+\beta\xi^2\varrho
+\varepsilon\xi^3\varrho^2
+\mathcal{O}(\varrho^3\xi)
\;.
\label{omega-taylor_1D}
\end{align}
In the following we focus therefore on finding the extremum of $E$
(minimum for $p>0$, maximum for $p<0$).

\subsubsection{Domain Splitting of the Integral}

We split the integral~\eqref{Ekappa_1D} into two parts, using again the
location $\xi=\nu\varrho$ where $\omega(\nu\varrho)=\kappa\varrho$ as
splitting point. By \eqref{omega-taylor_1D}, one has 
$\nu=\kappa+\mathcal{O}(\varrho^2)$.
We have then
\begin{align}
E(\kappa) &= F_-(\kappa)+F_+(\kappa)\;,
\\
F_-(\kappa)&=
\int\nolimits_{-1}^{\nu\varrho}
\bigl(\kappa\varrho-\omega(\xi)\bigr)^p\dd\xi
\;,
\\
F_+(\kappa)&=
\int\nolimits_{\nu\varrho}^1
\bigl(\omega(\xi)-\kappa\varrho\bigr)^p\dd\xi
\;.
\end{align}
By substituting the integration variables, one obtains
\begin{align}
F_{\mp} &= \int\nolimits_0^{1\pm\nu\varrho}
\bigl(\mp\omega(\mp\xi+\nu\varrho)\pm\kappa\varrho\bigr)^p\dd\xi
\label{Fmp_1D}
\end{align}
where again the upper and lower signs refer to $F_-$ and $F_+$, respectively.

\subsubsection{Evaluation of the Integrals}

The Taylor expansion for $\omega$ around $\nu\varrho$ is identical
with \eqref{omega-taylor-shifted}.
Inserting this into \eqref{Fmp_1D}, we have further
\begin{align}
F_{\mp} 
&= \int\nolimits_0^{1\pm\nu\varrho}
\bigl((1+2\,\beta\nu\varrho^2)\xi\mp\beta\varrho\xi^2+\mathcal{O}(\varrho^3\xi)
\bigr)^p\dd\xi
\notag\\
&= \int\nolimits_0^{1\pm\nu\varrho}
\xi^p\bigl(1+2\,\beta\nu\varrho^2\mp\beta\varrho\xi+\mathcal{O}(\varrho^3)
\bigr)^p\dd\xi
\notag\\
&= \int\nolimits_0^{1\pm\nu\varrho}
\xi^p\biggl(1+2\,p\beta\nu\varrho^2\mp p\beta\varrho\xi+
\frac{p(p-1)}2\beta^2\varrho^2\xi^2
+\mathcal{O}(\varrho^3)
\biggr)\dd\xi
\notag\\
&= 
\bigl(1+2\,p\beta\nu\varrho^2+\mathcal{O}(\varrho^3)\bigr)
\int\nolimits_0^{1\pm\nu\varrho} \xi^p\dd\xi
\mp p\beta\varrho
\int\nolimits_0^{1\pm\nu\varrho} \xi^{p+1}\dd\xi
\notag\\&\quad{}
+ \frac{p(p-1)}2\beta^2\varrho^2
\int\nolimits_0^{1\pm\nu\varrho} \xi^{p+2}\dd\xi
\;,
\end{align}
from which by evaluating the standard integrals, adding $F_-$ and $F_+$
and inserting $\nu=\kappa+\mathcal{O}(\varrho^2)$ we reach
\begin{align}
E(\kappa) &= F_-(\kappa)+F_+(\kappa)
= \mathrm{const}(\kappa) 
+ p\varrho^2\left(\kappa^2-2\frac{p-1}{p+1}\beta\kappa\right)
+ \mathcal{O}(\varrho^3) \;.
\end{align}
The extremum of $E$ is again found as the apex of the quadratic function on
the r.h.s., from which the claim for regular points follows.

For critical points, the reasoning from 
Appendix~\ref{proof-prop-pde-gmdisc-generic} applies.
\hfill$\Box$

\subsection{Proof of Proposition~\ref{prop-pde1d-mode}}

Assuming again that the regular location for the signal $u$ is $x_0=0$,
and $u$ is strictly monotonically increasing and Lipschitz
within $[-\varrho,\varrho]$,
the density of each value $u(x)$ for $-\varrho\le x\le\varrho$ is
proportional to $1/u'(x)$. The maximum of these values is reached at
$u(-\varrho)$ if $u$ is convex, or $u(\varrho)$ if $u$ is concave. 
This proves the claim for regular points. If $x_0$ is a local extremum,
the density has a pole at $u(x_0)$ and is finite
for all other values, making $u(x_0)$ the mode.
\hfill$\Box$

\subsection{Proof of Proposition~\ref{prop-pde3d-p}}
\label{proof-prop-pde3d-p}

\subsubsection{Regular Points: Ansatz via Taylor Expansion}

Let the volume image $u$ and mean order $p$ be given as in the proposition.
Assume w.l.o.g.\ that
the regular location $\boldsymbol{x}_0$ is $(0,0,0)$ with $u(0,0,0)=0$, 
and that the gradient of $u$ at $(0,0,0)$ is in the positive $x$ direction, 
i.e., $u_x>0$, $u_y=u_z=0$.
Let a neighbourhood radius $\varrho>0$ be given, and denote the
closed (Euclidean) $\varrho$-neighbourhood of $(0,0,0)$ by
$\mathrm{B}_{\varrho}$.

Using Taylor expansion of $u$ up to third order, we can write
for $(x,y,z)\in\mathrm{B}_\varrho$ the ansatz
\begin{align}
u(x,y,z) &= \alpha\bigl(x + \beta x^2 + \gamma_0 xy + \gamma_1 xz
+ \delta_0 y^2 + \delta_1 yz + \delta_2 z^2
+ \varepsilon_0x^3+\varepsilon_{10}x^2y+\varepsilon_{01}x^2z
\notag\\*&\qquad{}
+\varepsilon_{20}xy^2+\varepsilon_{11}xyz
+\varepsilon_{02}xz^2
+\varepsilon_{30}y^3+\varepsilon_{21}y^2z+\varepsilon_{12}yz^2
+\varepsilon_{03}z^3\bigr)
\notag\\*&\quad{}
+ \mathcal{O}\bigl((x+y+z)^4\bigr)
\;.
\label{u-taylor_3D}
\end{align}

We assume that $\varrho$ is chosen small enough such that $u_x$ is positive
everywhere in $\mathrm{B}_\varrho$,
each level set of $u$ within the ball
$\mathrm{B}_\varrho$ is either a smooth surface patch bounded by a 
closed regular curve on the boundary of the ball,
or one of two single points on the boundary
of $\mathrm{B}_\varrho$ where $u$ takes its maximum and minimum on
$\mathrm{B}_\varrho$, respectively.

The order-$p$ mean of $u$ within $\mathrm{B}_\varrho$ is the minimiser of
\begin{equation}
E_0(\mu) := \sgn(p)
\iiint\nolimits_{\mathrm{B}_\varrho}\lvert u(x,y,z)-\mu\rvert^p\dd z\dd y\dd x
\;.
\end{equation}
Rough estimates again ensure $\mu\sim\varrho^2$ for $\varrho\to0$.
Combining an appropriate rescaling with a transition to cylindrical coordinates
with the axis in gradient ($x$) direction, we substitute
\begin{gather}
x = \varrho\xi\;, \quad
y = \varrho\eta\cos\varphi\;, \quad
z = \varrho\eta\sin\varphi\;, \quad
\mu = \varrho^2\alpha\kappa\;,
\label{D1substitution_3D_1}
\\
u(x,y,z) = \varrho\alpha\omega(\xi,\eta,\varphi)
\label{D1substitution_3D_2}
\end{gather}
and obtain
\begin{align}
E_0(\mu) &= \sgn(p) \varrho^{p+3}\alpha^p E(\kappa) \;,
\\
E(\kappa) &= \iint\nolimits_{\mathrm{D}_1}\int\nolimits_{0}^{2\pi}
\lvert\omega(\xi,\eta,\varphi)-\kappa\varrho\rvert^p \eta
\dd\varphi\dd\eta\dd\xi \;,
\label{Ekappa3D}
\end{align}
where the integration in cylindrical coordinates has been written using
the disc $\mathrm{D}_1$ for the $\xi$, $\eta$ coordinates.
The Taylor expansion of $u$ transfers to
\begin{align}
\omega(\xi,\eta,\varphi)
&=\xi+\beta\xi^2\varrho+\gamma(\varphi)\xi\eta\varrho
+\delta(\varphi)\eta^2\varrho
\notag\\*&\quad{}
+\varepsilon_0\xi^3\varrho^2+\varepsilon_1(\varphi)\xi^2\eta\varrho^2
+\varepsilon_2(\varphi)\xi\eta^2\varrho^2
+\varepsilon_3(\varphi)\eta^3\varrho^2
+\mathcal{O}\bigl(\varrho^3(\xi+\eta)\bigr)
\label{omega-taylor_3D}
\;,
\\
\gamma(\varphi)&:=\gamma_0\cos\varphi+\gamma_1\sin\varphi\;,\\
\delta(\varphi)&:=\delta_0\cos^2\varphi+\delta_1\cos\varphi\sin\varphi
+\delta_2\sin^2\varphi\;,\\
\varepsilon_1(\varphi)&:=\varepsilon_{10}\cos\varphi
+\varepsilon_{01}\sin\varphi\;,\\
\varepsilon_2(\varphi)&:=\varepsilon_{20}\cos^2\varphi
+\varepsilon_{11}\cos\varphi\sin\varphi+\varepsilon_{02}\sin^2\varphi\;,\\
\varepsilon_3(\varphi)&:=\varepsilon_{30}\cos^3\varphi
+\varepsilon_{21}\cos^2\varphi\sin\varphi
+\varepsilon_{12}\cos\varphi\sin^2\varphi
+\varepsilon_{03}\sin^3\varphi\;.
\end{align}
We aim again at finding the extremum of $E$.

\subsubsection{Separation of the Integral}

Similar to Appendix~\ref{proof-prop-gmdisc-generic-3},
the integral $E$ from \eqref{Ekappa3D} can be reorganised into a nested
integration where the inner double integral (in polar coordinates) integrates
over a level surface of $\omega$ going through $(\xi,0,0)$, and the outer 
integral then integrates along the $\xi$ axis. We have
\begin{align}
E(\kappa) &= \int\nolimits_{-1}^{1} 
\left(\,
\int\nolimits_0^{2\pi}
\int\nolimits_0^{\eta^*(\varphi)}
\frac{\eta}
     {\frac{\partial\omega}{\partial\xi}
      \bigl(\tilde{\xi}(\eta),\eta,\varphi\bigr)}
\dd\eta
\dd\varphi
\right) 
\lvert\omega(\xi,0,0)-\kappa\varrho\rvert^p
\frac{\partial\omega}{\partial\xi}(\xi,0,0)
\dd\xi
\notag\\*&\quad{}
+\mathcal{O}(\varrho^3)
\label{EkappaLL_3D}
\end{align}
where $\tilde{\xi}$ is a function of $\eta,\varphi$ that describes the 
level set of $\omega$ which goes through $(\xi,0,0)$, and reaches the 
boundary of $\mathrm{B}_1$ at $(\eta^*(\varphi),\varphi)$. 
(Note that our initial assumptions on $u$ ensure that the level set 
can be described in this way.)

Analogously to Appendix~\ref{proof-prop-gmdisc-generic-3} we rewrite
\eqref{EkappaLL_3D} as 
\begin{align}
E(\kappa) &=
\int\nolimits_{-1}^1W(\xi)\dd\xi \;,
\\
W(\xi) &:=
\int\nolimits_0^{2\pi}
\int\nolimits_0^{\eta^*(\varphi)}
\eta\,
\frac{\frac{\partial\omega}{\partial\xi}(\xi,0,0)}
     {\frac{\partial\omega}{\partial\xi}
      \bigl(\tilde{\xi}(\eta),\eta,\varphi\bigr)}
\dd\eta
\dd\varphi
\;.
\label{EkappaLLi_3D}
\end{align}

\subsubsection{Evaluation of the Weight Integral}

Within any axial plane ($\varphi=\mathrm{const}$), \eqref{omega-taylor_3D}
is exactly \eqref{omega-taylor}.  
We can therefore transfer verbatim the analysis from
Appendix~\ref{proof-prop-gmdisc-generic-4}, which leads
to the expression \eqref{xiLL} for $\tilde{\xi}$,
the expression for $\eta^*_+$ from \eqref{etapm} for $\eta^*(\varphi)$, 
and \eqref{domega-dxi-axisLLquotient} for
$\frac{\partial\omega}{\partial\xi}(\xi,0,0)/\frac{\partial\omega}{\partial\xi}
(\tilde{\xi},\eta,\varphi)$.

Inserting \eqref{domega-dxi-axisLLquotient} into the inner integral
of \eqref{EkappaLLi_3D} leads to
\begin{align}
W(\xi,\varphi) &:=
\int\nolimits_0^{\eta^*(\varphi)}
\eta\,
\frac{\frac{\partial\omega}{\partial\xi}(\xi,0,0)}
     {\frac{\partial\omega}{\partial\xi}
      \bigl(\tilde{\xi}(\eta),\eta,\varphi\bigr)}
\dd\eta
\notag \\
&=
\int\nolimits_0^{\eta^*(\varphi)}\!\!
\eta\dd\eta
+
\bigl(-\gamma(\varphi)\varrho+2\beta\gamma(\varphi)\xi\varrho^2
-2\varepsilon_1(\varphi)\xi\varrho^2\bigr)
\int\nolimits_0^{\eta^*}\!\!\eta^2\dd\eta
\notag\\&\quad{}
+
\bigl(2\beta\delta(\varphi)+\gamma(\varphi)^2-\varepsilon_2(\varphi)\bigr)
\varrho^2
\int\nolimits_0^{\eta^*}\!\!\eta^3\dd\eta
+ \mathcal{O} (\varrho^3)
\notag\\
&=
\frac12
\eta^*(\varphi)^2
+
\frac16
(-\gamma_0+2\beta\gamma_0\xi\varrho-2\varepsilon_{10}\xi\varrho)
\varrho
\eta^*(\varphi)^3\cos\varphi
\notag\\&\qquad\qquad\quad{}
+
\frac16
(-\gamma_1+2\beta\gamma_1\xi\varrho-2\varepsilon_{01}\xi\varrho)
\varrho
\eta^*(\varphi)^3\sin\varphi
\notag\\&\qquad\qquad\quad{}
+
\frac14
(2\beta\delta_0+\gamma_0^2-\varepsilon_{20})\varrho^2
\eta^*(\varphi)^4\cos^2\varphi
\notag\\&\qquad\qquad\quad{}
+
\frac12
(2\beta\delta_1+\gamma_0\gamma_1-\varepsilon_{11})\varrho^2
\eta^*(\varphi)^4\cos\varphi\sin\varphi
\notag\\&\qquad\qquad\quad{}
+
\frac14
(2\beta\delta_2+\gamma_1^2-\varepsilon_{02})\varrho^2
\eta^*(\varphi)^4\sin^2\varphi
+ \mathcal{O} (\varrho^3)
\;.
\label{Wxiphi_3D}
\end{align}
To finally obtain $W(\xi)$, the latter expression needs to be integrated
over $\varphi$.
From \eqref{etapm} one obtains by lengthy but straightforward calculation
\begin{align}
\int\nolimits_0^{2\pi}\eta^*(\varphi)^2\dd\varphi
&= \pi\Bigl(
2(1-\xi^2)
+2(\delta_0+\delta_2)(1-\xi^2)\xi\varrho
+(\gamma_0^2+\gamma_1^2)\xi^4\varrho^2
\notag\\&\qquad{}
+\tfrac14(3\delta_0^2+\delta_1^2+3\delta_2^2+2\delta_0\delta_2)
+2\varPsi(\xi)(1-\xi^2)\varrho^2
)
\Bigr)
+\mathcal{O}(\varrho^3)\;,
\label{int_etastar2_3D}
\end{align}
where $\varPsi$ is a third-order polynomial in $\xi$ obtained by
integrating \eqref{polypsi} w.r.t.\ $\varphi$,
\begin{align}
\varPsi(\xi)&=\frac1\pi\int\nolimits_0^{2\pi}\psi(\xi,\varphi)\dd\varphi
=\varPsi_0+\varPsi_1\xi+\varPsi_2\xi^2+\varPsi_3\xi^3
\;.
\end{align}
Analogously one obtains
\begin{align}
\int\nolimits_0^{2\pi}\eta^*(\varphi)^3\cos\varphi\dd\varphi
&=3\pi\gamma_0(1-\xi^2)\xi^2\varrho
+\mathcal{O}(\varrho^2)
\;,\\
\int\nolimits_0^{2\pi}\eta^*(\varphi)^3\sin\varphi\dd\varphi
&=3\pi\gamma_1(1-\xi^2)\xi^2\varrho
+\mathcal{O}(\varrho^2)
\;,\\
\int\nolimits_0^{2\pi}\eta^*(\varphi)^4\cos^2\varphi\dd\varphi
&=\pi(1-\xi^2)^2
+\mathcal{O}(\varrho)
\;,\\
\int\nolimits_0^{2\pi}\eta^*(\varphi)^4\cos\varphi\sin\varphi\dd\varphi
&=0
+\mathcal{O}(\varrho)
\;,\\
\int\nolimits_0^{2\pi}\eta^*(\varphi)^4\sin^2\varphi\dd\varphi
&=\pi(1-\xi^2)^2
+\mathcal{O}(\varrho)
\;.
\label{int_etastar4ss_3D}
\end{align}
Inserting \eqref{Wxiphi_3D}--\eqref{int_etastar4ss_3D} into
\eqref{EkappaLLi_3D} yields after sorting terms, similarly to
\eqref{Wxi-final}, \eqref{w0thru3},
\begin{align}
W(\xi) &= \Bigl(
(w_{0,0} + w_{0,2}\varrho^2)
+(w_{1,1} + w_{1,2}\varrho)\varrho\xi
+(w_{2,0} + w_{2,2}\varrho^2)\xi^2
\notag\\&\qquad{}
+(w_{3,1} + w_{3,2}\varrho)\varrho\xi^3
+w_{4,2}\varrho^2\xi^4
+w_{5,2}\varrho^2\xi^5
\Bigr)\pi
+\mathcal{O}(\varrho^3)
\label{Wxi-final_3D}
\end{align}
with
\begin{equation}
\left.
\begin{aligned}
w_{0,0} &= %
    1 
\;, \\
w_{0,2} &= %
    \tfrac12\beta(\delta_0+\delta_2)
  + \tfrac14(\gamma_0^2+\gamma_1^2)
  - \tfrac14(\varepsilon_{20}+\varepsilon_{02})
  + \varPsi_0
\;, \\
w_{1,1} &= %
    \delta_0+\delta_2
\;, \\
w_{1,2} &= %
    \varPsi_1
\;, \\
w_{2,0} &= %
  - 1
\;, \\
w_{2,2} &= %
  - \beta(\delta_0+\delta_2)
  - \tfrac32(\gamma_0^2+\gamma_1^2)
  + \tfrac12(\varepsilon_{20}+\varepsilon_{02})
  + \tfrac18(3\delta_0^2+\delta_1^2+3\delta_2^2+2\delta_0\delta_2)
\\&\quad{}
  + \varPsi_2 - \varPsi_0
\;, \\
w_{3,1} &= %
  - (\delta_0+\delta_2)
\;, \\
w_{3,2} &= %
    \varPsi_3 - \varPsi_1
\;, \\
w_{4,2} &= %
    \tfrac12\beta(\delta_0+\delta_2)
  + \tfrac94(\gamma_0^2+\gamma_1^2)
  - \tfrac14(\varepsilon_{20}+\varepsilon_{02})
  - \tfrac18(3\delta_0^2+\delta_1^2+3\delta_2^2+2\delta_0\delta_2)
  - \varPsi_2
\;,\kern-1.0em \\
w_{5,2} &= %
  - \varPsi_3 \;.
\end{aligned}
\quad
\right\}
\label{w0thru5_3D}
\end{equation}

\subsubsection{Domain Splitting of the Outer Integral}

As the outer integral of \eqref{EkappaLL_3D} has the same structure as in
the 2D case, we use the same domain splitting \eqref{Ekappa-split}.

\subsubsection{Evaluation of the Outer Integral I}

The first steps in evaluating the integrals $F_{\mp}$ are as in the 2D case.
In \eqref{FmFp-intermed1}, the longer expansion \eqref{Wxi-final_3D} has
to be used for $W(\mp)$, which then leads to
\begin{align}
F_{\mp} 
&= \int\nolimits_{\sqrt{\vphantom{X}\varrho}}^1
\Bigl(
(w_{0,0} + w_{0,2}\varrho^2)
+(w_{1,1} + w_{1,2}\varrho)\varrho\xi
+(w_{2,0} + w_{2,2}\varrho^2)\xi^2
\notag\\&\qquad\qquad{}
+(w_{3,1} + w_{3,2}\varrho)\varrho\xi^3
+w_{4,2}\varrho^2\xi^4
+w_{5,2}\varrho^2\xi^5
+\mathcal{O}(\varrho^3)
\Bigr)\pi
\xi^p %
\notag\\&\qquad{}\times
\left(1
\pm p\kappa\frac{\varrho}{\xi} %
\mp p\beta\xi\varrho %
+\binom{p}{2}\kappa^2\frac{\varrho^2}{\xi^2} %
\mp\binom{p}{3}\kappa^3\frac{\varrho^3}{\xi^3} %
+p\varepsilon_0\xi^2\varrho^2 %
-2\binom{p}{2}\beta\kappa\varrho^2 %
\right.\notag\\&\qquad\quad\left.{}
+\binom{p}{2}\beta^2\xi^2\varrho^2 %
+\binom{p}{4}\kappa^4\frac{\varrho^4}{\xi^4} %
+ \mathcal{O}(\varrho^{5/2})
\right)
\dd\xi \;,
\label{FmFp-intermed1_3D}
\end{align}
which yields
\begin{align}
F_-+F_+ &=
2\pi\int\nolimits_{\sqrt{\vphantom{X}\varrho}}^1
(w_{0,0} + w_{2,0}\xi^2)
\xi^p
\left(1+\binom{p}{2}\kappa^2\frac{\varrho^2}{\xi^2} %
+p\varepsilon_0\xi^2\varrho^2 %
-2\binom{p}{2}\beta\kappa\varrho^2 %
\right.\notag\\&\qquad\qquad\qquad\qquad\qquad\qquad\left.{}
+\binom{p}{2}\beta^2\xi^2\varrho^2 %
+\binom{p}{4}\kappa^4\frac{\varrho^4}{\xi^4} %
+ \mathcal{O}(\varrho^{5/2})
\right)
\dd\xi
\notag \\ &\quad{} 
+2\pi\int\nolimits_{\sqrt{\vphantom{X}\varrho}}^1
( w_{0,2} + w_{2,2} \xi^2 + w_{4,2} \xi^4 ) \varrho^2
\xi^p
\left(1
+ \mathcal{O}(\varrho^{1/2})
\right)
\dd\xi
\notag\\&\quad{} 
+2\pi\int\nolimits_{\sqrt{\vphantom{X}\varrho}}^1
( w_{1,1} \xi + w_{3,1} \xi^3 ) \varrho
\xi^p
\left(
-p\kappa\frac{\varrho}{\xi} %
+p\beta\xi\varrho %
+ \mathcal{O}(\varrho^{3/2})
\right)
\dd\xi
\notag\\&\quad{} 
+2\pi\int\nolimits_{\sqrt{\vphantom{X}\varrho}}^1
( w_{1,2} \xi + w_{3,2} \xi^3 + w_{5,2} \xi^5 ) \varrho^2
\xi^p
\left(
-p\kappa\frac{\varrho}{\xi} %
+ \mathcal{O}(\varrho^{1/2})
\right)
\dd\xi
\;.
\label{Fpm-intermed1_3D}
\end{align}
Using the abbreviation
\begin{align}
J_q &:= \int\nolimits_{\sqrt{\varrho}}^1\xi^q\dd\xi  \;,
\end{align}
we can sort this into
\begin{align}
F_-+F_+
&= 
2\pi\Biggl(
  w_{0,0} \binom{p}{2}\kappa^2 \varrho^2 J_{p-2} 
\notag\\&\qquad\quad{}
+ \biggl(
  w_{0,0}
+ \Bigl(
- 2w_{0,0} \binom{p}{2}\beta\kappa
+ w_{2,0} \binom{p}{2}\kappa^2
+ w_{0,2} 
- w_{1,1} p\kappa \Bigr) \varrho^2
\biggr) J_p
\notag\\&\qquad\quad{}
+ \biggl(
  w_{2,0}
+ \Bigl(
  w_{0,0} p\varepsilon_0
+ w_{0,0} \binom{p}{2}\beta^2
- 2w_{2,0} \binom{p}{2}\beta\kappa 
+ w_{2,2}
\notag\\&\quad\qquad\qquad\qquad\qquad{}
+ w_{1,1} p\beta 
- w_{3,1} p\kappa \Bigr) \varrho^2
\biggr) J_{p+2}
\notag\\&\qquad\quad{}
+ \biggl(
  w_{2,0} p\varepsilon_0 
+ w_{2,0} \binom{p}{2}\beta^2 
+ w_{4,2}
+ w_{3,1} p\beta \biggr) \varrho^2 J_{p+4} 
\Biggr)
+ \mathcal{O}(\varrho^{5/2})
\;,
\label{Fpm-intermed_3D}
\end{align}
which by $J_q= \tfrac1{q+1}(1-\varrho^{(q+1)/2})$ for $q\ne-1$ 
(the special case $J_{-1}$ only occurs as $J_{p-2}$ for $p=1$ and
has then a vanishing coefficient) yields
\begin{align}
F_-+F_+
&= 
2\pi\Biggl(
  w_{0,0}\frac1{p+1}
+ w_{2,0}\frac1{p+3}
\notag\\&\quad\qquad{}
+ \biggl(
  w_{0,0} \frac{p}2\kappa^2
- 2w_{0,0} \frac{p(p-1)}{2(p+1)}\beta\kappa
+ w_{2,0} \frac{p(p-1)}{2(p+1)}\kappa^2
+ w_{0,2} 
- w_{1,1} \frac{p}{p+1}\kappa 
\notag\\&\qquad\qquad\qquad{}
+ w_{0,0} \frac{p}{p+3}\varepsilon_0
+ w_{0,0} \frac{p(p-1)}{2(p+3)}\beta^2
- 2w_{2,0} \frac{p(p-1)}{2(p+3)}\beta\kappa 
+ w_{2,2}\frac1{p+3}
\notag\\&\qquad\qquad\qquad{}
+ w_{1,1} \frac{p}{p+3}\beta 
- w_{3,1} \frac{p}{p+3}\kappa 
+ w_{2,0} \frac{p}{p+5}\varepsilon_0 
+ w_{2,0} \frac{p(p-1)}{2(p+5)}\beta^2 
\notag\\&\qquad\qquad{}
+ w_{4,2}\frac1{p+5}
+ w_{3,1} \frac{p}{p+5}\beta 
\biggr) \varrho^2
- w_{0,0}\frac1{p+1} \varrho^{(p+1)/2}
\notag\\&\quad\qquad{}
- \biggl(
  w_{0,0} \frac{p}{2}\kappa^2
+ w_{2,0}\frac1{p+3} \biggr) \varrho^{(p+3)/2}
\Biggr)
+ \mathcal{O}(\varrho^{5/2})
+ \mathcal{O}(\varrho^{(p+5)/2})\;.
\label{Fpm-final_3D}
\end{align}

\subsubsection{Evaluation of the Outer Integral II}

Starting with the
same substitution $\xi=\sqrt{\vphantom{X}\varrho}\,\zeta$ and\
Taylor expansion of $\omega$ in $\xi$ direction as in
Appendix~\ref{proof-prop-gmdisc-generic-7}, we evaluate
\begin{align}
G_\mp 
&\stackrel{\kern-1em\eqref{Wxi-final_3D}\kern-1em}{=}
\pi\varrho^{(p+1)/2}
\int\nolimits_0^{1\pm\nu\sqrt{\vphantom{X}\varrho}}
\bigl(
w_{0,0} + w_{2,0}\varrho\zeta^2
\mp w_{1,1}\varrho^{3/2}\zeta 
\mp 2w_{2,0}\nu\varrho^{3/2}\zeta
+\mathcal{O}(\varrho^2)
\bigr) 
\notag\\&\qquad\qquad\qquad{}\times
\left(1\mp\beta\zeta\varrho^{3/2}
+\mathcal{O}(\varrho^2)
\right)^p
\zeta^p
\dd\zeta
\notag \\
&=
\pi\varrho^{(p+1)/2}w_{0,0}
\int\nolimits_0^{1\pm\nu\sqrt{\vphantom{X}\varrho}}\zeta^p\dd\zeta
\mp\pi\varrho^{(p+4)/2}
(w_{1,1}\!+\!2w_{2,0}\nu\!+\!w_{0,0}p\beta)
\int\nolimits_0^{1\pm\nu\sqrt{\vphantom{X}\varrho}}\!\zeta^{p+1}\dd\zeta
\notag\\&\quad{}
+\pi\varrho^{(p+3)/2}w_{2,0}
\int\nolimits_0^{1\pm\nu\sqrt{\vphantom{X}\varrho}}\zeta^{p+2}\dd\zeta
+\mathcal{O}(\varrho^{(p+5)/2})
\notag \\
&=
\frac{\pi\varrho^{(p+1)/2}}{p+1} w_{0,0}
(1\pm\nu\sqrt{\vphantom{X}\varrho})^{p+1}
\mp\frac{\pi\varrho^{(p+4)/2}}{p+2}
(w_{1,1}+2w_{2,0}\nu+w_{0,0}p\beta)
(1\pm\nu\sqrt{\vphantom{X}\varrho})^{p+2}
\notag\\&\quad{}
+\frac{\pi\varrho^{(p+3)/2}}{p+3} w_{2,0}
(1\pm\nu\sqrt{\vphantom{X}\varrho})^{p+3}
+\mathcal{O}(\varrho^{(p+5)/2})
\notag \\
&=
\frac{\pi\varrho^{(p+1)/2}}{p+1}w_{0,0}(1+\nu^2\varrho)
\mp\frac{\pi\varrho^{(p+4)/2}}{p+2}(w_{1,1}+2w_{2,0}\nu+w_{0,0}p\beta)
\notag\\&\quad{}
+\frac{\pi\varrho^{(p+3)/2}}{p+3}w_{2,0}
+\mathcal{O}(\varrho^{(p+5)/2})
\;,
\\
G_-+G_+
&=
2\pi\varrho^{(p+1)/2}
\left(
\frac1{p+1}w_{0,0}
+\frac{p}{2}\nu^2\varrho w_{0,0}
+\frac1{p+3}\varrho w_{2,0}
\right)
+\mathcal{O}(\varrho^{(p+5)/2})
\;.
\label{Gpm-final_3D}
\end{align}

\subsubsection{Extremum of the Combined Integral}

When we finally combine
\eqref{Ekappa-split}, \eqref{Fpm-final_3D} and \eqref{Gpm-final_3D} and
apply \eqref{w0thru5_3D} and $\nu=\kappa+\mathcal{O}(\varrho^2)$,
we observe as in the 2D case than all terms originating from $G_-+G_+$
\eqref{Gpm-final_3D} cancel, and it remains
\begin{align}
E(\kappa)
&=
\mathrm{const}(\kappa)
+ \mathcal{O}\bigl(\varrho^{\min\{(p+5)/2,5/2\}}\bigr)
+ 
\left(
  p
- \frac{p(p-1)}{p+1}
\right)
\pi\varrho^2\kappa^2
\notag\\&\quad{}
+
\left(
- \frac{2p(p-1)}{(p+1)} \beta
+ \frac{2p(p-1)}{(p+3)}\beta
- \frac{2p}{p+1} (\delta_0+\delta_2) 
+ \frac{2p}{p+3} (\delta_0+\delta_2) 
\right)
\pi\varrho^2\kappa
\notag\\
&=
\mathrm{const}(\kappa)
+ \mathcal{O}\bigl(\varrho^{\min\{(p+5)/2,5/2\}}\bigr)
\notag\\&\quad{}
+\frac{2p}{p+1}\pi\varrho^2\kappa^2
-\left(\frac{4p(p-1)}{(p+1)(p+3)}\beta
+\frac{4p}{(p+1)(p+3)}(\delta_0+\delta_2)\right)\pi\varrho^2\kappa
\;.
\label{quadfnkappa_3D}
\end{align}

For $\varrho\to0$, the extremum of $E(\kappa)$ can again be found
as the apex of the quadratic function in \eqref{quadfnkappa_3D},
which yields
\begin{align}
\kappa 
&= 
\frac{
\frac{4p(p-1)}{(p+1)(p+3)}\beta
+\frac{4p}{(p+1)(p+3)}(\delta_0+\delta_2)}
{\frac{4p}{p+1}\pi\varrho^2}
+ \mathcal{O}(\varrho^{\min\{(p+1)/2,1/2\}})\
\notag\\
&=
\frac{p-1}{p+3}\beta
+\frac{1}{p+3}(\delta_0+\delta_2)
+ \mathcal{O}(\varrho^{\min\{(p+1)/2,1/2\}})\
\;.
\label{kappaapex_3D}
\end{align}

\subsubsection{Conclusion of the Proof}

From \eqref{kappaapex_3D}
the claim of the proposition for regular points follows by substituting back
$\kappa\alpha\varrho^2=\mu$, $\alpha\beta=u_{xx}/2$,
$\alpha\delta_0=u_{yy}/2$, $\alpha\delta_2=u_{zz}/2$
and noticing that by our ansatz $u_x>0$, $u_y=u_z=0$
the coordinates $x$, $y$, $z$ coincide with the geometric coordinates
$\eta$, $\xi$, $\chi$ as used in the proposition.

For critical points, the reasoning from 
Appendix~\ref{proof-prop-pde-gmdisc-generic} applies.
\hfill$\Box$

\subsection{Proof of Proposition~\ref{prop-pde3d-mode}}
\label{proof-prop-pde3d-mode}

Analogous to Appendix~\ref{proof-prop-pde-gmdisc-mode}, we calculate
\begin{equation}
V(\xi) = \pi(1+(\delta_0+\delta_2)\xi\varrho-2\beta\xi\varrho)(1-\xi^2)
  +\mathcal{O}(\xi^2\varrho^2) \,.
\end{equation}
The relevant solution of $V'(\xi)=0$ yields up to higher order terms 
$\omega(\xi)=\tfrac12(\delta_0+\delta_2-2\beta)\varrho$.
\hfill$\Box$

\section{Continuous Order-$p$ Means and Mode: A Toy Example}
\label{app-toy}

To understand the behaviour of order-$p$ mean filters for $p>-1$, $p\ne0$
and their relation to the mode of a continuous density, we consider the
following simple example.
Let $z$ be a real random variable with (non-normalised) density
\begin{equation}
\gamma(z) = 
\begin{cases}
1-\lambda (z-m)^2\;, &-1\le z\le 1\;,\\
0 & \text{otherwise.}
\end{cases}
\end{equation}
Here, $0<m<\!\!<1$ is a fixed parameter, and $0<\lambda\le(1+m)^{-2}$ to
ensure that $\gamma(z)\ge0$ for all $z$.
Obviously, the mode of $z$ is the maximum of $\gamma$, i.e., $m$.

For any $p>-1$, $p\ne0$ the order-$p$ mean of $z$ is given by the
minimiser of $E(\mu)$ where
\begin{align}
\kern2em&\kern-2em
\sgn(p) E(\mu) = \int\nolimits_{-1}^1 \gamma(z)\,\lvert z-\mu\rvert^p\dd z
\notag \\
&= 
   \int\nolimits_{-1}^\mu \bigl(1-\lambda (z-m)^2\bigr) (-z+\mu)^p\dd z
  +\int\nolimits_\mu^1 \bigl(1-\lambda (z-m)^2\bigr) (z-\mu)^p\dd z
\notag \\
&= 
   \int\nolimits_0^{1+\mu} \bigl(1-\lambda (z-\mu+m)^2\bigr)\, z^p\dd z
+  \int\nolimits_0^{1-\mu} \bigl(1-\lambda (z+\mu-m)^2\bigr)\, z^p\dd z
\notag \\
&=
   \bigl(1-\lambda(\mu-m)^2\bigr)
     \left(
       \int\nolimits_0^{1+\mu} z^p \dd z
      +\int\nolimits_0^{1-\mu} z^p \dd z
     \right)
\notag\\&\quad{}
+  2\lambda (\mu-m)
     \left(
       \int\nolimits_0^{1+\mu} z^{p+1} \dd z
      -\int\nolimits_0^{1-\mu} z^{p+1} \dd z
     \right)
\notag\\&\quad{}
-  \lambda
     \left(
       \int\nolimits_0^{1+\mu} z^{p+2} \dd z
      +\int\nolimits_0^{1-\mu} z^{p+2} \dd z
     \right)
\notag \\
&=
   \frac1{p+1} \, \bigl(1-\lambda(\mu-m)^2\bigr)
     \left( (1+\mu)^{p+1} + (1-\mu)^{p+1} \right)
\notag\\&\quad{}
+  \frac{2\lambda}{p+2} \, (\mu-m)
     \left( (1+\mu)^{p+2} - (1-\mu)^{p+2} \right)
-  \frac{\lambda}{p+3} 
     \left( (1+\mu)^{p+3} + (1-\mu)^{p+3} \right)
\notag \\
&=
   \frac2{p+1} \, \bigl(1-\lambda(\mu-m)^2\bigr)
     \left( 1 + \binom{p+1}{2}\mu^2 + \mathcal{O}(\mu^4) \right)
\notag\\&\quad{}
+  \frac{4\lambda}{p+2} \, (\mu-m)
     \left( (p+2) \mu + \mathcal{O}(\mu^3) \right)
-  \frac{2\lambda}{p+3} 
     \left( 1 + \binom{p+3}{2}\mu^2 + \mathcal{O}(\mu^4) \right)
\notag \\
&=
   \frac2{p+1} + p\mu^2 
   - \frac{2\lambda}{p+1} \mu^2 + \frac{4\lambda}{p+1} m\mu
   - \frac{2\lambda}{p+1} m^2 
\notag\\&\quad{}
   - p\lambda m^2\mu^2
+  4\lambda\mu^2 - 4\lambda m\mu
-  \frac{2\lambda}{p+3} - (p+2)\lambda\mu^2
+ \mathcal{O}(\mu^3)
\notag \\
&= \mathrm{const}(\mu)
   + \left( \frac{4\lambda}{p+1} -4\lambda
     \right) m \mu
\notag\\&\quad{}
   + \left( p - \frac{2\lambda}{p+1} - p\lambda m^2 + 4\lambda - (p+2)\lambda
     \right) \mu^2
+ \mathcal{O}(\mu^3)
\notag \\
&= \mathrm{const}(\mu)
   + \frac{-4p\lambda}{p+1}
     m \mu
   + \left( p - \frac{p(p-1)}{p+1}\lambda - p\lambda m^2
     \right) \mu^2
+ \mathcal{O}(\mu^3)
\end{align}
from which the minimiser $\mu^*$ of $E(\mu)$ can be read off as the apex of
the quadratic function of $\mu$ as
\begin{align}
\mu^* 
&= 
-\frac{-2p\lambda}{p+1}\,m\bigg/
      \left( p - \frac{p(p-1)}{p+1}\lambda - p\lambda m^2 \right)
=
\frac{2\lambda}{(p+1)-(p-1)\lambda}\,m + \mathcal{O}(m^3) \;.
\end{align}
For any fixed $\lambda\in\bigl(0,(1+m)^{-2}\bigr)$, 
the minimiser $\mu^*$ goes to $m$ for $p\to-1$, 
but approaches $2\lambda m / (1+\lambda)<m-m^2/3$ for $p\to0$. 
Moreover, if we send $\lambda$ to $0$, making the density
more and more uniform, for any $p>-1$ we have $\mu^*\to0$ which comes as
no surprise as for a flattening out density, any penalisation where the
penaliser increases with distance will end up in the symmetry centre of the
support interval $[-1,1]$.

\end{document}